\documentclass[acmsmall,screen,table,nonacm]{acmart}
\AtBeginDocument{%
  \providecommand\BibTeX{{%
    \normalfont B\kern-0.5em{\scshape i\kern-0.25em b}\kern-0.8em\TeX}}}

\setcopyright{acmcopyright}
\copyrightyear{2024}
\acmYear{2024}
\acmDOI{XXXXXXX.XXXXXXX}

\acmConference[Preprint]{Make sure to enter the correct
  conference title from your rights confirmation emai}{November 2024}{Preprint}
\acmBooktitle{Preprint} 
\acmPrice{15.00}
\acmISBN{978-1-4503-XXXX-X/18/06}

\usepackage{listings}
\usepackage{courier}
\usepackage{xcolor}
\usepackage[shortlabels]{enumitem}

\begin{document}

\title[Howzat? Expert Judgement of Human and AI Hints]{Howzat?  Appealing to Expert Judgement for Evaluating Human and AI Next-Step Hints for Novice Programmers}
\titlenote{Howzat is a contraction of ``How's that?'', used in cricket to appeal to the umpire for a judgement on whether a batter is out.}

\author{Neil C. C. Brown}
\orcid{0000-0001-6086-2479}
\email{neil.c.c.brown@kcl.ac.uk}
\affiliation{%
    \institution{King's College London}
    \streetaddress{Strand}
    \city{London}
    \country{UK}
    \postcode{WC2R 2LS}
}

\author{Pierre Weill-Tessier}
\email{pierre.weill-tessier@kcl.ac.uk}
\orcid{0000-0002-6999-8767}
\affiliation{%
    \institution{King's College London}
    \streetaddress{Strand}
    \city{London}
    \country{UK}
    \postcode{WC2R 2LS}
}

\author{Juho Leinonen}
\email{juho.2.leinonen@aalto.fi}
\orcid{0000-0001-6829-9449}
\affiliation{
    \institution{Aalto University}
    \city{Espoo}
    \country{Finland}
}

\author{Paul Denny}
\email{paul@cs.auckland.ac.nz}
\orcid{0000-0002-5150-9806}
\affiliation{
    \institution{University of Auckland}
    \city{Auckland}
    \country{New Zealand}
}

\author{Michael K\"{o}lling}
\email{michael.kolling@kcl.ac.uk}
\orcid{0000-0003-0544-2003}
\affiliation{%
    \institution{King's College London}
    \streetaddress{Strand}
    \city{London}
    \country{UK}
    \postcode{WC2R 2LS}
}

\renewcommand{\shortauthors}{Brown et al.}

\begin{abstract}

\noindent\textbf{Motivation:} Students learning to program often reach states where they are stuck and can make no forward progress -- but this may be outside the classroom where no instructor is available to help.  In this situation, an automatically generated next-step hint can help them make forward progress and support their learning.  It is important to know what makes a good hint or a bad hint, and how to generate good hints automatically in novice programming tools, for example using Large Language Models (LLMs).

\noindent\textbf{Method and participants:} We recruited 44 Java educators from around the world to participate in an online study.  We used a set of real student code states as hint-generation scenarios.  Participants used a technique known as comparative judgement to rank a set of candidate next-step Java hints, which were generated by Large Language Models (LLMs) and by five human experienced educators.  Participants ranked the hints without being told how they were generated.  The hints were generated with no explicit detail given to the LLMs/humans on what the target task was.  Participants then filled in a survey with follow-up questions.  The ranks of the hints were analysed against a set of extracted hint characteristics using a random forest approach.

\noindent\textbf{Findings:} We found that LLMs had considerable variation in generating high quality next-step hints for programming novices, with GPT-4 outperforming other models tested. When used with a well-designed prompt, GPT-4 outperformed human experts in generating pedagogically valuable hints. A multi-stage prompt was the most effective LLM prompt.  We found that the two most important factors of a good hint were length (80--160 words being best), and reading level (US grade 9 or below being best).   Offering alternative approaches to solving the problem was considered bad, and we found no effect of sentiment.  

\noindent\textbf{Conclusions:} Automatic generation of these hints is immediately viable, given that LLMs outperformed humans -- even when the students' task is unknown.  Hint length and reading level were more important than several pedagogical features of hints.  The fact that it took a group of experts several rounds of experimentation and refinement to design a prompt that achieves this outcome suggests that students on their own are unlikely to be able to produce the same benefit. The prompting task, therefore, should be embedded in an expert-designed tool.
\end{abstract}

\ccsdesc[500]{General and reference~Empirical studies}
\ccsdesc[300]{General and reference~Evaluation}
\ccsdesc[300]{Computing methodologies~Artificial intelligence}
\ccsdesc[300]{Computing methodologies~Classification and regression trees}
\ccsdesc[500]{Social and professional topics~Computer science education}
\keywords{LLMs, AI, Java, Next-step hints, comparative judgement}


\maketitle

\section{Introduction}

Students who are learning programming often get into a stuck state where they cannot make progress~\cite{Whalley2023}.  This may be because they cannot solve a compiler error~\cite{Prather2017}, a run-time error~\cite{Smith2019,Garner2005}, or other more general issues with problem solving~\cite{Prather2018,Castro2020}.  There has been work to try to offer hints to students based on intelligent tutors~\cite{Crow2018} or crowd-sourced hints~\cite{Glassman2016} or explanations~\cite{Guo2020}, but the new growth of generative Artifical Intelligence (AI) tools offers new possibilities for generating these hints.

Offering hints to students is a subtle art~\cite{Marwan2019b,Suzuki2017}.  Just giving the answer offers little or no pedagogical benefit, but being too coy or obscure is not helpful and may frustrate the student further.  Choosing the right level of hint is typically more difficult than offering the actual solution.  Human teachers have the advantage of often knowing more context about the student and rich knowledge from the student's reaction when attempting an explanation -- but teachers are often in large classes and cannot be present at every moment (including when the student works separately outside class) to give hints, so automated approaches to hinting are of interest to provide scale and constant availability.

Generative AI systems such as Large Language Models (LLMs) offer ways to aid students~\cite{denny2024desirable}, such as generating hints.  Students can already directly interact with such LLM systems, but this has two key problems.  The first is that students will often ask for the answer, not for a hint, which is less pedagogically beneficial.  The second is that students who struggle may be ill equipped to write good prompts to the LLM~\cite{Denny2024}.  Therefore it may be best for a tool, such as an Integrated Development Environment (IDE), to provide the prompt on behalf of the student, in order to generate a hint~\cite{pozdniakov2024large}.  It is this approach that we investigate in this paper.

The task we are setting the LLM here is more complex than that of solving a programming problem (which LLMs have been shown to be capable of~\cite{Kiesler2023,Mahon2023}).  The model has to work out what the intended solution is, and it has to do this from limited information: in our context, instead of being given a specification of the programming task, the input consists solely of an erroneous, work-in-progress snapshot of an inexperienced programming novice's source code.  The task has to be inferred.  When the LLM has deduced the solution, it is expected to \textit{not give it to the student}, but instead devise a pedagogically useful hint that creates a learning experience in which the student makes progress towards a solution.

Generating the hint automatically leads to a set of questions: What makes a good prompt for our purpose?  And even with the best prompts and the best LLM: Are the hints generated good enough to be worth showing to novices?  Can we consistently generate hints of good enough quality that they help students more than they confuse them?  Attempts to investigate these issues reveal two fundamental questions we need to answer as part of this work:  What makes a good hint, anyway?  And how do we judge whether a hint is ``good enough''?

To answer the first of these two questions, we use a method called \textit{comparative judgement} to rank a number of hints (see \autoref{sec:method}) according to their quality.  This allows us to extract characteristics of good hints in general.  To judge whether the hints are of sufficient quality to be shown to students, we use a benchmark: we compare generated hints to those given by expert humans.  If the generated hints are judged better than those from humans, they provide an improvement on the status quo and are therefore useful.  The details of the methodology are described in \autoref{sec:method}.

In summary, this work provides the following major contributions:

\begin{enumerate}
    \item We evaluate \textbf{which LLM} currently performs best in providing next-step hints for novice programmers.
    \item We determine \textbf{the best prompts} for generating hints using the state-of-the-art LLMs at the time of the experiment.
    \item We investigate whether optimal LLM/prompt combinations can \textbf{perform as well as} (or better than) \textbf{humans} in generating hints.
    \item We describe and demonstrate \textbf{a repeatable method} for determining the best prompts and evaluating their performance, which could be re-used when LLM systems update in future.
    \item We investigate \textbf{whether comparative judgement is a viable method} for providing rankings in computing education research studies.
    \item We summarise the characteristics of the best hints (as rated by educators) to \textbf{determine what makes a good hint}: what length, what kind of language, what pedagogical features.
\end{enumerate}

\section{Related work}
\label{sec:related-work}

We divide related work into two main parts: prior work on hint-generation that did not use LLMs, and the use of LLMs in programming education.  We also review related work on when hints should be given, as that is a basis for selecting data for our paper.

\subsection{Non-AI hinting for novice programmers}
\label{sec:related-non-ai-hints}

The idea of giving automated hints to stuck novice programmers has a long history that pre-dates LLMs, and thus there have been multiple reviews on the topic.  \citet{Crow2018} conducted a review of intelligent programming tutors, from the 1980s to 2018, and found that they were very varied in the features they provide, including whether hinting support is present or not.  \citet{Keuning2018} performed a systematic literature review of feedback generation in general in programming education.  They found that relatively little feedback generation focused on next-step hints.  \citet{McBroom2021} surveyed hint generation systems from 2014--2018 and introduced a framework that synthesised work on hint generation.  Interestingly, it is not clear that LLM-based hinting would fit into \citeauthor{McBroom2021}'s framework (which revolves around constructing hints by starting with sets of hint data to narrow down or transform), suggesting that generative AI is quite distinct from existing work on automated hint generation.  Perhaps surprisingly, little work is available on human hint generation, or investigating what makes a good hint independent of automation -- the closest work is \citet{Suzuki2017} which categorised types of hints, but without evaluating their usefulness.  Most of the hint literature is concerned with discussing the various ways to automate the activity.

In terms of non-AI techniques to generate hints, one way to generate them is to imitate the hints that teachers would give~\cite{Suzuki2017,Jeuring2022}.  Another is to use techniques such as program repair to generate fixes~\cite{Phothilimthana2017,Ahmed2020}, or hand-written rules~\cite{Ichinco2018,Wiggins2021}.  Existing solutions to a specific programming problem can be used to infer hints for future attempts~\cite{Obermueller2021}.

Several aspects of automated hinting see no overall agreement in the literature, and evidence is inconclusive or contradictory.  While some studies are largely positive about the value of automated hint generation, other studies provide some evidence to suggest that hints themselves may not be useful~\cite{Price2020}, or that they may not help learning~\cite{Marwan2023}.  One approach to hinting is to show multiple possible hints, at the different appropriate points in the code where each hint could be enacted, although some students found this overwhelming~\cite{Price2021} -- and similar research on suggested fixes for errors found that students tended not to use the fixes even when they were appropriate~\cite{Brown2023}.  Students report difficulties using hints that are vaguer~\cite{Price2021}.  Hints which have an explanation are perceived as more helpful and more interpretable but do not necessarily result in better performance or learning~\cite{Marwan2019,Marwan2019b}.
  Overall, it appears that hints must be carefully designed in order to be useful.

\subsection{Timing of hints}

There is a large body of work to identify struggling students across the duration of a whole course~\cite{Hellas2018,Falkner2012,Ahadi2015}.  However, this is a distinct problem from trying to identify which students need hints and, crucially, \textit{when} precisely they would benefit from a hint.

A programming environment can provide a hint when students attempt to run the program~\cite{Phothilimthana2017} or when they receive a compiler error~\cite{Ahmed2020}.  The most common approach is to wait until the student explicitly requests a hint~\cite{Ichinco2018,Obermueller2021,Wiggins2021,Hellas2023}.  Alternatively, some tools suggest hints automatically as the user enters code~\cite{Gusukuma2017,Prather2023}.

\citet{Jeuring2022} performed a study asking experts when they would intervene to provide hints, and found ``a frequent conflict caused by different
pedagogical approaches: (a) an early intervention prevents a student
from writing unnecessary code and spending extra time on an
assignment, which may lead to student confusion and frustration,
versus (b) a delayed intervention gives a student a chance to struggle productively, which may improve student learning.''  In a follow-up study, \citet{Lohr2024} found similarly mixed results over when educators chose or chose not to provide feedback: ``sometimes one expert uses a reason at a step to explain why they do intervene, and another expert uses the same reason at the step to not intervene.''  Thus there is no clear recommendation from the literature for when is the best time to give a hint.

\subsection{Generative AI and Large Language Models in programming education}

Since the release of ChatGPT in late 2022 there has been an explosion of interest in LLMs, including in programming education teaching and research.  The result has been a dizzying rate of publication; all of the LLM studies cited in this section (of which there are more than thirty) were published in the last two years.  In an early 2023 study, educators were found to be split between those who wanted to resist AI tools and those who wanted to embrace them~\cite{Lau2023}.  However, the recent explosion in popularity seems to imply that resistance may well be futile~\cite{Joshi2023,Ghimire2024}, and it may be best to consider adapting our pedagogy~\cite{Tyson2023,Sheard2024,denny2024computing} and assessment~\cite{Raman2022} instead.  In this section we survey different aspects of LLMs in programming education in turn: students using them directly, students using them via tools, and their use in generating hints and explanations.

\subsubsection{Students' direct use of Large Language Models}

A natural first step in studying LLMs in programming education was to study what happens when students used LLMs directly, without scaffolding, in a manner of their choosing.

\citet{Prather2023} performed a study of how novices worked with Copilot, a generative AI tool that is designed to aid in program construction.  They found that novices struggle to understand and use the tool.  A later study by \citet{Prather2024} suggested that LLMs may widen the divide between students at the top and bottom of the class.
\citet{Zamfirescu2023} observed users without experience in LLM prompts as they designed a chatbot.  They found that the users struggled to modify LLM prompts to achieve the desired effect, although they were not using the LLM to directly modify program code.  \citet{Fiannaca2023} found that users could struggle with what made an effective prompt, and worried about syntax issues within prompts such as line breaks, and the presence or absence of question marks.
\citet{Denny2024} explored how students write prompts in order to solve programming exercises, when the exercises themselves could not simply be copy-and-pasted into the prompt.  They found that ``many students, even ones many years into their programming education, do not necessarily understand how to write effective prompts [for LLM systems].''  \citet{Nguyen2024} similarly found that many novices struggled to write effective prompts when using LLMs for the first time.

Thus the initial research in the area suggests that novices struggle to directly use LLMs in an effective manner.  This chimes with other research considering the pedagogical implications: \citet{Xue2024} and \citet{Kazemitabaar2023} found that direct use of LLMs did not produce any significant effect on learning (although the latter suggest that students with higher prior knowledge may have received greater benefits from using the generator than students with less prior knowledge), while \citet{Mailach2024} concluded that ``we cannot just give vanilla [LLM] chatbots to students as tools to learn programming, but we additionally need to give proper guidance on how to use them—otherwise, students tend to use it mainly for code generation without further reflection on or evaluation of generated code.''

\subsubsection{Student use of tools powered by Large Language Models}

An alternative mode of use suggested by several researchers~\cite{Scholl2024,Xiao2024,Nam2024} is to build tools powered by LLMs.  This can avoid issues with students' inability to create prompts, and give more control over the tools' output.

\citet{Liffiton2023} created a tool where users could fill in four items: which programming language is being used, the relevant code, the error (if any), and the question they want help with.  This is then structured into a single larger prompt to the LLM, and the response is shown in the tool.  They concluded that students liked the tool, including the fact that it did not just ``give away the answer''.  In a follow-up study, \citet{Sheese2024} found that students tended to ask for help with their immediate problem and would not typically ask more general queries, such as seeking understanding of a wider concept.

\citet{Birillo2024} combined LLMs with static analysis in order to create a tool that provides next-step hints.  A brief evaluation with students suggested that the tool showed promise.  \citet{denny2024desirable} studied students' use of an LLM-powered assistant, and found that students engaged it with extensively, and also found that students preferred scaffolding and guidance rather than simply being told the final answer.  This suggests that hint-generation may be a more useful and more popular tool for students than simply providing correct program code.

\subsubsection{Use of Large Language Models for hinting and explanation}

Several studies have investigated the use of LLMs for feedback, explanation or hinting -- the latter being the precise topic of the current research.

\citet{Leinonen2023b} used some early LLMs to generate enhanced programming error message explanations.  The researchers rated the error message explanations as high quality.  More recently, \citet{Cucuiat2024} investigated secondary school teachers' views on LLM-generated programming error message explanations.  They used feedback literacy theory to analyse interviews and found that educators preferred LLM explanations that \textit{guided} and \textit{developed understanding} rather than \textit{tell} (emphasis indicates terms defined by feedback literacy theory~\cite{Mclean2015}).

\citet{Hellas2023} investigated the use of LLMs to solve historic student help requests. They used data from a course where students could ask for help from human teachers.  They used the code that students asked for help on, combining it with a boilerplate AI prompt, then analysed the responses that came back from the AI, in terms of features such as ``identifies at least one actual issue'' and ``includes code'' in order to compare two different AIs (Codex and GPT-3.5), each in English and in Finnish.  They found that the AI would frequently provide code despite being instructed not to, and that LLMs could make the same mistakes as students when trying to help them.  \citet{Roest2024} created a tool to give next-step hints for novice Python programmers, by providing the problem description and current code, and evaluated them with three students and two educators.  They similarly found that it was difficult to control the LLM output and that the hints were sometimes misleading.  \citet{Kiesler2023b} also used a similar technique of using historic incorrect student code submissions and analysed the responses using the feedback categorisation of \citet{Keuning2018}.  They again found LLMs could be misleading but that the quality was generally good.

\citet{macneil2023experiences} included code explanations generated by LLMs into an online course and found that students rate AI-generated explanations useful for their learning, and that students preferred concise, high-level explanations over line-by-line explanations of the code.
\citet{Leinonen2023} found that LLMs produced explanations of program code that were rated as more accurate and easier to understand than explanations generated by students on the course.  This suggests that AI hints may be very valuable for learners.

\citet{Pankiewicz2024} investigated students' affective states when receiving hints from GPT for solving compiler error messages, and found an increase in focus and decrease in confusion, although more generally they found a mixed pattern as to whether students reported that the results were useful for not, and a mixed result in performance.  \citet{Xiao2024} investigated students' opinions of using an LLM-powered tool to generate hints.  They found that the hints were high quality, but students commented on the lack of flexibility in the interface, and they were confused by some of the higher-level hints which they could find vague.
\citet{Nguyen2024b} used few-shot learning to train GPT-4 to generate hints and had two instructors evaluate them for accuracy and usefulness, finding that the model performed well and was useful.

Overall, previous studies have suggested that LLMs can produce good quality hints, but there are challenges to tightly control the output (in particular, to avoid giving too much program code or too much of the solution) and avoid misleading hints.  As we will detail in the next section, our study differs from previous work in the following ways: 

\begin{itemize}
    \item We focus on stuck students with no information about the problem description that they are working on -- a harder but more flexible domain than when the problem is known.
    \item We use a larger-scale educator evaluation in contrast to the prior evaluations that use 2--3 educators (often the researchers themselves).
    \item We provide a detailed assessment of hint characteristics in multiple dimensions (length, readability, pedagogy, correctness, sentiment) that combine the disparate subsets into one.
    \item We ask the educators to use comparative judgement in order to form a ranking of hints which allows us to infer links between these hint characteristics and relative hint quality.  This provides useful information about hint attributes that is orthogonal to how they were generated.
    \item We include human-generated hints to allow us to compare performance of human experts to LLMs.
    \item We assess multiple prompts with multiple models to see how much effect prompts have on model performance.
\end{itemize}

\section{Method}
\label{sec:method}

From our survey of prior work in \autoref{sec:related-non-ai-hints}, it seems that work on non-AI hints has shown mixed results as to its effectiveness.  Many hint approaches rely on knowledge derived from existing solutions to the problem and thus target that specific problem.  Generative AI, LLMs in particular, has the potential to be more flexible and powerful when generating a hint for an unseen problem.  Studies of LLMs find that students can struggle to formulate prompts, with an alternative approach emerging of using a tool to construct the prompt based on constrained, guided information from the user.

Our approach in this paper is to use source code data from real-world programming sessions from stuck students.  We will use the term \textit{Snapshot} for a code sample of a student during a programming session at a point in their work when they were stuck.

We then present each snapshot to a set of \textit{Generators}.  We use the term \textit{Generator} to refer to the producer of the hint: this is either a combination of a specific LLM with a specific prompt, or an individual human.

Next, we show these resulting hints to a set of experienced educators.  The educators tell us which hints they consider to be the best.  This gives us five results: the best hints, the best-performing LLMs, the prompt templates to generate the best hints, a comparison of human and LLM hints, and the general characteristics of the best hints.

\begin{figure*}
    \centering
    \includegraphics[width=0.9\textwidth]{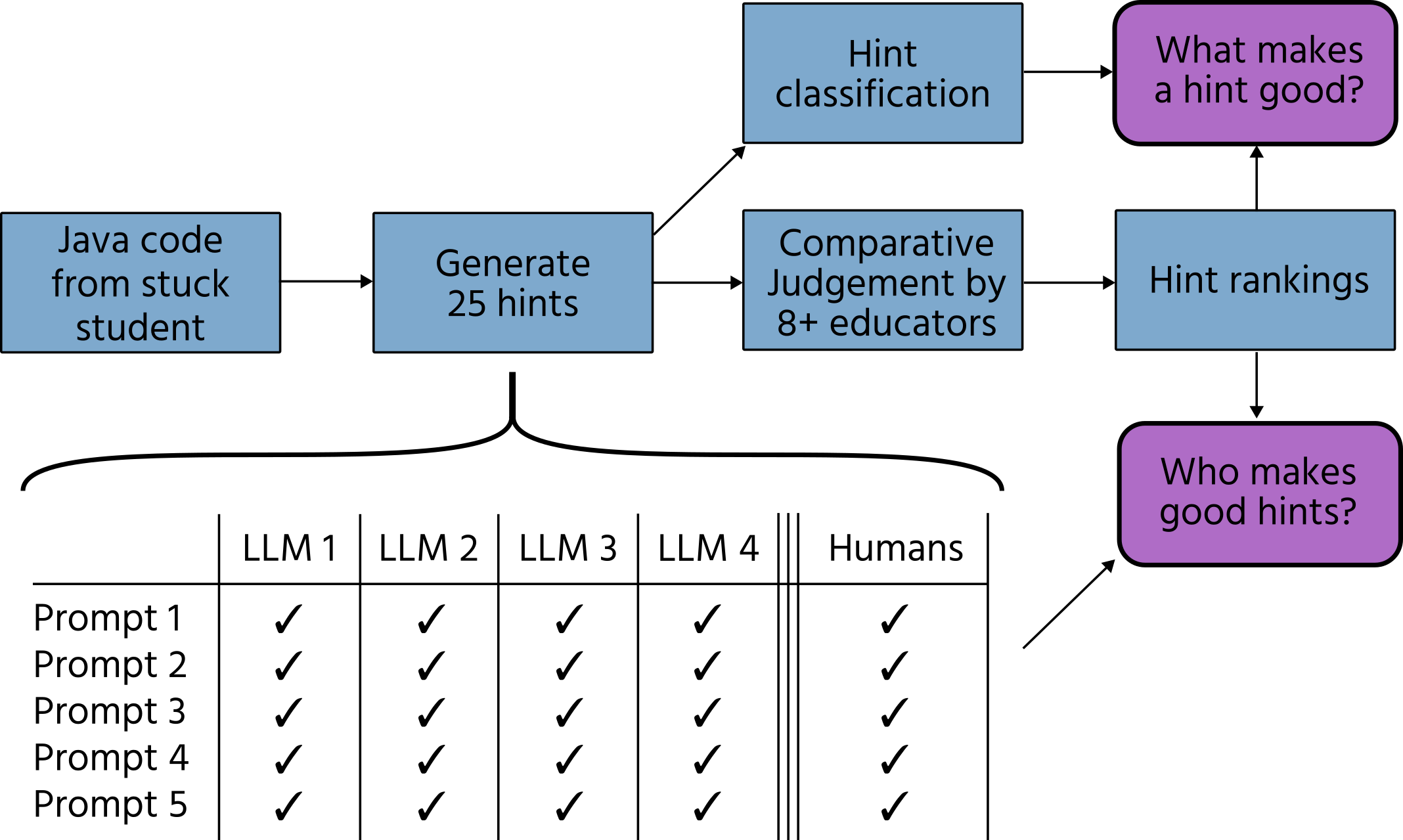}
    \Description{Diagram showing the study design: we take Java code from a stuck student and generate 25 hints (5 prompts fully crossed  with 4 LLMs, plus five humans).  We then get 8+ educators to rank these hints, and classify the hint attributes.  From this we can infer what makes a hint good, and who makes good hints.}
    \caption{The design of the study: We take a Snapshot of student code from Blackbox, generate 25 hints for this Snapshot (four LLMs each with five prompts, plus a hint from each of five human experts), and then ask eight or more expert educators to compare the hints to form a ranking.  \textbf{We repeat this whole process with four different Snapshots and different educators, resulting in 100 hints ranked by more than 32 educators.}  This allows us to infer what characteristics make a good hint, and which Generator (LLM+prompt, or human) generates the best hints.}
    \label{fig:process}
\end{figure*}

Our method thus has four main parts: the acquisition and selection of the student Snapshots; the manner of creation of the set of LLM prompts to evaluate, leading to the generation of the hints; the manner to evaluate the resulting hint quality using human experts; and the evaluation of the attributes of interest of these hints.  The overall method is shown in \autoref{fig:process} and explained further in the following subsections.

\subsection{Student Snapshots}

To find Snapshots (stuck student states) for our study, we randomly sampled sessions from the Blackbox dataset~\cite{Brown2014} from an arbitrary week-day late in the typical northern hemisphere first semester (21st November 2023), manually selecting states where we inferred that the students were stuck and in need of a hint, as evidenced by making no productive progress for some time after that point.  Previous research has found that educators disagree about when is the best time to intervene~\cite{Jeuring2022,Lohr2024}, so in lieu of clear recommendations from the literature, we used our own judgement.  The exact point chosen is not crucial to the study, as long as it provides an interesting case for which to generate hints.  As described later in \autoref{sec:results-stimuli-evaluation}, multiple participants  commented on the choice of student code being good, and realistic.

One example of a Snapshot is shown later in the paper in \autoref{fig:task1}  on page~\pageref{fig:task1}.

\subsection{Prompt formation}
\label{sec:prompt-formation}

Current literature does not yet suggest potential successful prompt templates for hint generation.  Furthermore, as AI models evolve, which prompts are effective may change over time.  To try to future-proof our methodology as much as possible when models continue to evolve, we used the following approach.  Five researchers, who are also experienced programming educators with significant teaching experience, experimented with four LLMs to formulate prompts.  The four models were Mixtral-8x7B, GPT-3.5, GPT-4 and Gemini.  Each model was used by a different researcher, but with the most recent model, GPT-4, used by two.  As suggested by research into brainstorming~\cite{Rietzschel2006}, the researchers first brainstormed multiple prompts individually by using some sample Snapshots (these were distinct from those used in the actual study).  Then the prompts were combined and refined experimentally, during iterative cycles of collaborative discussion, resulting in a final set of five distinct prompts.

\subsection{Human-generated hints}

Alongside the AI-generated hints, we also supplied human-generated hints.  Each of the five researchers was tasked with constructing one hint for each Snapshot in the experiment.  This was an attempt to produce the optimal hint that they, as experienced educators, could provide to the student at that point, and it provided us with a benchmark in our results: we can not only compare automatically generated hints against each other, but compare them to hints produced by experienced humans. 
Constructing these hints was done independently; no researchers saw another researcher's hints until they had completed writing their own.

\subsection{Educator evaluation}

One way to rank hints is to ask educators to rate each one on an absolute scale, say 1--10.  It can be difficult to evaluate how good a hint is on such an absolute scale.  Is a particular hint a 5/10 or a 6/10?  Is everything just 7/10?  Can participants remain internally consistent, and can the scale be consistent between participants?

An alternative method to produce a ranking is a technique called \textit{comparative judgement}.  The key idea behind comparative judgement is that people (termed judges) produce more reliable judgements when repeatedly asked to compare two items and pick the best one, than to rate each one individually on an absolute scale.  Instead of ``rate this hint 1--10'', the problem becomes ``which of these two hints is better''.  By asking judges to pick the best item from a set of random pairs, the judges act like the comparison function in a bubble sort, to sort the hints into an ordered list from best to worst. This form of judgement is more consistent across different judges, as it does not rely on an abstract absolute scale that would be influenced by different standards of individual participants.

Comparative judgement has been improved upon in an algorithm known as adaptive comparative judgement which minimises the number of comparisons needed~\cite{Pollitt2012}.  Intuitively, if one hint is chosen several times as always worse than others, you can leave it near the bottom of the list and focus on the more borderline comparisons, to sort the list with fewer comparisons.  Adaptive comparative judgement has been widely used for assessment in education and found to be reliable~\cite{Bartholomew2018}, and has been used in other areas of education research~\cite{Jones2023}.

Each educator (judge) is first shown the Snapshot, and then asked to repeatedly rate which of two presented hints are better in this circumstance, with the pairs generated by an adaptive comparative judgement algorithm.  \citet{Verhavert2019} performed a meta-analysis on non-adaptive comparative judgement and found that the number of judges did not impact reliability, so we will not specify a minimum sample size of judges.  If the judges are experts, \citet{Verhavert2019} found that for 90\% reliability, 26 to 37 presentations per item are needed.  Since each comparison is between two items, ranking of $N$ items requires $13N$ judgements to achieve 26 presentations of each item.
Given that adaptive comparative judgement aims to reduce the number of comparisons, fewer should be needed.  We use the No More Marking~\citep{Kolen2016} platform to present the hints and collect the judges' choices.  This platform uses a Progressive Adaptive Comparative Judgement algorithm\footnote{See \url{https://blog.nomoremarking.com/progressive-adaptive-comparative-judgement-dd4bb2523ffe}, visited 4 November 2024.} and recommends 10 comparisons per item\footnote{See \url{https://blog.nomoremarking.com/using-comparative-judgement-in-different-subjects-at-ks3-4415195f8947}, visited 4 November 2024}. In our study, we ranked 25 hints for each Snapshot, so we required $\approx 250$ comparisons, which we split across 8 judges doing 31 comparisons each.

For our experiment, we needed the expert educators who evaluated the hint quality to familiarise themselves with the Snapshot for which the hints were evaluated.  To reduce the overhead incurred by educators, we chose to expose each participant to only one Snapshot.  In order to help the participants all understand the Snapshot (which we took to be a prerequisite of the task of judging the hints, rather than something left to the educators to succeed or fail at) we provided brief notes with our interpretation of what the problem with the student code was.  These notes were not given to the LLMs when generating the hints.

We chose not to provide a detailed rubric for judging what makes the best hint.  Our instruction to educators was ``Imagine that you are helping a student who is somewhere within their first year of programming instruction, around the ages of 16-18. They are working on a problem and have become stuck and asked for help. Imagine that the computer they are working on could give them an automatic hint at this stage. We want you to determine which is the best hint to give in each circumstance."

\subsection{Measuring characteristics of hints}

To characterise the best hints, we measure the following attributes: length of the hint (in number of words), complexity of the vocabulary (i.e. reading level), (these first two are suggested in guidelines by \citet{Denny2021} as used by \citet{Prather2023b} for error message readability), sentiment, correctness, and type of feedback (as per \citet{Cucuiat2024}).

\subsection{Pre-registration and ethical approval}

The study was approved according to the ethics procedures of King's College London, approval number MRA-23/24-41449.

We pre-registered the design of this study on the Open Science Foundation (OSF) website\footnote{See \url{https://osf.io/x8u3t/?view_only=2cdfa22b8dc542d6850e0cd8ce0ad6ff} -- this link is anonymised and is safe for anonymous review.}.  The main changes since pre-registration are:
\begin{itemize}
  \item We decided to show each educator-participant only one Snapshot rather than several, to reduce the load on each participant.
  \item We refined the exact process that we used to generate the hints, as described in \autoref{sec:prompt-formation} and had the idea to add human-generated hints.
  \item We chose a new way to characterise hints using feedback literacy after reading the paper by \citet{Cucuiat2024} that was published after we pre-registered.
\end{itemize}

\section{Results}
\label{sec:results}

The study was carried out in 2024, with the hints generated in April 2024 and educators recruited to perform the comparative judgement in August and September 2024.

\subsection{Open data}
\label{sec:osf-repository}

All of the [anonymised] materials and data from this study are available publicly in an OSF repository: \url{https://osf.io/p436s/?view_only=048063e28b474fa7a8d5fa776985f39b} \footnote{This is an anonymous link for review which is safe for reviewers to visit without revealing the identity of the authors.}
including: all of the stages of prompt creation and merging, all of the Snapshots, all of the generated hints, all of the processing performed on the hints, our participant instructions, our survey design, all of our survey results and all of the results of the hint comparison, plus the full processing pipeline for all of our statistical analysis and figures.  We hope that this is useful for anyone interested in verifying or replicating our work.

\subsection{Prompt creation}

The five LLM prompts, shown in \autoref{tab:prompts}, were created as described in \autoref{sec:prompt-formation}.  Two of these are multi-stage prompts, which involves asking the LLM for an answer and then feeding it back to the LLM.  This may seem odd to those unfamiliar with using LLMs, but LLMs are not idempotent: when asked for information or an answer, and subsequently asked to use or improve it, -- even without giving any new information in the second request -- an LLM will generate a different and potentially improved answer.

\begin{table}
    \centering
    \begin{tabular}{l|p{9cm}}
    \# & Prompt \\ \hline \hline
        1 & I'm learning to program. Without giving the solution, can you guide me into the next step to fix the problem in this code? \$CODE \\ \hline \hline
        2 & You are an experienced programming instructor teaching a course in Java. I will provide you student code and your task is to generate a hint for the student. Please do not provide the full solution, but try to generate a hint that would allow the student to proceed in the task. Please address your response to the student. Here is the code: \$CODE \\ \hline \hline
        3a & You are an experienced programming instructor teaching a course in Java. I will provide you student code and your task is to try infer what the task the student is working on is.  Here is the code: \$CODE\\ \hline
        3b & You are an experienced programming instructor teaching a course in Java. I will provide you student code and an explanation of the task they are working on, and your task is to generate a hint for the student. Please do not provide the full solution, but try to generate a hint that would allow the student to proceed in the task. Please address your response to the student.  Explanation of the task: \$PREVIOUS\_ANSWER Here is the code: \$CODE\\ \hline \hline
        4 & You are a tutor who is helping a student who is stuck while writing code for an introductory programming course. To help the student, you must generate a hint that will identify their mistake and help them make progress continuing with their code. If there is a serious error, please point that out first. Please be helpful and encouraging, but do not reveal the answer - instead, make one concrete suggestion to help them progress. Here is their code: \$CODE  Please identify the most serious error in this code and suggest a hint to the student to help them. Just show the "Hint", suitable for immediate display to the student. Please do not include any other explanatory text.\\ \hline \hline
        5a &  I have written the following code: \$CODE  I need help. Act like a good teacher and give me some help. Respond in no more than three sentences. \\ \hline
        5b & Do you think the hint you just gave was a good hint? Critique it, including consideration of accuracy, friendly tone, helping the student to learn and not giving too much of the answer away.\\ \hline
        5c & In light of your criticism, improve the response you first gave. \\ \hline \hline
    \end{tabular}
    \caption{The five prompts given to the AI to generate the hints.  The Snapshot is substituted wherever ``\$CODE'' appears.  Prompts are separated by double-lines.  Some (3a+3b, 5a+5b+5c) are multi-stage prompts, separated by a single line; each prompt part is given in sequential order; the marker ``\$PREVIOUS\_ANSWER'' indicates that the full answer to the preceding part of the prompt should be inserted.}
    \label{tab:prompts}
\end{table}

\subsection{Hint generation}

The prompts were fed to the models in April 2024 to generate the hints.  We manually unified the formatting between the resulting hints, to make sure the code snippets were all highlighted in the same manner regardless of which model generated them.  We also made one minimal pass to remove boilerplate greetings from the very start or end of the hint, but we did not process the hint any further, to retain authenticity of machine generation.  Examples of things we \textit{did} remove:
\begin{itemize}
  \item Phrases responding to our prompt request such as ``Certainly!'' or ``Absolutely!'', or in the case of multi-stage prompts: ``Here's an improved version of my initial response''.
  \item Leading salutations such as ``Hey there'' or ``Alright, student!''\footnote{One of the authors plans to use the latter to communicate with their students in future.}.
  \item Trailing salutations such as ``Best regards.'' or final remarks such as ``If you need anything else, just ask.''
\end{itemize}

We felt that all of these phrases could be removed automatically in future by refining the prompt or using a second processing pass, and they distracted from the hint content we wanted to evaluate.

Examples of things we \textit{did not} remove:
\begin{itemize}
  \item Encouraging phrases such as ``Good luck!'' or ``Keep it up.''
  \item Emoji, e.g. an airplane emoji in a response about a student stuck calculating air miles points.
  \item Use of first-person phrases, e.g. ``I noticed that there are a few things worth looking into''.
  \item Cases where the AI has addressed the educator such as ``Now, to assist your student further, I recommend...''
\end{itemize}

The final item is essentially the prompt ``misfiring'' and we felt it was important to penalise the prompt and model for this behaviour.

All of the exact changes made to the generated versus presented hints can be seen in our OSF repository.

The final set of hints included 25 per Snapshot: four AI models (Mixtral-8x7B, Gemini, GPT-3.5, GPT-4) with all combinations of the 5 constructed prompts (see \autoref{tab:prompts}), plus five human-generated hints.  All hints were formatted similarly and given arbitrary identifiers to obscure how they might have been generated.

\subsection{Judges}

We recruited participants to act as judges via mailing lists, forums and social media.  We asked for ``Java educators'' to complete an online task plus survey.  There was no reward for participation.  85 participants signed up; three started but did not finish the task, 41 completed the task, and 35 of those completed the survey (we discuss these numbers further in the threats to validity in \autoref{sec:threats-validity}).  Participants were assigned in a round-robin fashion to try to ensure that as many Snapshots as possible had the necessary 8 completions, which proved difficult with the low completion rate.  Ultimately, four Snapshots reached the required 8 completions (with 8, 10, 8, 9 completions).  Two more Snapshots had only three completions each so are excluded from all the analysis of the comparative judgements and hint rankings -- but survey results of the judges of these Snapshots are retained for the purpose of analysing judges' reflections on performing the task.

\subsection{Validity of Judging}
\label{sec:infit}

Given that our results are reliant on the comparative judgement task, it is important to check that the participants took the task seriously.  We employed several metrics for this purpose.  

\begin{figure}
    \centering
    \includegraphics[width=0.7\textwidth]{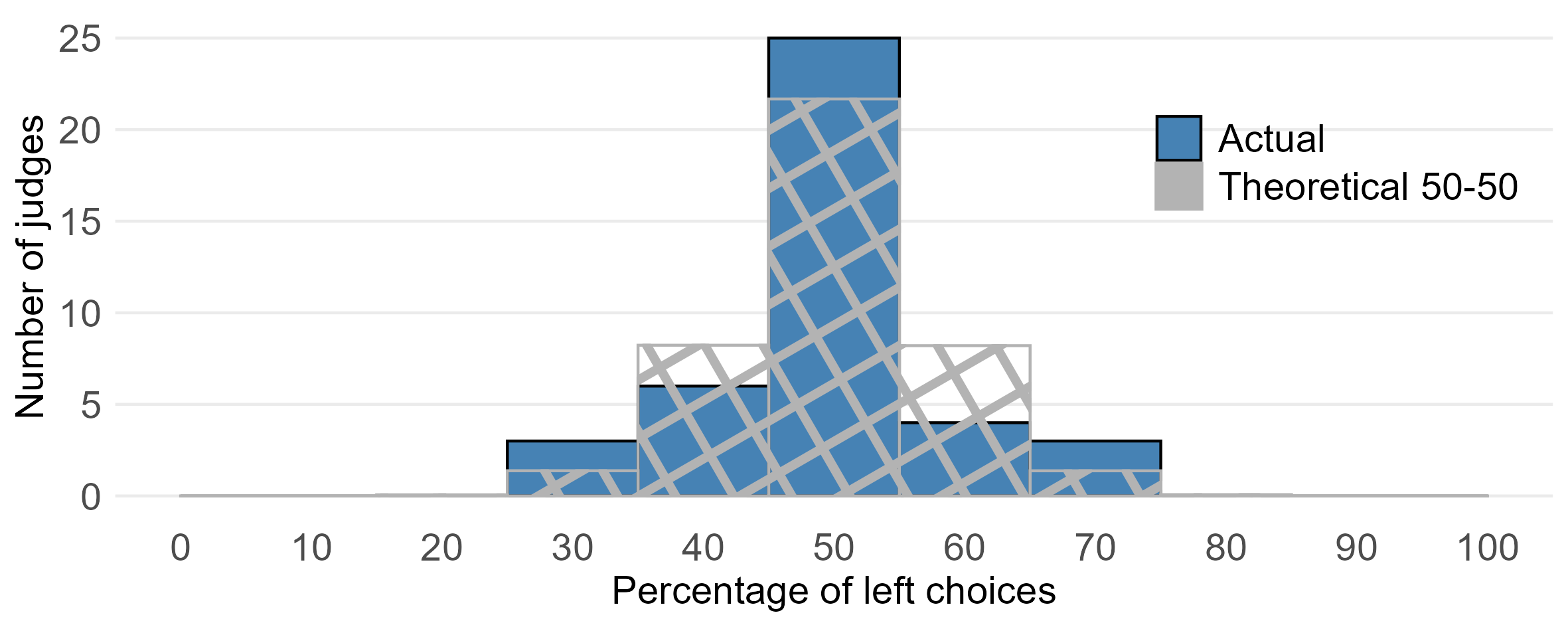}
    \Description{A histogram showing the actual left-vs-right clicks, and theoretical.  It shows slightly higher than expected in the centre (at 50\%), and slightly higher than expected at the tails, and thus lower than expected just adjacent to the centre.}
    \caption{A frequency histogram based on how many participants made a given percentage of ``left'' decision when asked to make a left-vs-right judgement.  The blue bars show the actual values; the grey hashed bar overlay is the theoretical distribution if the decisions were completely random (i.e. drawn from a binomial distribution with probability of 50\% for each decision).}
    \label{fig:left-clicks}
\end{figure}

The first check was how often the participants selected the left option as the best hint when given the left-vs-right decision to pick the best hint of a presented pair.  If any/many participants consistently just clicked the same button it could indicate boredom during the task.  \autoref{fig:left-clicks} shows this data, compared to the theoretical binomial distribution that should occur if the percentage was 50-50\%.  A chi-squared test ($\chi^2=44.0, p = 0.06$) between the two was not significant, suggesting the observed data was not significantly different from a 50-50 distribution of left-right clicks -- there was no bias.

\begin{figure}
    \centering
    \includegraphics[width=0.6\textwidth]{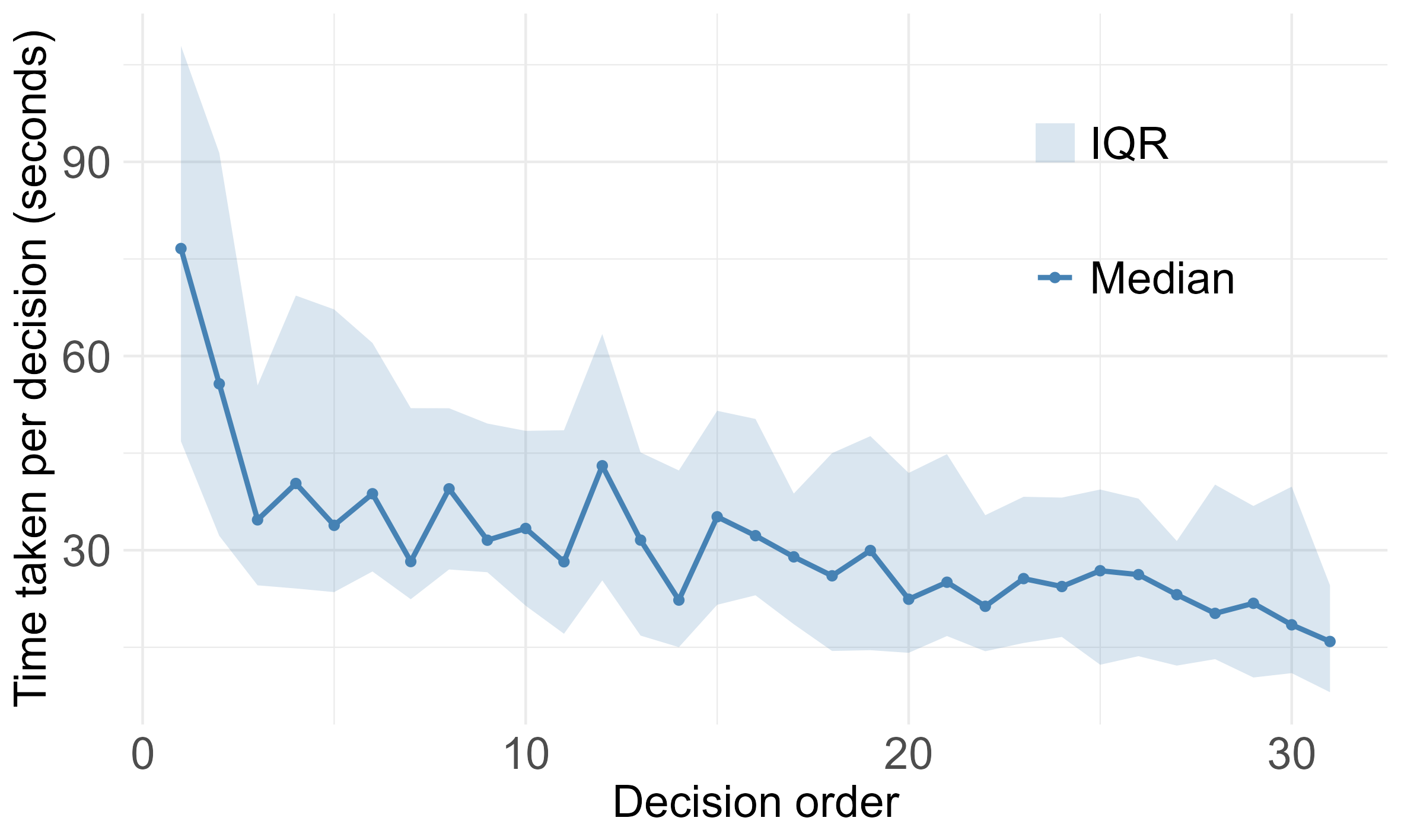}
    \Description{A line chart showing time taken vs decision order.  The pattern is spiky/noisy but overall it is a sharp decline (in time taken) at the beginning from the first to the third point, followed by a gradual further decline until the end.}
    \caption{The median (line) and interquartile range (shaded area) of time taken to make each judgement, ordered by the decision order (first judgement made on the left, through to the last on the right).}
    \label{fig:time-taken}
\end{figure}

The second check was how long participants took to make the judgements, to see if this suddenly dropped off during the judging, which would again suggest boredom during the task.  \autoref{fig:time-taken} shows these results.  The participants take a long time for their first ($\approx$75 seconds) and second ($\approx$50 seconds) judgements as they acclimatise to the task, and this is followed by a gradual speeding up (from around 40 seconds to around 20 seconds) as they work through the task.  This gradual pattern suggests becoming more accomplished at the task rather than giving up.

\begin{figure}
    \centering
    \includegraphics[width=0.75\textwidth]{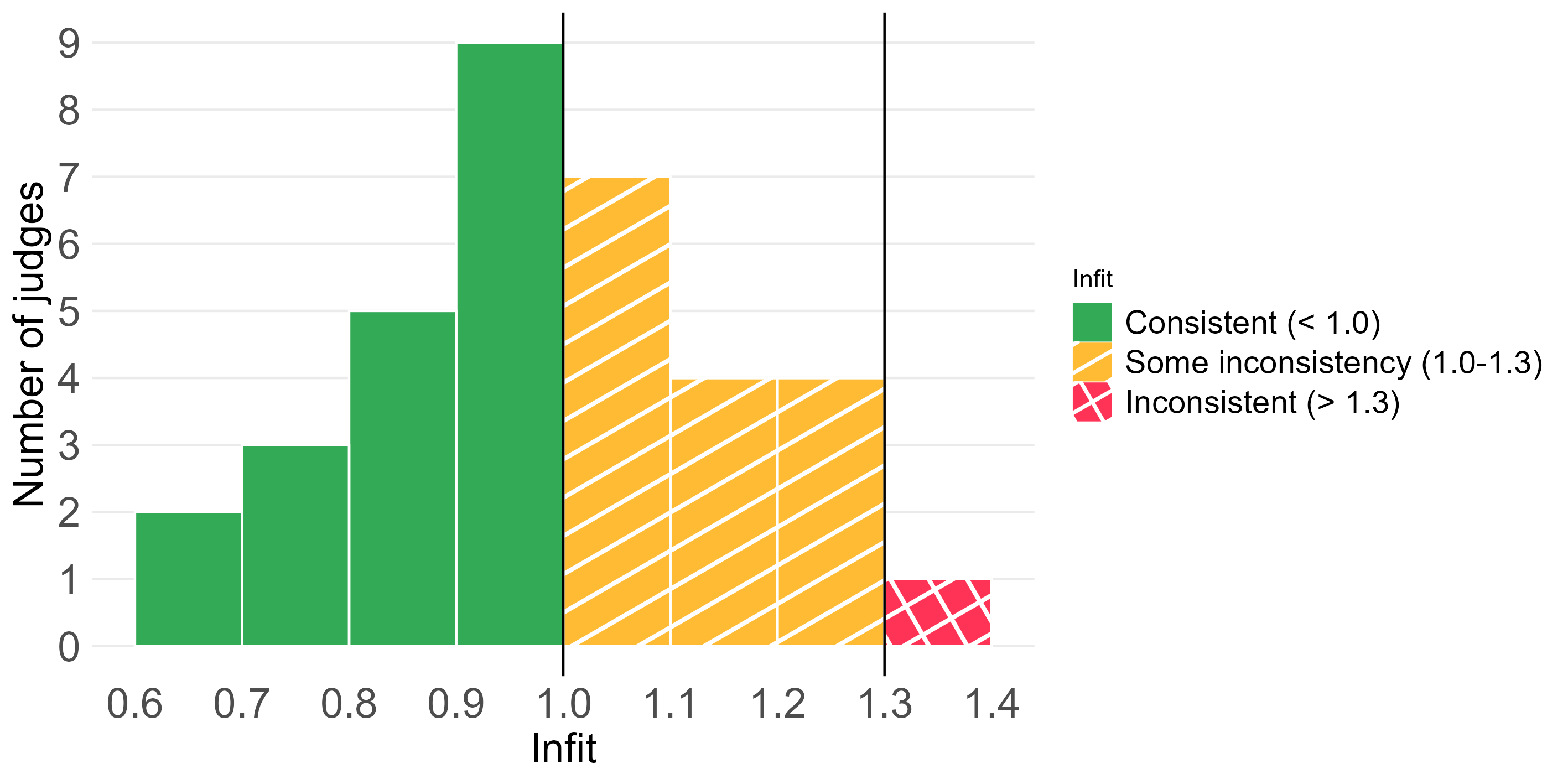}
    \Description{A bar chart showing frequencies of different Infit ratings.  Approximately 60\% of them are in the ``Consistent'' range, about 40\% in ``Some inconsistency'' and a single point in ``Inconsistent''.}
    \caption{The frequencies of different Infit ratings, one rating per participant.}
    \label{fig:judges-infit}
\end{figure}

\begin{figure}
    \centering
    \includegraphics[width=0.75\textwidth]{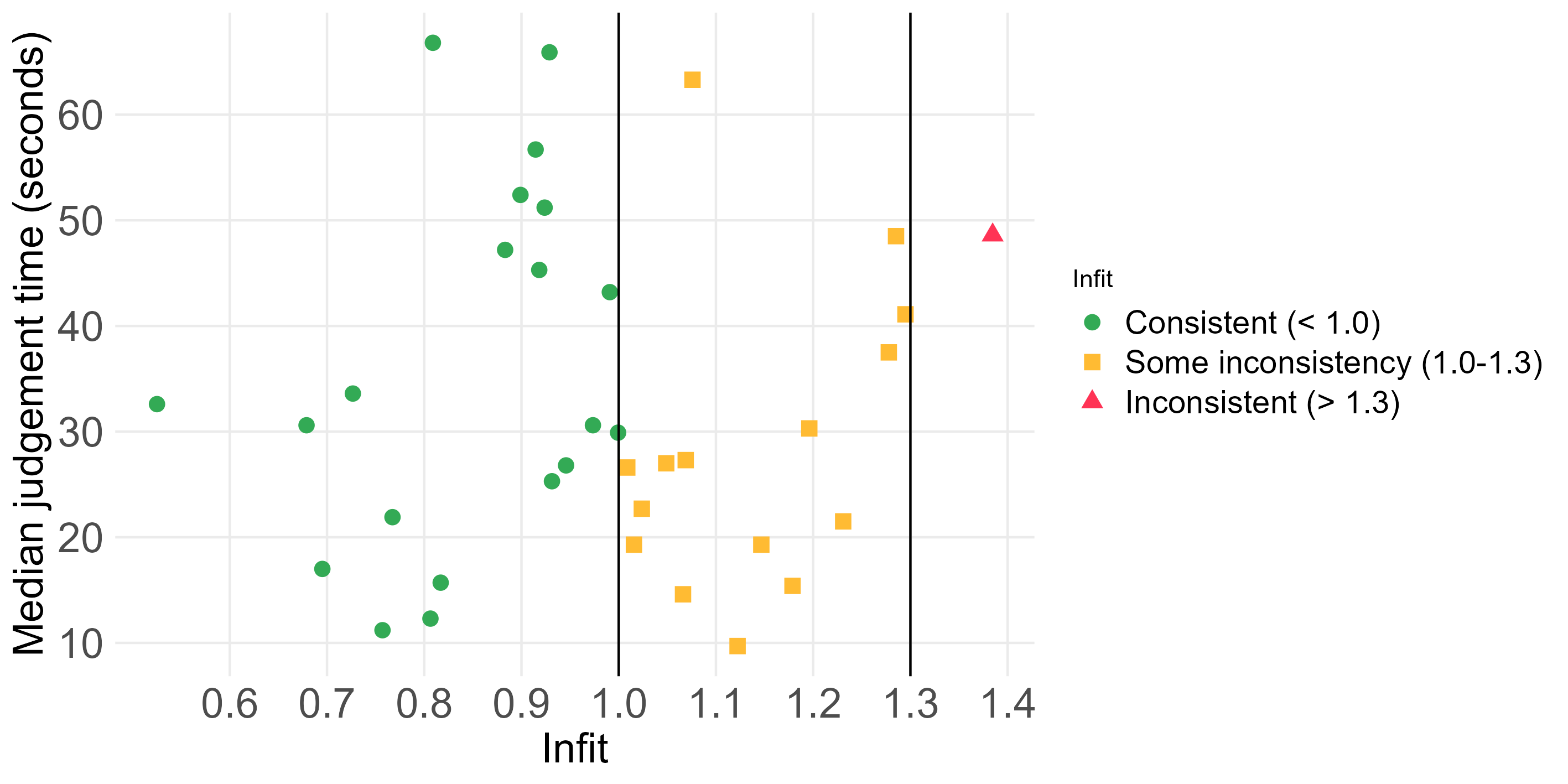}
    \Description{A scatter plot of Infit versus Medial judgement time.  The chart shows no discernable relation between the two variables.}
    \caption{A plot of Infit ratings (each plotted point is a participant) against median judgement time, to see if less consistent judges were judging faster.}
    \label{fig:judges-infit-by-speed}
\end{figure}

Those results suggest that judges took the task seriously, but we must also examine a core assumption of comparative judgement: that there is an underlying ranking of items that is shared among the judges.  In a perfect case, such as a task asking judges to pick the largest number, it should be possible (excepting human error) to perform a perfect ordering.  Most real-world tasks, however, will have some expected disagreements just because people have differing opinions.  This is acceptable as long as the disagreements are not too wide-scale.  To investigate this, two measures for inter-rater reliability in comparative judgement may be used: Reliability (a per-task measure) and Infit (a per-judge or per-item measure).

Our Reliability scores for the four Snapshots were 0.73, 0.76, 0.76 and 0.60.  A reliability of 0.7 is generally considered sufficient~\cite{Verhavert2019} so the majority of our Snapshots showed outcomes with a high level of inter-rater consistency.  Our Infit scores are shown in \autoref{fig:judges-infit}.  Note that inconsistency in Infit scores refers to \textit{inter}-judge consistency not \textit{intra}-judge: it is about whether the judge agrees with their peers, not whether their own decisions were self-consistent.  An inconsistent judge in this sense does not necessarily mean a ``wrong'' or ``bad'' judge, just one whose opinions differ from the other judges.  We therefore did not exclude these judges.  We did check if inconsistency was associated with faster judgements (suggesting a kind of speed-accuracy trade-off; maybe the inconsistent judges were choosing arbitrarily in a rush) but this was not the case (see \autoref{fig:judges-infit-by-speed}, confirmed by a linear regression being non-significant, $p = 0.693$).

Given that several previous studies of educators~\cite{Jeuring2022,Lohr2024,Brown2017} have found that educators do not always reach good agreement on pedagogical issues, it was slightly surprising that the agreement level was so high.

Finally, we checked the speed of the hint judging against reading speed.  A statistical model (linear regression of log-transformed time taken vs combined word count and participant, $p < 0.001$ for the combined word count factor) found that the average time taken to choose between two hints by combined word count was as follows:
\begin{itemize}
    \item 100 words: 28 seconds
    \item 200 words: 31 seconds
    \item 300 words: 36 seconds
    \item 400 words: 41 seconds
    \item 500 words: 46 seconds
\end{itemize}
Given that the average reading speed for non-fiction is 238 words per minute~\cite{Brysbaert2019}, it is likely that participants were not fully reading most of the hints.  However, this does not mean the judging was invalid, for several reasons:
\begin{itemize}
    \item It is valid for a participant to immediately assess that a hint is too long and that students will not have the patience to read it, without reading it themselves.
    \item Participants saw the same hints multiple times, so they may have begun to recognise hints without reading them through a second time.
    \item The participants are likely highly educated, and therefore used to skimming complex text effectively.
\end{itemize}

In summary, all our checks suggest that the judges took the task seriously, and that there is good if imperfect agreement among educators as to which hints are good.

\subsection{Method and stimuli evaluation}
\label{sec:results-stimuli-evaluation}

In our survey we asked participants whether they found the comparative judgement task easy or hard via a free-text response.  18 of 35 said they found it easy, 7 indicated a medium difficulty, 8 indicated they found it hard.  Only one participant mentioned boredom.

We asked participants for their observations about the Snapshot example that they saw, in a free-text response.  Not all participants gave a response.  Five mentioned that the Snapshots were well-chosen, eight said they thought they were realistic, and two said they thought they were unrealistic or outliers.

\subsection{Hint characteristics}
\label{sec:human-categorisation}
\label{sec:hint-attributes}

To evaluate which characteristics of hints were associated with their ranking, we extracted various attributes of the hints, shown in \autoref{tab:hint-features}.

\begin{table}
    \begin{tabular}{p{3.5cm}|l|p{8cm}}
        \textbf{Attribute} & Type & \textbf{Method of extraction} \\ \hline
        Word count & Integer & Count number of words, including words in code, but ignoring punctuation, symbols and emoji. \\
        Reading level & Real & Flesch-Kincaid Reading Grade Level~\cite{Kincaid1975}.\\
        Sentiment & Real & VADER Sentiment Analyzer~\cite{Hutto2014}, compound polarity score.\\
        Model & Category & The name of the AI model or the term ``Human'' to indicate how the hint was generated. \\
        Telling & Boolean & Feedback-literacy-inspired category.  See \autoref{sec:human-categorisation}.\\
        Guiding & Boolean & Feedback-literacy-inspired category.  See \autoref{sec:human-categorisation}.\\
        Developing understanding & Boolean & Feedback-literacy-inspired category.  See \autoref{sec:human-categorisation}.\\
        Opening up a new perspective & Boolean & Feedback-literacy-inspired category.  See \autoref{sec:human-categorisation}.\\
        Partially incorrect & Boolean & Flag indicating whether a hint was partially incorrect.\\
    \end{tabular}
    \caption{All of the hint attributes that were examined, and how the attributes were extracted.}
    \label{tab:hint-features}
\end{table}

The word count is straightforward, while the sentiment analysis~\cite{Hutto2014} and reading level~\cite{Kincaid1975} used existing techniques.  We did not initially plan to use Model as a factor given that the participants should be unaware of which model was used.  However, since the Mixtral-8x7B model generated longer and less readable hints in a particular style, we were unsure if an effect of word count would be related to the word count, the readability, or the model's individual style, so we included Model as a factor to evaluate if it had an effect not captured by the other factors.

All of the other attributes were categorised by the researchers.  The ``Partially incorrect'' flag is a relatively unambiguous technical check for incorrect parts of the hints (if any part was incorrect this flag was true, even if the rest was correct).  Consider, for example, this [erroneous] student code from one of our Snapshots in the study:

\begin{lstlisting}
    String letter=sc.nextLine();
    if (letter.equals(a|| c|| e ||g)))    
\end{lstlisting}

Multiple hints suggested a fix that included code similar to the following:

\begin{lstlisting}
    if (letter.equals('a') || letter.equals('c') ...
\end{lstlisting}

However, because \lstinline$letter$ is a String and \lstinline$'a'$ is a character, they will never be equal in Java even if \lstinline$letter$ has the value \lstinline$"a"$, because the types (String and Character) do not match.  Most of the incorrectness was subtle in this way, but we included it as a factor to see if it affected the relevant hints' placing.

The final four factors in \autoref{tab:hint-features} are inspired by feedback literacy theory, introduced by \citet{Mclean2015} and made known to us by its use in the study of programming error messages by \citet{Cucuiat2024}.  We adapted the four themes into categories of the same name by creating a set of definitions that applied the concepts to next-step hints, as follows.
Two researchers initially tagged some hints which were unused for the study, in order to calibrate.  Then they both categorised all 100 hints that were judged by educators for the four completed Snapshots.  All hints were tagged as yes/no for the four dimensions, giving $2^4=16$ possible overall categories.  In their initial independent tagging the researchers reached an agreement of 65\%.  The disputes were resolved in a meeting between the two researchers, and primarily revolved around clarifying the definition of the ``Developing understanding'' tag.    The resulting definitions after clarification are shown in \autoref{tab:codebook}.

\begin{table}
    \centering
    \begin{tabular}{p{4cm}|p{9cm}}
        Concept & Description \\ \hline
        Telling & The hint contains instructions on exactly what to change in the code.  This could be actual code (e.g. ``insert x = 0;'') or words that reach the same outcome (e.g. ``You need to assign 0 to x before the loop'' or ``you should start the loop at 0 not 1'').  The feedback requires no or little extra thought from the student other than to follow the instruction.  The instruction is either exact (delete this line) or close-to-exact (move this line to after the loop).  It does NOT include feedback that requires more thought or has ambiguity in exactly what needs doing (e.g. ``You need to update the loop variable somewhere within the loop'' or "The function call should not be inside the loop"). \\ \hline
        Guiding &  The hint contains explicit feedback about the original code above and beyond just direct accompaniment of what to change.  So if it says ``your loop is incorrect; you should start from 0'' this is only telling, not guiding.  This might include statements like ``your loop will run forever'' or ``you are not updating x anywhere after its declaration'' or ``consider whether the loop will terminate'' -- provided it does not also include an ensuing exact instruction on the fix (which would be categorised just as telling).  It can include positive specific feedback like ``Your loops are the correct structure''.  It does NOT include \textit{generic} feedback like ``you're so close'' or ``this is a great start'' which could be applied to any piece of code. \\ \hline
        Developing understanding & The hint contains a more general explanation of a concept or of the needed change to the code.  For example, it might explain why List cannot be indexed with square brackets in Java.  It may also point to places where students could find out more information for themselves (e.g. a URL or what concept to research).  The key is that it is explaining the general rule which would also apply to future coding, not just the specific issue with the current code (which would only be guiding).  This may be phrased as a general tip or tip-for-the-future, the key is that it is more general than just this specific code example. \\ \hline
        Opening up a different perspective & The hint contains a suggestion of a different approach to solving the problem (e.g. using a find method rather than manually looping through the list to search, or using a different programming language altogether).  This does NOT include cases where the student has not even made a coherent start; it needs to be in contrast to the student's current thinking or approach to the problem.  If they have done nothing, the suggestion might be telling or guiding depending how specific it is. \\ \hline
    \end{tabular}
    \caption{The complete definitions of the four concepts derived from feedback literacy theory~\cite{Mclean2015} that were used to manually characterise the hints.}
    \label{tab:codebook}
\end{table}

\begin{table}[ht]
\centering
\begin{tabular}{ccccr}
  \hline
Telling & Guiding & Developing & OpeningUp & Frequency \\ 
  \hline
    & \cellcolor{cyan!75}\checkmark &   &   &  56 \\ 
    & \cellcolor{cyan!75}\checkmark & \cellcolor{cyan!75}\checkmark &   &  17 \\ 
  \cellcolor{cyan!75}\checkmark & \cellcolor{cyan!75}\checkmark &   &   &   7 \\ 
  \cellcolor{cyan!75}\checkmark &   &   &   &   6 \\ 
  \cellcolor{cyan!75}\checkmark &   & \cellcolor{cyan!75}\checkmark &   &   3 \\ 
  \cellcolor{cyan!75}\checkmark & \cellcolor{cyan!75}\checkmark & \cellcolor{cyan!75}\checkmark & \cellcolor{cyan!75}\checkmark &   3 \\ 
  \cellcolor{cyan!75}\checkmark &   &   & \cellcolor{cyan!75}\checkmark &   2 \\ 
    & \cellcolor{cyan!75}\checkmark &   & \cellcolor{cyan!75}\checkmark &   2 \\ 
  \cellcolor{cyan!75}\checkmark &   & \cellcolor{cyan!75}\checkmark & \cellcolor{cyan!75}\checkmark &   2 \\ 
    &   & \cellcolor{cyan!75}\checkmark &   &   1 \\ 
  \cellcolor{cyan!75}\checkmark & \cellcolor{cyan!75}\checkmark & \cellcolor{cyan!75}\checkmark &   &   1 \\ 
    &   &   &   &   0 \\ 
    &   &   & \cellcolor{cyan!75}\checkmark &   0 \\ 
  \cellcolor{cyan!75}\checkmark & \cellcolor{cyan!75}\checkmark &   & \cellcolor{cyan!75}\checkmark &   0 \\ 
    &   & \cellcolor{cyan!75}\checkmark & \cellcolor{cyan!75}\checkmark &   0 \\ 
    & \cellcolor{cyan!75}\checkmark & \cellcolor{cyan!75}\checkmark & \cellcolor{cyan!75}\checkmark &   0 \\ 
   \hline
\end{tabular}
\caption{The frequencies of all possible different combinations of the four feedback literacy concepts (see \autoref{tab:codebook}) in the 100 hints, ordered by their frequency. \checkmark indicates the presence of the concept.}
\label{tab:fl-combinations}
\end{table}

The frequencies of the different combinations of concepts are shown in \autoref{tab:fl-combinations}.  More than half the hints contain ``Guiding'' with no other concepts.  With the hints being so similar in this regard, it will naturally reduce the discriminability that these feedback categories can provide.

\subsection{Hint scoring}
\label{sec:hint-scoring-metrics}

\begin{figure}
    \centering
    \includegraphics[width=0.8\textwidth]{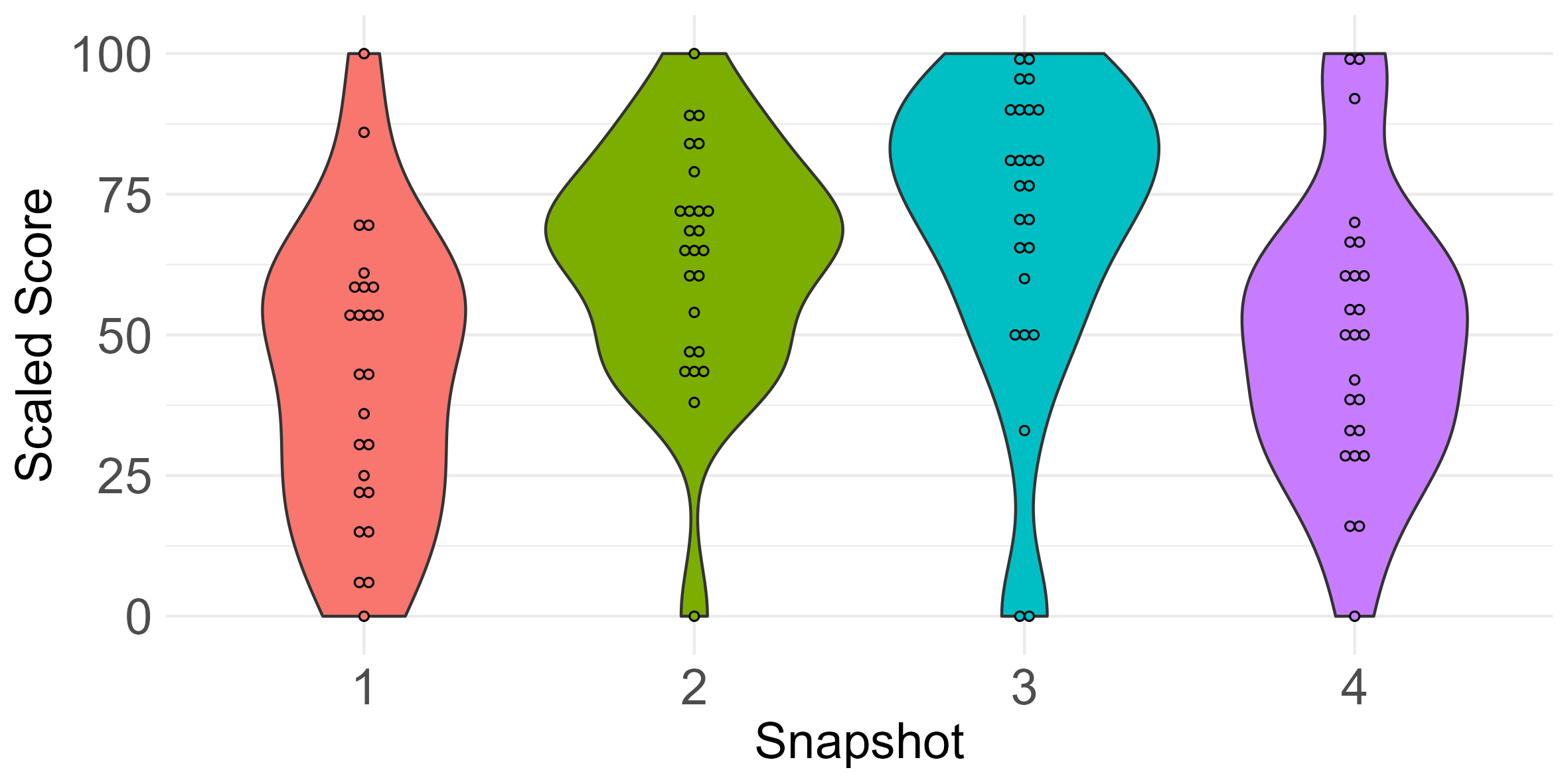}
    \Description{Four violin plots, with Scaled Score on the Y axis, and split by Snapshot on the X axis.  The 2nd and 3rd violins have a clear outlier at 0 on the graph, with the rest of the points at 30 or above.  This pattern is not present in the 1st and 4th violins.}
    \caption{The scaled scores of the hints, split by Snapshot on the X-axis.  Each dot is a hint, and the violin plot shows the same data as the hint dots, but is added to aid visualisation.  Snapshot 2 (and to a lesser extent Snapshot 3) have clear outliers at the bottom of the graph.}
    \label{fig:nmm-scaledscore}
\end{figure}

To arrive at a ranking, all Generators are applied across all Snapshots, and hints are scored for each Snapshot.  
Comparative judgement, as implemented in the NoMoreMarking platform that we used, supports two different ways of scoring hints.  One is a simple ranking: best hint, second-best, and so on.  The other is a ``scaled score'' which estimates how far apart the items are on a normalised scale (0 being the worst item, 100 being the best).  We had intended to use the latter as it was more informative.  However when we looked at the data we realised a problem.
As \autoref{fig:nmm-scaledscore} shows, in the case of Snapshot 2 (and to a lesser extent 3), one hint (or three hints) are so poor that the rest of the hints have their scores pushed up the scale.  Therefore a weak performance on Snapshot 2 is penalised less than on Snapshots 1 or 4 when we collapse a Generator's performance across Snapshots.  This would artificially skew the results, and thus we opted against using the scaled scoring.

This problem is to some extent present even in the hint rankings.  Since we have no cross-Snapshot normalisation (each participant only judged one Snapshot) we cannot determine whether, for example, all the hints on Snapshot 1 are better than all the hints on Snapshot 2.  But since each Generator provided a hint in all contexts, it at least counter-balances across the Snapshots.  We felt that using the ranks was a better choice than scaled score to minimise the effect of this warping within Snapshot.

\subsection{Hint rankings}

Based on the reasoning in the previous section, we thus used the ranks of the hints within their Snapshot as our dependent variable.  The ranks were mapped to ascending scores to be more intuitive, so that a rank score of 25 was the best hint for a Snapshot, 24 the second-best, down to 1 for the worst hint.  Thanks in part to comparative judgement being a forced-choice paradigm, there were no ties.  The scores are thus ``zero sum''; there are 4 scores of 1 (one for each Snapshot), 4 scores of 2, etc, up to 4 scores of 25.  The scores are all relative: if one hint is better, the others must be correspondingly worse.  The midpoint (both mean and median) hint is thus 13 by definition, and what is of interest is which factors lead to higher placings.  This metric is termed \textit{RankScore} and is the way we compare the quality of the hints for the remainder of the paper.

The rankings of the specific hints by themselves are not of direct interest in this paper (although the full set of hints and their ranks can be seen in our OSF repository); the interest lies in the accompanying factors, such as which Generator produced the best hints, or which attributes (see \autoref{sec:human-categorisation}) were associated with high-ranking hints.  To give a sense of the hints, \autoref{fig:task1} shows the code and the best and worst hint for Snapshot 1.

\begin{figure}
    \raggedright
    \hrule
    \vspace{2mm}
    \textit{Student code:}
    \begin{lstlisting}
    public static boolean isPalindrome(String input)
    {
        int length = input.length()-1;
        String palindrome="";
        for (int i = length; i == 0; i--) {
            System.out.println(i);
            char letter = input.charAt(i);
            palindrome = palindrome + letter;
            System.out.println(i);
        }
        if (palindrome == input)
        return true;
        else
        return false;
    }
    \end{lstlisting}
    \vspace{2mm}
    \hrule
    \vspace{2mm}
    \textit{Notes:}\\  There are two core problems in this code. One is that the for-loop condition for continuing is wrong: it should be \lstinline$i >= 0$ rather than \lstinline$i == 0$, which means the loop will not run at all for most inputs. The other problem, even if the loop did run, is that the code uses \lstinline$==$ to compare strings after the loop, rather than \lstinline$.equals()$.
    \vspace{2mm}
    \hrule
    \vspace{2mm}
    \textit{Best hint (GPT-4, Prompt 2):}\\ There are a few things you might want to reconsider in your code.
    \begin{enumerate}[1.]
\item Look at your for loop condition. You've set it to \lstinline`i == 0`, which means the loop will only execute if \lstinline`i` is equal to \lstinline`0`. Is that what you intended? 
\item You're comparing strings using \lstinline`==` operator. In Java, \lstinline`==` checks if two references point to the exact same object, not their equality in terms of content. You might want to use a different method for string comparison.
\item Also, consider the case sensitivity of your input string. Does your method handle both lower and upper case letters?
\end{enumerate}
Keep these points in mind and try to revise your code. Good luck!
    \vspace{2mm}
    \hrule
    \vspace{2mm}
    \textit{Worst hint (GPT-3.5, Prompt 5):}\\ Your loop condition should be \lstinline`i >= 0` to iterate over the characters correctly. Remember to use \lstinline`.equals()` method instead of \lstinline`==` for string comparison, as \lstinline`==` checks for reference equality. Additionally, consider using \lstinline`StringBuilder` for efficiency when building the palindrome string inside the loop. This will help you understand why string comparison doesn't work as expected and improve the performance of your code.
    \vspace{2mm}
    \hrule
    \Description{}
    \caption{The top shows the code from Snapshot 1 (which was reproduced exactly, including the student's original spacing and indentation), followed by the notes which we gave to participants to explain the problems with the code.  Below that we show the best hint and the worst hint (and their source) as ranked by the participants in our study.}
    \label{fig:task1}
\end{figure}

\subsection{Hint rankings by generation mechanism}

\begin{figure}
    \centering
    \includegraphics[width=0.7\textwidth]{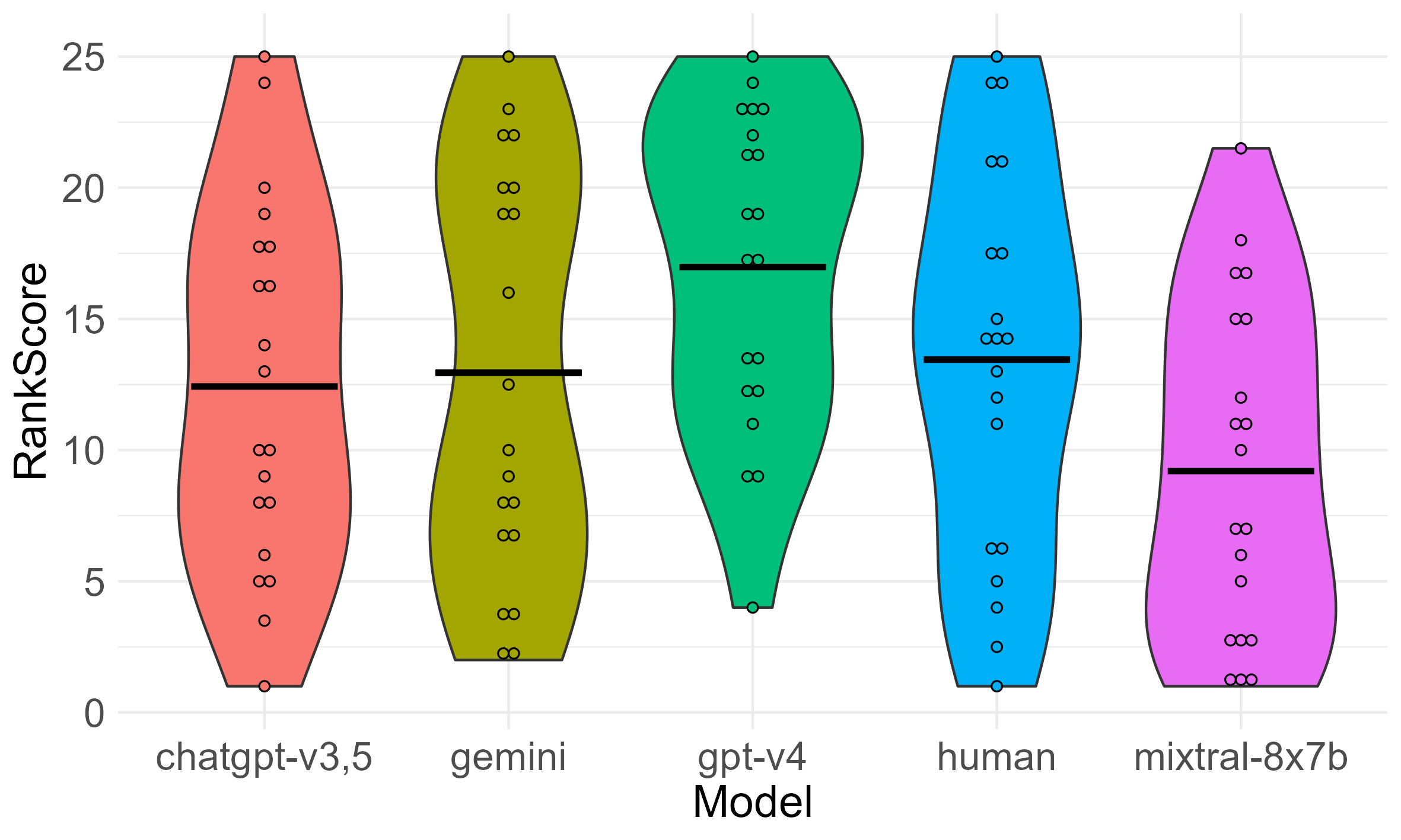}
    \Description{A violin plot with RankScore on the Y axis and Model on the X axis.  The violins all have points spanning almost the whole vertical range of the graph (1--25), although GPT-4's lowest point is 4 and second lowest is 8.  There are black lines in each violin showing the mean rank, and GPT-4 is about 17, followed by human at about 13, followed closely by Gemini then GPT-3.5, with Mixtral-8x7B lower at about 9.}
    \caption{The rank score of the hints (higher is better) split by the model that produced them, with different prompts averaged to a single value for each LLM.  Each dot is a hint; the violin plot shows the same data as the hint dots and is added to aid visualisation.  The black horizontal line is the mean rank of the hints generated by that model.}
    \label{fig:rank-by-model}
\end{figure}

\begin{figure}
    \centering
    \includegraphics[width=0.7\textwidth]{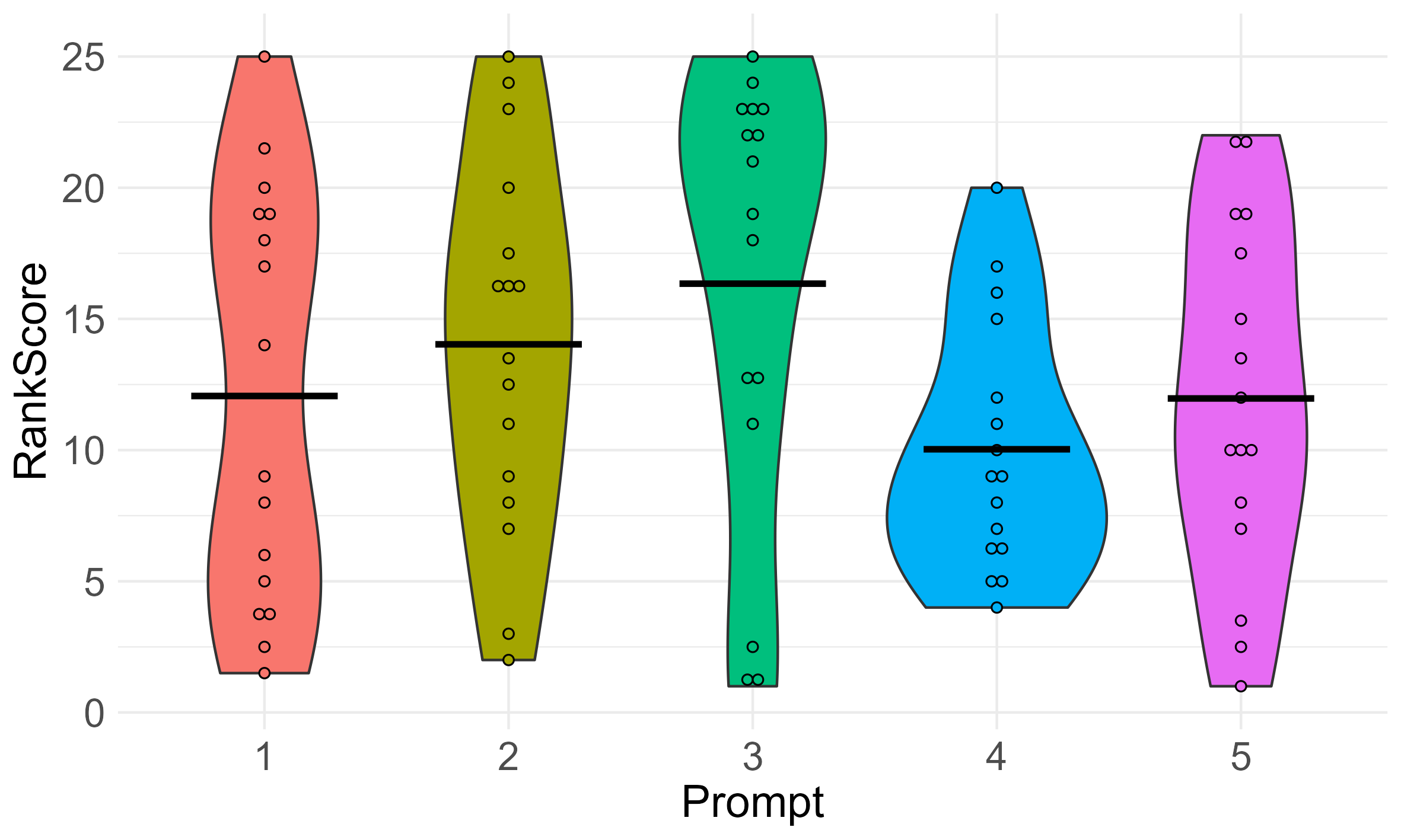}
    \Description{A violin plot with RankScore on the Y axis and Prompt on the X axis.  Most violins span the full range from 1--25, except Prompt 4 which is only in the range 4--20.  Black horizontal lines indicate the mean rank, and are around 16 for Prompt 3, then followed (with equal spacing) by Prompts 2, 1 and 5 equal, and 4, with Prompt 4 having mean rank of 10.}
    \caption{The rank score of the AI-generated hints (higher is better) split by the prompt used to generate them.  Each dot is a hint; the violin plot shows the same data as the hint dots and is added to aid visualisation.  The black horizontal line is the mean rank of the hints generated by that model.}
    \label{fig:rank-by-prompt}
\end{figure}

\autoref{fig:rank-by-model} shows the RankScore of the hints grouped by the Generator that produced them (with all five prompts collapsed into a single score for each LLM).  As can be seen, GPT-4 has the highest mean rank, followed by humans, with Mixtral-8x7B the worst.
For the AI-generated hints, \autoref{fig:rank-by-prompt} shows the rank scores split by the prompt used to generate them.  The prompt numbers correspond to \autoref{tab:prompts}.  Prompt 3 produces the best results, and prompt 4 produces the worst results.

\begin{figure}
    \centering
    \includegraphics[width=0.8\textwidth]{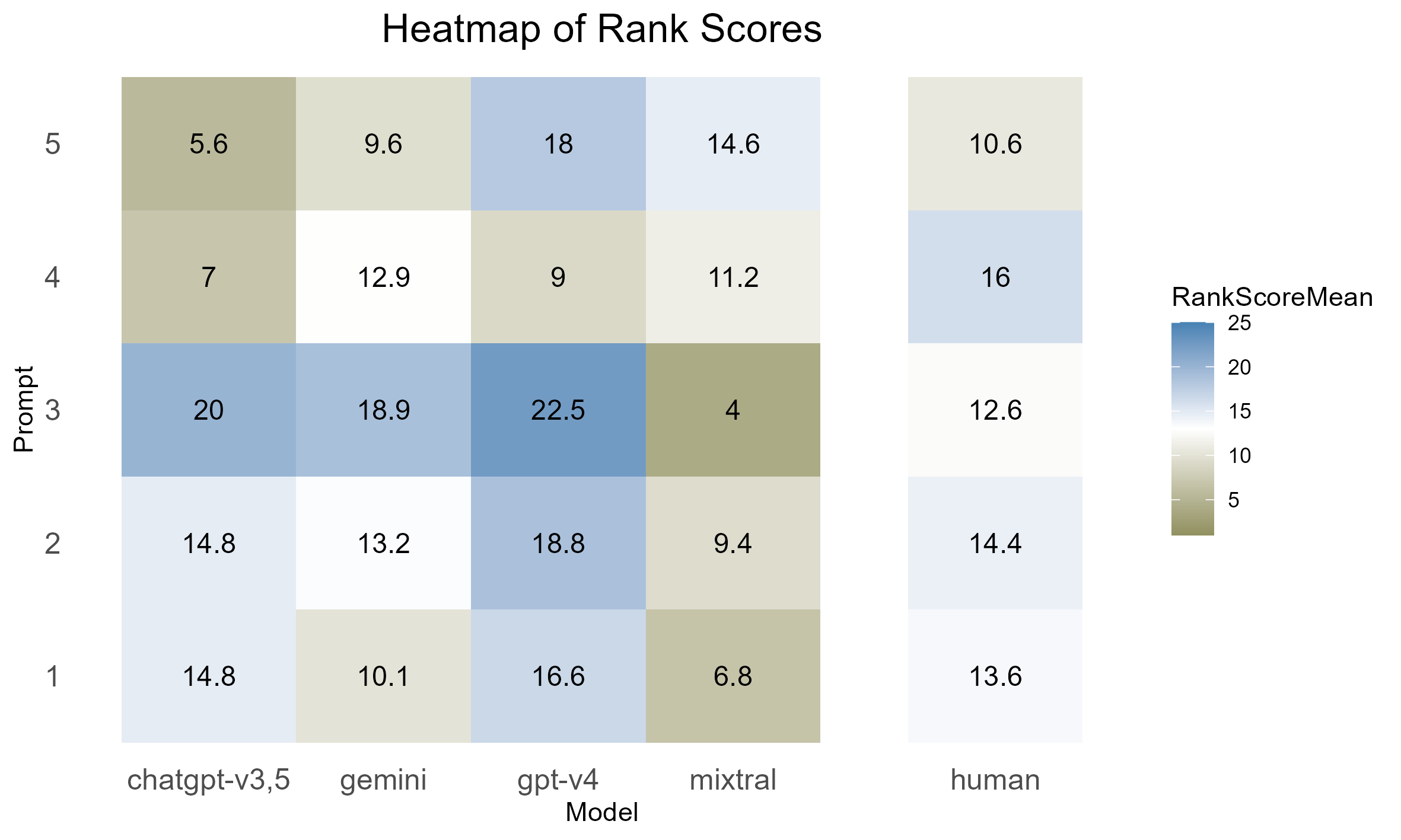}
    \Description{A 5 by 4 grid of heatmap numbers for prompt and LLM, plus an adjacent 5 by 1 heatmap for human.  Prompt 3 is clearly best (mean ranks 20, 18.9, 22.5) except when paired with the Mixtral model (mean rank 4).  Similarly GPT-v4 is clearly best (mean ranks 16.6, 18.8, 22.5, 18) except when paired with Prompt 4 (mean rank 9).  The humans are all middling (mean ranks 10.6--16).}
    \caption{A heatmap of AI model vs prompt (plus the five humans), showing the mean rank for that Generator.  Each entry is derived from only four points (one per Snapshot) so it should be interpreted with caution.}
    \label{fig:full-heatmap}
\end{figure}

We have 100 hints overall: (5 AI models $\times$ 4 prompts + 5 human hints = 25 Generators) $\times$ 4 Snapshots.  Thus for each AI model we have 20 data points (5 prompts $\times$ 4 Snapshots), for each prompt we similarly have 20 data points (5 models $\times$ 4 Snapshots) but for each Generator we only have 4 data points (one per Snapshot).  Therefore we must be very cautious in interpreting this data because we may be interpreting noise; nevertheless it is shown in \autoref{fig:full-heatmap}.  This seems to indicate there may be an interaction between prompt and model as to which is best.  For example, although prompt 3 is generally best, it interacts very poorly with the Mixtral model.  Similarly, GPT-4 is the best model but does poorly with prompt 4.  The five humans who generated hints are shown in the same graph; it would appear that there is less variation between humans than between models or between prompts.  

\subsection{Hint rankings by hint characteristics}

The analysis by hint characteristics is potentially completely orthogonal to the analysis of the hint generation mechanism.  Here we are interested in attributes of the hints themselves regardless of how they were generated.  We used the hint attributes described in \autoref{sec:hint-attributes} to see what effect they had on the hints' RankScore.

To perform the analysis we used random forests~\cite{Ho1995}.  Decision trees are a classic data mining method where the data is used to create a binary tree that classifies the outcome variable, by splitting the values around a point (e.g. perhaps hints having a word count above 300 leads to a lower RankScore).  The problem with decision trees is that they tend to overfit the data.  Random forests solve this problem by extracting 500 random subsets of the data and then fitting a decision tree to each subset, yielding 500 trees.  The results of these 500 trees are then averaged in a ``forest'' to form a classifier.  This classifier could potentially be used to predict new rank scores based on a new hint, but here we are solely interested in introspecting which hint attributes were important for the classification of best hints and in what way (e.g. is higher better).

The advantage of a random forest over classical statistical methods is that they can identify complex non-monotonic patterns.  Regressions are generally monotonic: for example, longer hints are worse or longer hints are better, but not a pattern in a U-shape or other non-linear pattern.  Random forests can identify arbitrary variation in patterns.

The first output to check in a random forest is the \textit{importance} of each input attribute.  Importance (technically, the percentage increase in mean squared error of the outcome when the input factor is omitted from the model, higher means the factor is more important, 0 or negative means totally unimportant) tells us which factors most influenced the outcome variable, although it does not indicate whether it was a positive or negative or mixed influence.  The ranking by importance is given in \autoref{tab:rf-importance}.

By far the two most important factors were word count and reading level, followed by the feedback literacy item of ``Opening up a different perspective''.  Note that Model was relatively unimportant, suggesting there are few lasting effects of how the hint was generated, once the other factors are taken into account.

\begin{table}
    \centering
    \begin{tabular}{lr}
Input factor & Importance \\ \hline
WordCount&               23.9\\
FleschKincaidGradeLevel& 17.9\\
OpeningUp&               10.7\\
Guiding&                  5.6\\
PartiallyIncorrect&       5.5\\
Model&                3.0\\
Telling&                  0.5\\
Sentiment&                0.2\\
DevelopingUnderstanding& -0.7\\
\end{tabular}
    \caption{The factors in the random forest model (outcome variable: RankScore) and their importance (percentage increase in mean squared error of the outcome if omitted from the model).  Higher importance means the attribute was more important in predicting the RankScore of each hint.  Zero or negative means that the factor was unimportant.}
    \label{tab:rf-importance}
\end{table}

\begin{figure}
    \centering
    \begin{tabular}{cc}
    \includegraphics[width=0.47\textwidth]{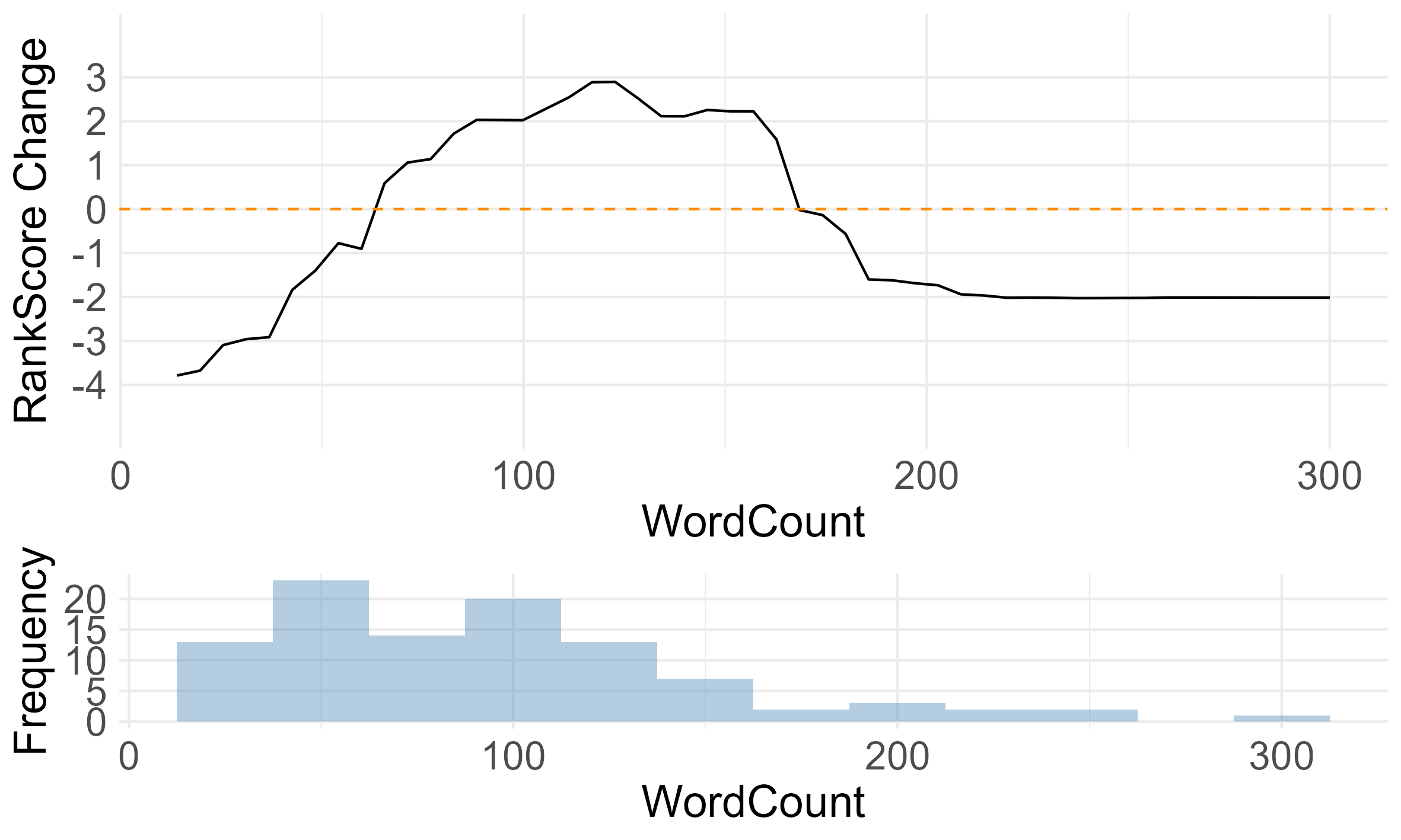}&
    \includegraphics[width=0.47\textwidth]{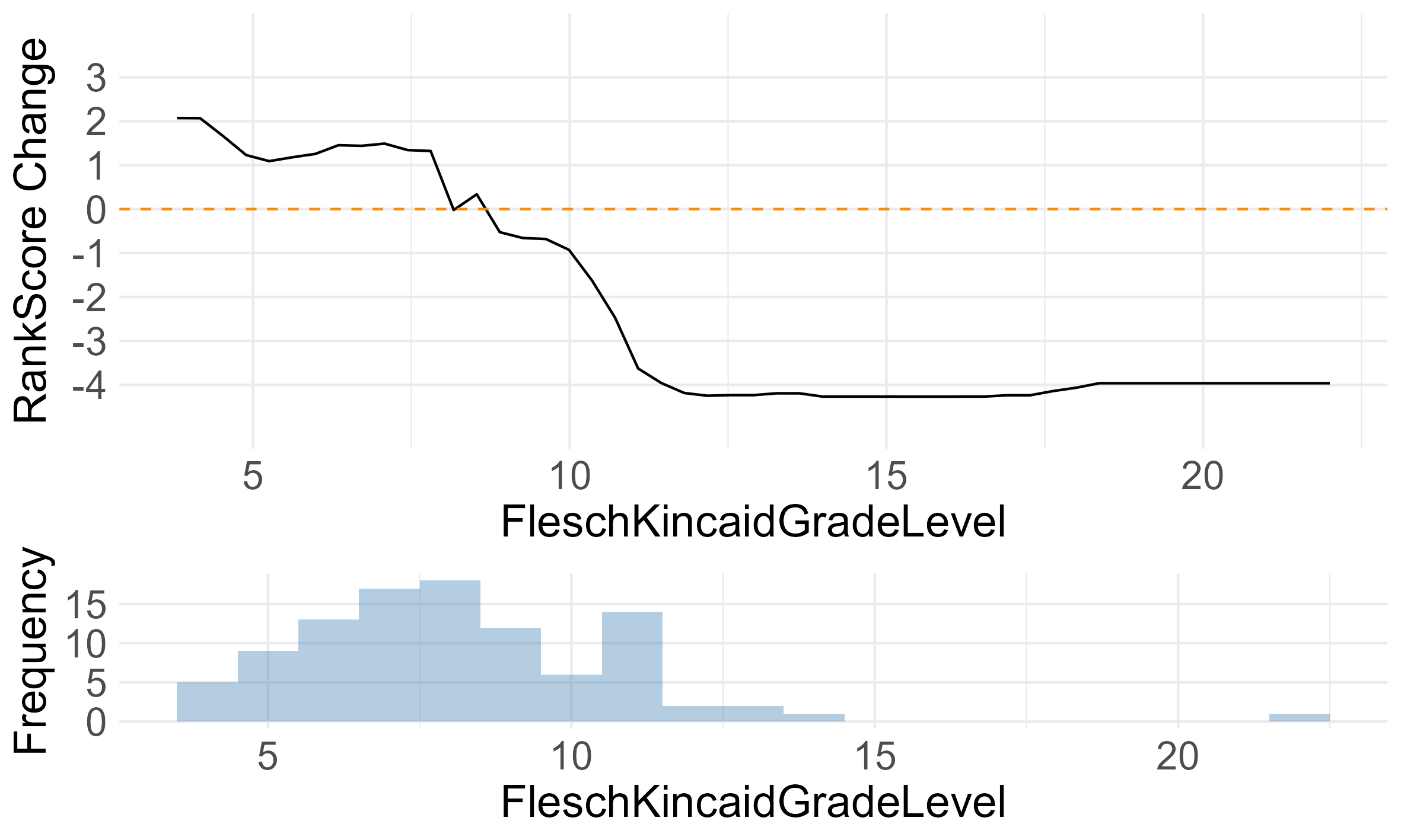}\\~\\~\\
    \includegraphics[width=0.47\textwidth]{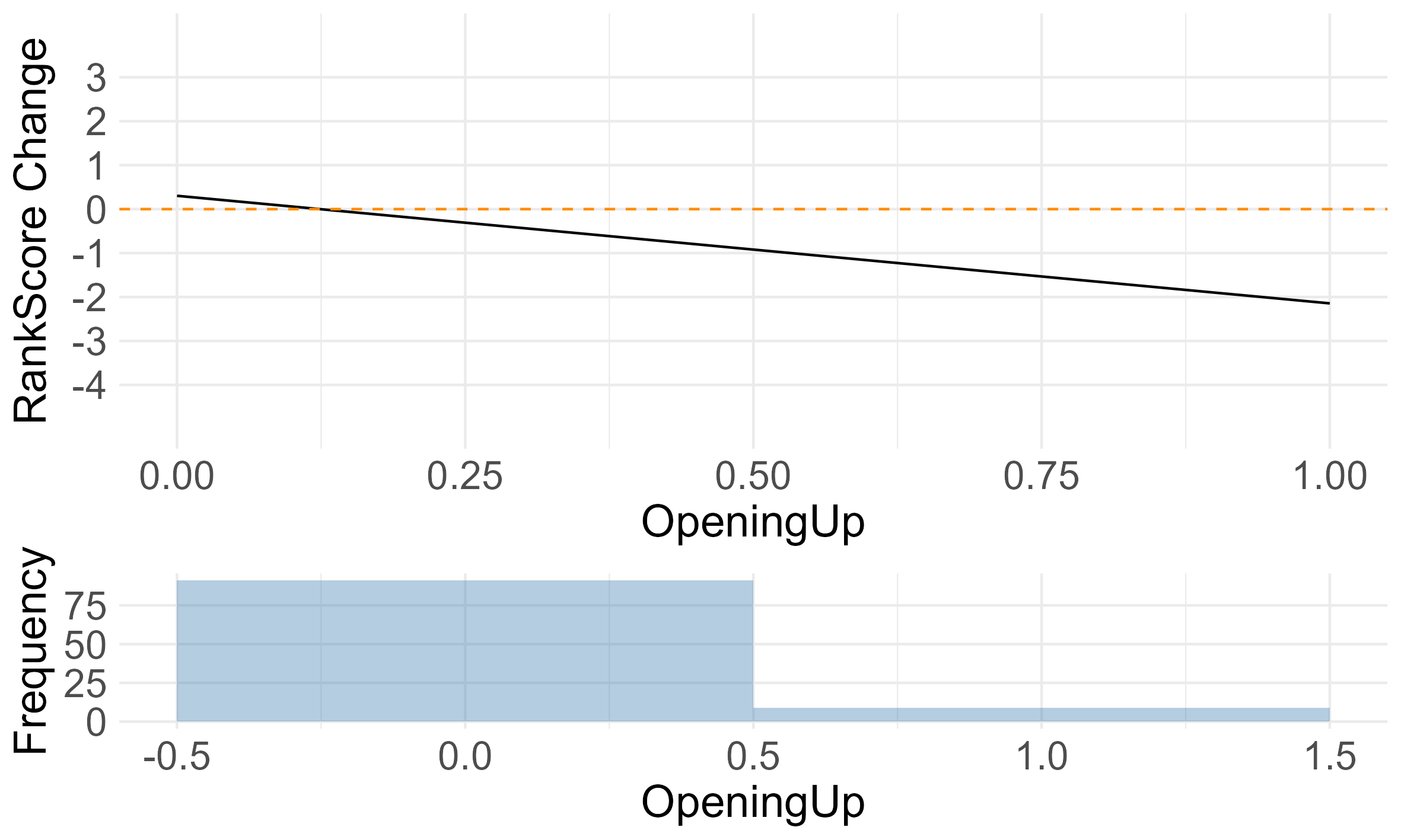}&
    \includegraphics[width=0.47\textwidth]{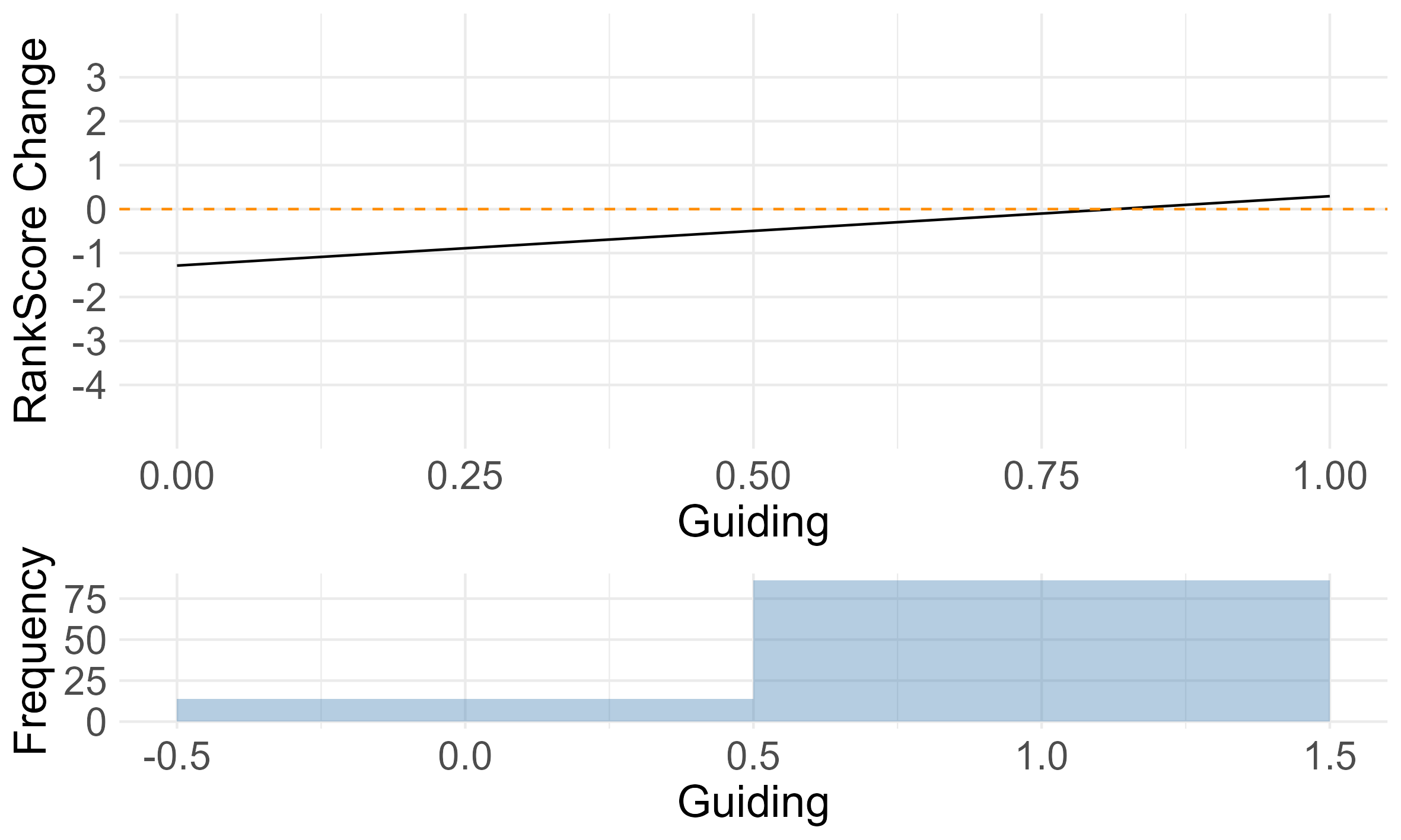}\\
    \end{tabular}
    \Description{Four plots of factors: Word Count, Flesch Kincaid Grade Level, Opening Up and Guiding.  Word count has an inverted U-shape, with the best region (the tip of the inverted U) at 80--160 words.  Reading grade level has a cliff-edge pattern, with a gradually descending line that suddenly drops around grade 9 or 10.  Opening up is worse for value 1 and Guiding is better for value 1 (both are booleans with values 0 or 1, so the plot shows a straight line between these two points).}
    \caption{Partial dependency plots for the four most important factors in predicting which hints are best.  The plots are in vertical pairs; the top-plot shows the effect on the RankScore on the Y axis (above the orange line: better than average) by value of the attribute on the X axis.  The bottom plot of each pair is a histogram showing the frequencies of those values in the 100 hints.}
    \label{fig:rf-pdp}
\end{figure}

To visualise the effect of the important attributes, in \autoref{fig:rf-pdp} we graph partial dependency plots, which show the effect on RankScore for each value of the attribute.  The dotted line across each shows the baseline, with each value of the hint attribute potentially increasing RankScore (better hint, above the dotted line) or decreasing it (worse hint, below the dotted line).

The results for word count reveal a ``sweet spot'' where hints that are 80--160 words long are ranked highly, around 4--5 places higher than hints with word counts below or above this range.  Short hints are rated particularly poorly.

The results for reading level show that a lower grade reading level is better.  The grade level scale here corresponds to grade levels in US schools, so for example a reading grade level of 9 (where the hint quality suffers a sudden drop) corresponds to 14 year-olds.  So any hints not understandable by fourteen year olds are rated around 5 places lower than hints understandable by thirteen year olds (grade 8) and younger students.

The results for \textit{opening up different perspectives} show that hints which suggest an alternative approach to the problem are ranked 2 places \textit{lower} than hints which do not.  Whereas \textit{guiding} hints which offer additional guidance beyond exactly what to change are ranked around 2 places higher.  See \autoref{tab:codebook} for the definitions of these items.

\section{Educator survey}

As well as performing the comparative judgement task, we asked participants to complete a short survey.

We asked about their experience of teaching Java.  Our past experience in other studies had suggested that a simple numeric field (e.g. ``How many years have you taught Java?'') was insufficient to capture the wealth and variety of experience.  We asked them for a free text entry describing their experience of teaching Java.  We then ranked these responses (using comparative judgement, but with the researchers as judges) on a loosely defined ``Java educator experience'' basis.  This allowed us to sort the participants by experience and thus we can summarise their experience with an upper quartile, median and lower quartile:
\begin{itemize}
    \item Upper quartile:  ``I have started teaching Java in 1998 and
have 30+ years of teaching experience as a TA, scientific researcher, and educator. Most of it was done in Java.''
\item Median experience: ``Teaching Java for more than twenty years,  have taught Pascal, C, C++ before...''
\item Lower quartile: ``I have taught Java programming to High
School students about 8 years. I also teach
Scratch, HTML, Javascript...''
\end{itemize}

\begin{figure}
    \centering
    \includegraphics[width=0.7\linewidth]{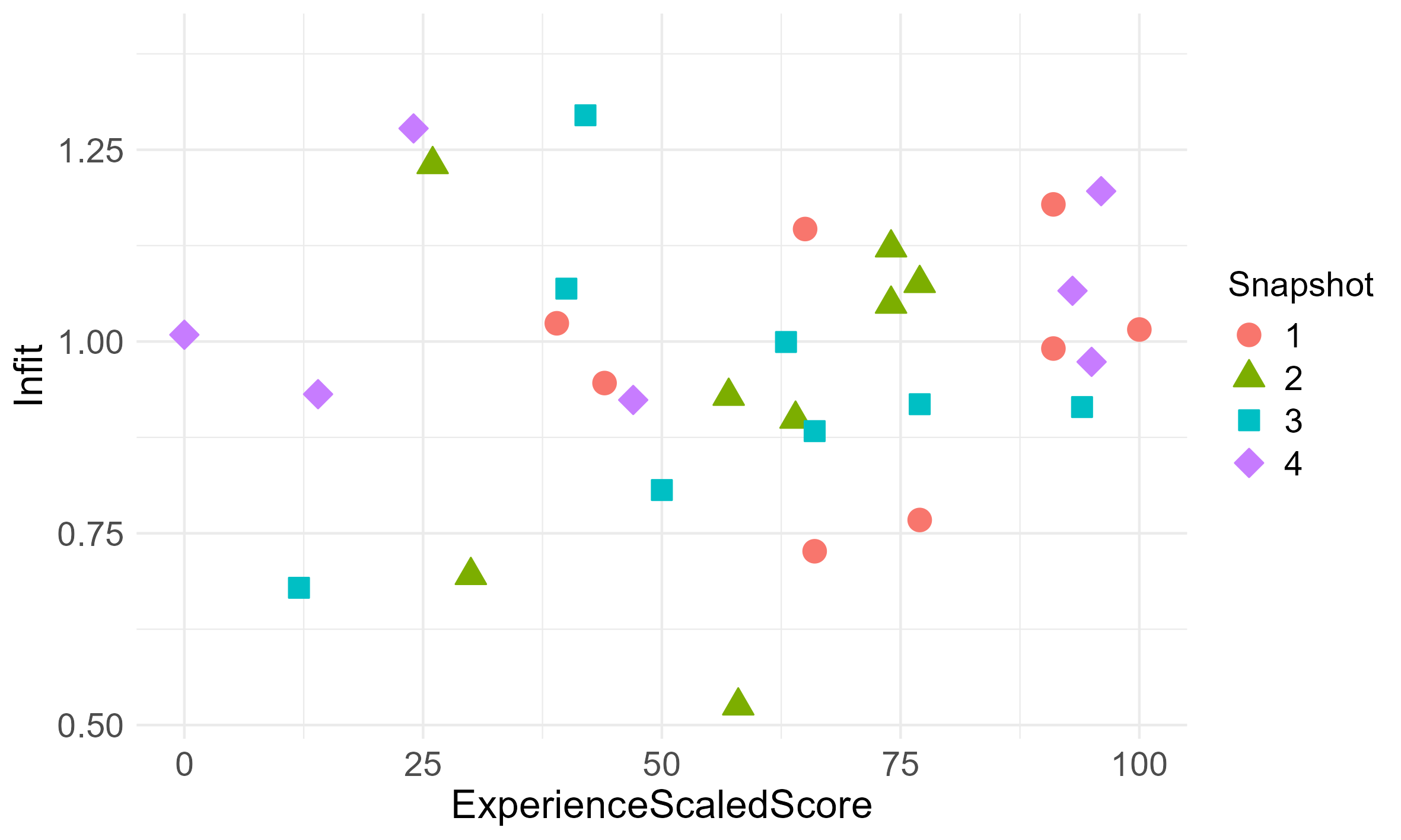}
    \Description{A scatter plot showing Infit against the experience scaled score.  The graph shows no discernable pattern.}
    \caption{A participant's experience (a scaled score ranked by researchers using comparative judgement) against their Infit (how much they agree with fellow participants: lower values of Infit indicate higher agreement with their peers).}
    \label{fig:infit-by-experience}
\end{figure}

The comparative judgement also gave us a scaled score (as described earlier in \autoref{sec:hint-scoring-metrics}) from 0 to 100.  We could then plot this against the (described earlier in \autoref{sec:infit}) Infit to see if there was a relationship between experience and agreement with peers, as shown in \autoref{fig:infit-by-experience}.  A linear regression confirmed there was no effect ($p = 0.60$) of experience on Infit.

\begin{figure}
    \centering
    \includegraphics[width=0.8\linewidth]{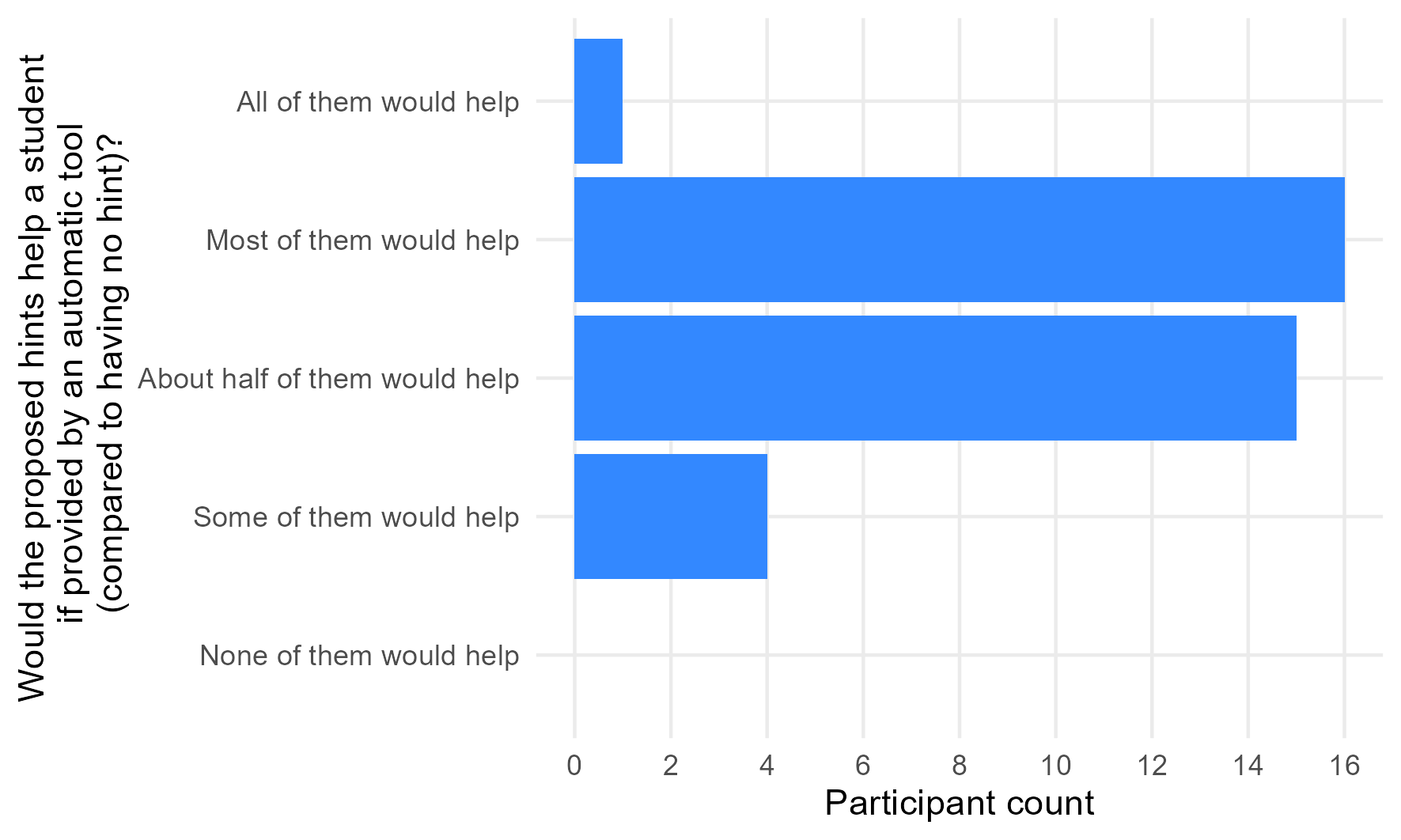}
    \Description{A bar chart showing the Likert results.  1 participant stating ``All of them would help'', 16 said ``Most of them would help'', 15 said ``About half of them would help'' 4 said ``Some of them would help''.  Zero said ``None of them would help''.}
    \caption{Results of asking the participants whether the hints would better than having no hints on a 5-point Likert scale.}
    \label{fig:judges-helpful}
\end{figure}

One important survey question related to the overall opinions of the hints.  Because our comparative judgement task is entirely relative, it cannot tell us whether all the hints were good or all the hints were bad, or somewhere inbetween.  For this purpose we asked the participants how the hints compared to having no hint, and the results are shown in \autoref{fig:judges-helpful} -- the judges generally thought that most of the hints would be helpful.  We asked the participants why they thought the hints were helpful (or not) as a free-text response, and performed a miniature thematic analysis to analyse these responses -- the counts of different themes are given in \autoref{tab:hint-why-opinions}.

\begin{table}
    \centering
    \begin{tabular}{p{9cm}|r}
    Theme & Participant count \\ \hline
    \textbf{Help not hinder independent learning:} Participants mentioned that hints should aid independent learning (e.g. by giving cause for the students to think) rather than hinder it (e.g. by providing the exact solution to the student with no need for further thought). & 10 \\ \hline
    \textbf{Context matters:} Participants mentioned that they needed to understand more about the context of the student receiving the hint in order to decide whether the hint was appropriate or whether it needed further adjustment. & 6 \\ \hline
    \textbf{One at a time:} Participants expressed a dislike for hints which addressed many errors at once, and stated they would prefer a hint which identified and focused on solving only one problem with the code. & 6 \\ \hline
    \textbf{Too long or complicated:} Participants stated that some hints were too long or complicated to provide any benefit to a student, and expressed doubt that the students would read and/or understand such hints. & 5 \\ \hline
    \end{tabular}
    \caption{The themes we identified in participants' responses about why the hints they saw would (or would not) be better than having no hint, plus the count of unique participants (out of 35) who mentioned this theme.  The table is sorted by frequency.}
    \label{tab:hint-why-opinions}
\end{table}

Participants could also offer their opinion on what they thought was important in a good hint, as a free text response.  We similarly performed a miniature thematic analysis to analyse these responses, and the counts of different themes are given in \autoref{tab:hint-attribute-opinions}.

\begin{table}
    \centering
    \begin{tabular}{p{9cm}|r}
    Theme & Participant count \\ \hline
    \textbf{Conciseness:} Participants preferred short, concise, to-the-point hints. & 22 \\ \hline
    \textbf{Hint not solution:} Participants wanted hints that did not provide the exact solution, but rather a pointer or suggestion or thought-provocation that would involve the student thinking further. & 19 \\ \hline
    \textbf{Not over-praising:} Participants disliked hints that over-emphasised praise or positive language. & 11 \\ \hline
    \textbf{Specificity:} Participants preferred hints that were specific rather abstract and vague. & 9 \\ \hline
    \textbf{Correctness:} Participants mentioned wanting correct, accurate hints (usually mentioned  because they had spotted a hint they felt was incorrect). & 9 \\ \hline
    \textbf{Positive tone:} Participants liked a positive or encouraging tone to the hint. & 8\\ \hline
    \textbf{Not too short:} Participants mentioned disliking very short hints as being unhelpful or lacking in useful detail. & 5\\ \hline
    \textbf{Unhelpful summary:} Participants mentioned disliking the tendency for hints to contain a summary of what the code was doing or trying to do, because they felt this was unhelpful. & 4 \\ \hline
    \end{tabular}
    \caption{The themes we identified in participants' responses about which characteristics of hints were important to their ranking choices, plus the count of unique participants (out of 35) who mentioned this theme.  The table is sorted by frequency.}
    \label{tab:hint-attribute-opinions}
\end{table}

\begin{figure}
    \centering
    \includegraphics[width=0.8\linewidth]{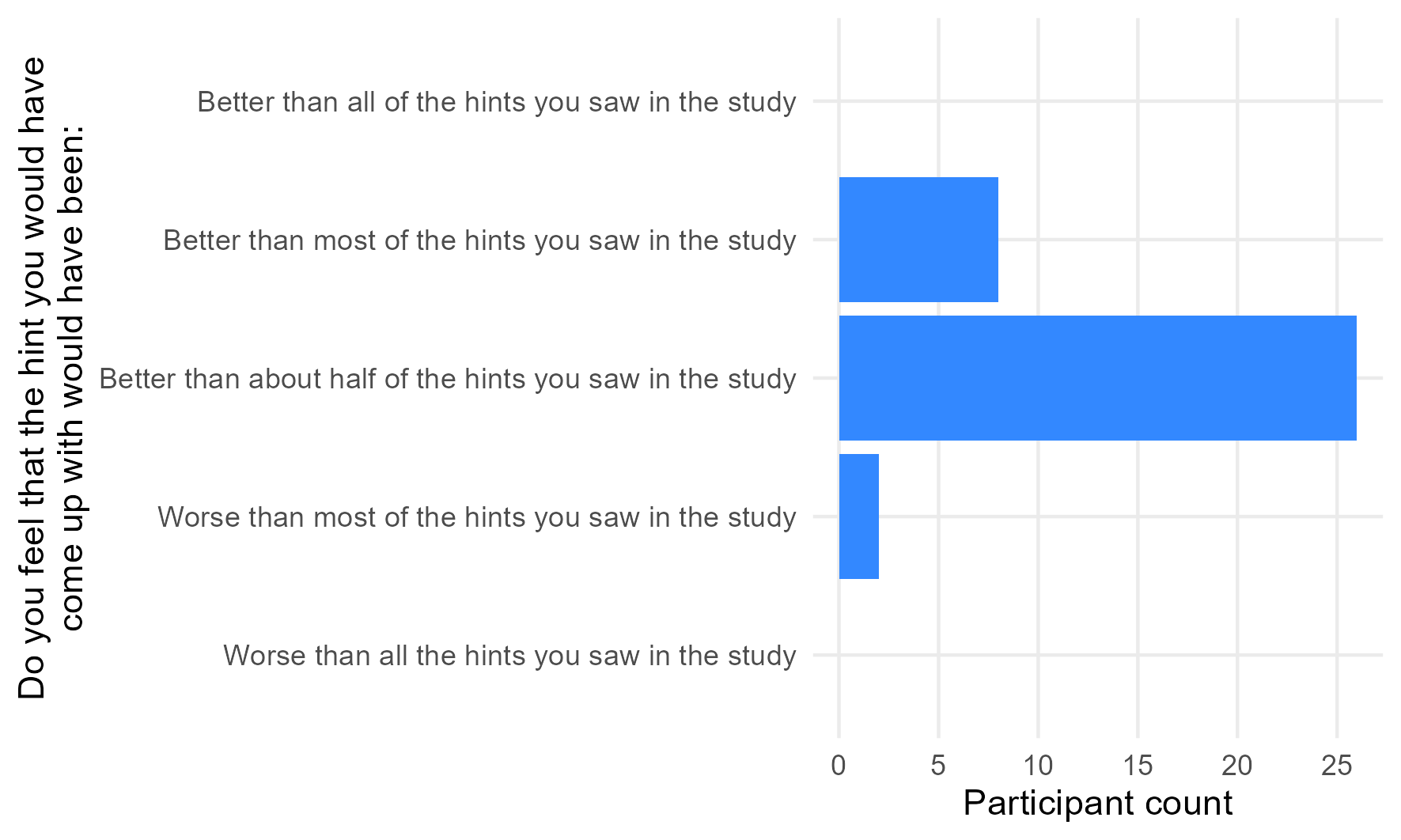}
    \Description{A bar chart showing the Likert results.  Zero participants stated ``Better than all of the hints you saw in the study'', 7 or 8 said ``Better than most of the hints you saw in the study'', 26 said ``Better than about half of the hints you saw in the study'' 2 said ``Worse than most of the hints you saw in the study''.  Zero said ``Words than all the hints you saw in the study''.}
    \caption{Results of asking the participants whether they felt they could make better hints than those in the study, on a 5-point Likert scale.}
    \label{fig:judges-better}
\end{figure}

We also asked participants to state whether they felt they could do better themselves.  The results are shown in \autoref{fig:judges-better}.  It is important to interpret this finding in light of the experience result described earlier in this section.  Over half of our participants had the equivalent of 20+ years of Java teaching experience, and yet the vast majority of them felt their hints would be around the median hint in the study.  This matches with our results which show that the human-generated hints from the researchers (several of whom would be in the top half of experience in the study) were around the median.  We interpret that the hints in the study were generally considered high quality.

In a slight oversight on our part, we did not explicitly ask the participants whether they thought the hints were AI-generated.  This was not part of our research questions but in retrospect it may have been useful to ask.  Four participants spontaneously made reference to AI or LLMs in their responses; one said the hints ``feel like more of what an AI might respond with'', one said ``One hint seemed to contain a bit of a [LLM] prompt.'', one said ``LLM / ChatGPT levels of positivity would be irritating over time'', and one suggested that another research group ``is also experimenting with AI-generated hints''.  We suspect that most participants inferred or assumed that the hints were AI-generated; the surprise for them might instead have been that some were human-generated, rather than all being AI-generated.

\section{Discussion}
\label{sec:discussion}

This study has findings in multiple dimensions, which we will discuss in turn.

\subsection{Hint characteristics}

We analysed the ranking of hints against their characteristics in order to investigate which characteristics of the hints were most important.  We found that the two most important aspects were length of the hint (with 80--160 words being ideal) and the reading level (with US grade level 9 or lower, i.e. understandable by 14 year-olds or younger being ideal).  Pedagogical aspects of the hints, based on feedback literacy theory~\cite{Mclean2015,Cucuiat2024} were less important; inclusion of alternate approaches to the solution were found to \textit{decrease} a hint's rating, while including guidance beyond stating the answer \textit{increased} a hint's rating -- but the last item was the least influential of the four.  We found no effect of sentiment on the hints' rating (most hints were positive, but in a wide range from slightly to very positive), and whether the hint highlighted a general rule (e.g. why the \lstinline$==$ operator cannot be used for string comparison in Java) also had no effect.

These results can provide useful lessons for educators and tool-makers about the best kind of hints to provide in contexts where short written hints are appropriate.

\subsection{Hint generation}

We asked educators to compare AI-generated hints from different AI models and different AI prompts, as well as human hints, without knowing which hints were generated by which method.  We found that the best model in our study, GPT-4, produced hints that were rated more highly than hints produced by the [human!] researchers.  This is promising for future research into adding hints in novice programming environments.

We found that there was as large a variation among prompts as there was among AI models.  This is important for the automatic generation of hints, but also has implications for students' individual use of LLMs for help.  Previous research~\cite{Zamfirescu2023} has found that non-experts can struggle to design prompts, so students may struggle to create a prompt themselves that produces a hint as good as the best hints in this study.  This suggests that there is room for tools to ``package up'' pre-written prompts and automatically deliver hints using these, rather than exposing the raw LLM prompt interface to users.

AI is currently undergoing rapid development, with new models being introduced every few months.  In that regard, the specific models will outdate, and some of our results along with it.  To ensure ours is a lasting contribution, we have detailed a reproducible method that allows a replication of the study to be run in future.  Specifically: we described our process of hint creation, we detailed a methodology of running multiple prompts with multiple models, and how to use comparative judgement to evaluate these hints.  All of our analysis scripts are in our OSF repository (see \autoref{sec:osf-repository}) to allow for easy replication.

Neither the researchers nor the AI models knew the exact context of the Snapshot (since this is not available in Blackbox), i.e. what precisely the student was aiming to do.  All of them inferred the student's task based solely on the student's source code.  This is in contrast to large portions of the hint-generating literature which rely on knowing the problem context to provide hints~\cite{McBroom2021,Crow2018,Keuning2018}.


\subsection{The missing student perspective}
\label{sec:missing-student-perspective}

This study has only looked at educators' ranking of hints.  Naturally, it would make sense to also investigate the student perspective of hints since they would be the ultimate consumers.  We did not feel that students were likely to be able to rate hints in the same way given the complexity of the task: participants first need to read someone else's poorly-written code, understand what the code does and what the code is trying to do, understand the notes on the current problems with the code, and then read two hints and compare them to decide which would be the most useful.  Our educator participants seemed able to complete this task (backed by their decades of experience).  We were, however, not convinced that students could do the same, so we did not ask students to also perform this task.

A better design for evaluating hint quality for students might be to ask them to complete a given programming task, and when they get stuck, to be able to ask for a hint.  Students could be shown two hints and be asked to select the preferred one.  This would remove the complexity of understanding someone else's code.  Students would already know what they are trying to do, making evaluation of the usefulness of the hint more straightforward.  It is possible to build the best prompt and model from this study into a tool to conduct such a second study.

With our current study design, it remains a possibility that our experienced educators are not very good at the task of deciding which hints would be most useful to students.  For example, it may be that students prefer even shorter hints, or perhaps they favour more explanation.  Perhaps students prefer being told the answer to receiving a hint.  This is a classic educational conflict: who is best-placed to decide which hint is better?  A student who perhaps wants the easy option of being told the answer or given a detailed explanation, or the educator who believes they are best-served by receiving a more circumspect hint?  It is not obvious that either the student or the educator alone can provide the perfect assessment of which hint is best.  In this study we have provided a large piece of the puzzle by asking educators.

\subsection{The nature of hints}

In this study we have chosen to focus on ``one-shot'' next-step hints which are provided to the student by a programming environment to help them move forwards.  We have not considered any aspects of interface design, for example whether students should be able to manually request these hints at any time or whether they should be automatically offered or at what point.  We consider these aspects to be outside the scope of this work, but they may be investigated in separate research.  \citet{Jeuring2022} and \citet{Lohr2024}, for example, investigated when to provide hints, but found low agreement among educators.

We have also not considered the possibility of an ongoing dialogue between the student and the hint mechanism.  One of the distinctive features of systems such as ChatGPT is the ability to converse with the LLM, asking for more detail, for clarification, or working in tandem together.  Although part of our analysis was taken from feedback literacy theory~\cite{Mclean2015}, there are entire dimensions we omitted which are important for human contact but non-applicable in this kind of one-shot hint generation work, such as agency, direction, and temporality.  All of these might be relevant in an ongoing dialogue.  This is a potential avenue for future research.

\subsection{The personal connection}

The idea of an ongoing dialogue leads us back to the personal touch.  We have focused on a very specific context: one-off next-step hint generation.  Although we found that humans were not as good as the best AI on this task, this does not mean that human educators are redundant.  The personal connection in education is still important.  Portions of the hype around AI in education are reminisicent of the excitement around Massive Open Online Courses (MOOCs) a decade ago.  Having the educational resources freely and widely available did not lead to a massive uptake or improvement in education, or to educators losing their jobs.  There remains value in formal education based around human contact.

\subsection{Relation to prior work}

Our work provides interesting contrasts with some prior work.  Compared to much previous work we found few issues with incorrect or misleading hints being generated by LLMs, which may reflect a difference in how we prompted the LLMs, or more likely general technological advancement in LLMs.  Like prior work, we did find that LLMs did not always obey our instructions.  Despite asking for only a hint, several ``hints'' provided exact solutions, and many would list out all the problems in the code despite explicitly being asked for a single hint relating to a one selected problem.

Our prompts were quite different to some previous work.  For example, \citet{Roest2024} used very minimal requests of 1--2 sentences alongside the problem description and code.  Our prompts are much longer, but also have no problem description to work with, which is in contrast to the majority of previous work.

In their analysis of enhanced error explanations using feedback literacy theory, \citet{Cucuiat2024} found that explanations using \textit{guiding} were preferred to \textit{telling} (feedback literacy theory term~\cite{Mclean2015}) which matched with our results.  However they found that educators considered \textit{developed understanding} as positive, but under our operationalisation (where this meant that the hint explained the general rule, e.g. why you cannot use the \lstinline$==$ operator for string comparison in Java) it made no difference to how the hints were evaluated.  This is particularly interesting because \citet{Sheese2024} found that students would not typically seek out the general rule for themselves, and educators seem to think that it is also not worthwhile to include it in a next-step hint.

\textit{Opening up a different perspective}, the highest level of abstraction in feedback literacy theory as used by \citeauthor{Cucuiat2024}, was considered negative in our hints. This may suggest that educators, when considering concrete examples of next-step hints, consider this too overwhelming to be helpful overall.

\subsection{Comparative judgement as a research method}

There were multiple research methods which could have been used to get educators' opinions on the hints.  One option would be to interview them, as done by \citet{Cucuiat2024}.  This has the advantage of getting deeper answers about the why, but it would also not have allowed us to end up with precise guidance over the hint length or the reading level.  The use of comparative judgement (plus a survey to fill in or corroborate the why) plus a random forest for analysis allowed us to exact specific guidelines on what made a good hint.  Our checks verified that, at least for this length of task, participants took the task seriously, showed essentially no signs of boredom or giving up, and produced reliable results.  No participant mentioned being confused by the requirements of the task.  We believe comparative judgement may be a useful research method for the future for asking participants to rank a set of stimuli.

Another advantage of asking participants to rank hints is as follows.  It is possible that there is a discrepancy of an educator's preference expressed in the abstract and their opinion when confronted with a concrete representation of the concept.  For example, educators may express a general preference to \textit{opening up perspectives} when interacting with students, but rate hints attempting to do just that lower, because the associated drawbacks (length, complexity, distraction) become more obvious.  In this light, comparative judgement as used here can offer more concrete results, by observing what participants ``do'' (how they rank) rather than what they ``say'' (when asked in an interview).

\section{Threats to Validity}
\label{sec:threats-validity}

Our completion rate was relatively low: only around 50\% of those who signed up to participate in the ranking task went on to complete it.
One possible explanation is that the task itself was too boring or hard for participants.  However, the comparative judgement platform allows us to see who began the task, and only three participants began the task and did not finish, so the non-completers did not even start the task.  

It is possible the task description was off-putting. However, it consisted of only 2-3 relatively sparse pages (available in our OSF repository, see \autoref{sec:osf-repository}).  We believe the most likely explanation is the time of year we recruited and the general business of academics and teachers.  We had aimed to recruit before teaching began but we ended up recruiting in August and September when many high school and university teachers in the northern hemisphere are very busy with the start of their teaching terms.

In this study we have only surveyed educators for their opinions on hints and not students.  As described in \autoref{sec:missing-student-perspective} we felt that students may struggle to understand someone else's code, the problem(s) with it, and then judge which hints they would prefer in that case.  It is possible that educators' opinions on which hints would be best for students may not accord with students' opinions.  Of course, it is not clear that the student opinion is necessarily better than the educator opinion even if that were the case: Students may prefer hints that optimise efficiency of creating the solution, while educators may prefer hints that maximise the learning effect.

The LLMs we used will naturally outdate as technology progresses, but we have tried to mitigate this through our research design that used multiple models and multiple prompts.  Our findings can also inform the hints literature independently of the technology used to generate the hints for the study.

\section{Future work}

One clear future direction is to ask students to evaluate hints.  We believe the best study design will be to implement automatic hint generation in a novice programming environment and then ask students to use it and rate or rank the hints they are given, and/or monitor their programming activity immediately after receiving the hint.

Although we believe we have shown that comparative judgement is a viable experimental technique, the fact remains that recruiting participants (especially busy teachers and academics) is a difficult process; we had eight Snapshots prepared but only recruited enough completing participants to evaluate four of them.  Some recent work~\cite{Lippert2024,Ashokkumar2024} has investigated whether LLMs can emulate human participants in social science experiments in order to generate synthetic data.  On the one hand, this risks AI ``marking its own work'', with LLMs evaluating the output of LLMs (as done, for example, by \citet{Koutcheme2024,Koutcheme2024b}).  On the other hand there may be ways to use this technique to boost evaluation of new behavioural interventions such as identifying better hints.

The hint generation in this work was done with a set of brainstormed prompts.  We not only know which prompt produced the best hints, we also now have extra information about the characteristics of best hints, in terms of length, reading level and other aspects (e.g. that offering alternative approaches is rated negatively).  This opens the possibility to design a new prompt that takes these insights into account in order to improve hint generation.  This minor step forward is one possible avenue to deploy LLM evaluation, rather than recruiting 85 human participants again just to test a minor improvement on the prompts.

One further direction for related work is to analyse our existing data under other hint classifications.  Although we chose to focus on feedback literacy theory, other classifications have been proposed, such as by \citet{Keuning2018} and separately by \citet{Suzuki2017}.  Categorising the existing hints using those schemes and relating the results to our ranking is a possible next step.  With our data being open, this can also be done by other research teams.

\section{Conclusion}
\label{sec:conclusion}

In this paper we asked human educators to rank sets of hints generated by AI models and human researchers using comparative judgement.  This provided findings in several different dimensions.

One finding is that GPT-4 was found to produce better hints than experienced humans.  It is particularly important to note that the hints were generated without providing any context of the task that the student was performing.  The data was taken from the Blackbox dataset, which provides examples from arbitrary novice programmers, without knowledge of the exact task being performed.  In this GPT-4-better-than-humans sense, this paper is one in a recent line of ``LLMs beat humans at $<$ programming education task $>$''. Prior results in this line examined creating code explanations~\cite{Leinonen2023}, small programming exercises~\cite{Kiesler2023} or full programming exams~\cite{Mahon2023}.  However, we provide several further contributions beyond this latest LLM feat.

Some do relate specifically to the operation of LLMs.  We evaluated five different LLM prompts which are quite different in their construction, including two multi-stage prompts.  Although the prompts do have an interaction with the choice of LLM, one was clearly better than the others (prompt 3 in \autoref{tab:prompts}).  This prompt first asked the LLM to summarise the task the student was inferred to perform and then fed the result back to a second request to provide a hint.  (Our prompts, methods and analysis scripts are all open to allow easy replication in future as LLMs advance.)  Furthermore, although GPT-4 was better than humans, all the other models were not.  There is a pronounced effect of model, prompt and their interaction, which showed much greater variation in average performance than we found between the five human researchers who created hints.  It is still the case that LLMs beat humans only with the right model and the right prompt.

We also provide contributions that are entirely orthogonal to AI.  Our study can be seen solely as an investigation into the characteristics of hints that are most important to judge the hint useful -- with the fact that some hints were generated by AI merely acting as a convenient artifact generation mechanism.  We have found that the most important attributes predicting a hint's ranking was its length and reading level.  Experienced Java educators (more than half with an equivalent of over 20+ years of experience) rated hints most highly where the word count was 80--160 words, the reading level was typically understandable by those in US grade 9 (age 14) or below, where guidance was provided beyond just stating the answer, but alternative approaches to solving the problem were avoided.  We found no effect of sentiment or of explaining a more general underlying principle on the perceived quality of a hint.

We have also demonstrated the use of comparative judgement (previously primarily used for assessing writing skills) as a research methodology, showing that, at least for a 20 minute task, participants took it seriously and did not get bored, and the results produced were reliable and interpretable.  Comparative judgement is useful when participants are required, individually or collectively, to rank a number of experimental stimuli than can be placed alongside each other on one screen.  There are several free comparative judgement websites; we used NoMoreMarking\footnote{\url{https://www.nomoremarking.com/}}, with details on how we set up the task available in our OSF repository (along with all of our data and analysis code) as described in \autoref{sec:osf-repository}.

\begin{acks}
We are very grateful to all of the participants in the study.  We are additionally grateful to Arto Hellas for advice during the design stage, to Jane Waite for advice on feedback literacy theory, to Barbara Ericson for her help with recruitment, and to the NoMoreMarking team for supporting research projects on their platform.
This research was supported by the Research Council of Finland (Academy Research Fellow grant number 356114).
\end{acks}

\bibliographystyle{ACM-Reference-Format}
\bibliography{refs}


\begin{thebibliography}{83}


\ifx \showCODEN    \undefined \def \showCODEN     #1{\unskip}     \fi
\ifx \showDOI      \undefined \def \showDOI       #1{#1}\fi
\ifx \showISBNx    \undefined \def \showISBNx     #1{\unskip}     \fi
\ifx \showISBNxiii \undefined \def \showISBNxiii  #1{\unskip}     \fi
\ifx \showISSN     \undefined \def \showISSN      #1{\unskip}     \fi
\ifx \showLCCN     \undefined \def \showLCCN      #1{\unskip}     \fi
\ifx \shownote     \undefined \def \shownote      #1{#1}          \fi
\ifx \showarticletitle \undefined \def \showarticletitle #1{#1}   \fi
\ifx \showURL      \undefined \def \showURL       {\relax}        \fi
\providecommand\bibfield[2]{#2}
\providecommand\bibinfo[2]{#2}
\providecommand\natexlab[1]{#1}
\providecommand\showeprint[2][]{arXiv:#2}

\bibitem[Ahadi et~al\mbox{.}(2015)]%
        {Ahadi2015}
\bibfield{author}{\bibinfo{person}{Alireza Ahadi}, \bibinfo{person}{Raymond Lister}, \bibinfo{person}{Heikki Haapala}, {and} \bibinfo{person}{Arto Vihavainen}.} \bibinfo{year}{2015}\natexlab{}.
\newblock \showarticletitle{Exploring Machine Learning Methods to Automatically Identify Students in Need of Assistance}. In \bibinfo{booktitle}{\emph{Proceedings of the Eleventh Annual International Conference on International Computing Education Research}} (Omaha, Nebraska, USA) \emph{(\bibinfo{series}{ICER '15})}. \bibinfo{publisher}{Association for Computing Machinery}, \bibinfo{address}{New York, NY, USA}, \bibinfo{pages}{121–130}.
\newblock
\showISBNx{9781450336307}
\urldef\tempurl%
\url{https://doi.org/10.1145/2787622.2787717}
\showDOI{\tempurl}


\bibitem[Ahmed et~al\mbox{.}(2020)]%
        {Ahmed2020}
\bibfield{author}{\bibinfo{person}{Umair~Z. Ahmed}, \bibinfo{person}{Nisheeth Srivastava}, \bibinfo{person}{Renuka Sindhgatta}, {and} \bibinfo{person}{Amey Karkare}.} \bibinfo{year}{2020}\natexlab{}.
\newblock \showarticletitle{Characterizing the Pedagogical Benefits of Adaptive Feedback for Compilation Errors by Novice Programmers}. In \bibinfo{booktitle}{\emph{Proceedings of the ACM/IEEE 42nd International Conference on Software Engineering: Software Engineering Education and Training}} (Seoul, South Korea) \emph{(\bibinfo{series}{ICSE-SEET '20})}. \bibinfo{publisher}{Association for Computing Machinery}, \bibinfo{address}{New York, NY, USA}, \bibinfo{pages}{139–150}.
\newblock
\showISBNx{9781450371247}
\urldef\tempurl%
\url{https://doi.org/10.1145/3377814.3381703}
\showDOI{\tempurl}


\bibitem[Bartholomew and Yoshikawa-Ruesch(2018)]%
        {Bartholomew2018}
\bibfield{author}{\bibinfo{person}{S Bartholomew} {and} \bibinfo{person}{Emily Yoshikawa-Ruesch}.} \bibinfo{year}{2018}\natexlab{}.
\newblock \showarticletitle{A systematic review of research around adaptive comparative judgement (ACJ) in K-16 education}.
\newblock \bibinfo{journal}{\emph{Council on Technology an Engineering Teacher Education: Research Monograph Series}} \bibinfo{volume}{1}, \bibinfo{number}{1} (\bibinfo{year}{2018}).
\newblock
\urldef\tempurl%
\url{https://doi.org/10.21061/ctete-rms.v1.c.1}
\showDOI{\tempurl}


\bibitem[Birillo et~al\mbox{.}(2024)]%
        {Birillo2024}
\bibfield{author}{\bibinfo{person}{Anastasiia Birillo}, \bibinfo{person}{Elizaveta Artser}, \bibinfo{person}{Anna Potriasaeva}, \bibinfo{person}{Ilya Vlasov}, \bibinfo{person}{Katsiaryna Dzialets}, \bibinfo{person}{Yaroslav Golubev}, \bibinfo{person}{Igor Gerasimov}, \bibinfo{person}{Hieke Keuning}, {and} \bibinfo{person}{Timofey Bryksin}.} \bibinfo{year}{2024}\natexlab{}.
\newblock \showarticletitle{One Step at a Time: Combining LLMs and Static Analysis to Generate Next-Step Hints for Programming Tasks}. In \bibinfo{booktitle}{\emph{Proceedings of the 24th Koli Calling International Conference on Computing Education Research}} \emph{(\bibinfo{series}{Koli Calling '24})}. \bibinfo{publisher}{Association for Computing Machinery}, \bibinfo{address}{New York, NY, USA}, Article \bibinfo{articleno}{9}, \bibinfo{numpages}{12}~pages.
\newblock
\showISBNx{9798400710384}
\urldef\tempurl%
\url{https://doi.org/10.1145/3699538.3699556}
\showDOI{\tempurl}


\bibitem[Brown and Altadmri(2017)]%
        {Brown2017}
\bibfield{author}{\bibinfo{person}{Neil C.~C. Brown} {and} \bibinfo{person}{Amjad Altadmri}.} \bibinfo{year}{2017}\natexlab{}.
\newblock \showarticletitle{Novice Java Programming Mistakes: Large-Scale Data vs. Educator Beliefs}.
\newblock \bibinfo{journal}{\emph{ACM Trans. Comput. Educ.}} \bibinfo{volume}{17}, \bibinfo{number}{2}, Article \bibinfo{articleno}{7} (\bibinfo{date}{May} \bibinfo{year}{2017}), \bibinfo{numpages}{21}~pages.
\newblock
\urldef\tempurl%
\url{https://doi.org/10.1145/2994154}
\showDOI{\tempurl}


\bibitem[Brown et~al\mbox{.}(2023)]%
        {Brown2023}
\bibfield{author}{\bibinfo{person}{Neil C.~C. Brown}, \bibinfo{person}{Jamie Ford}, \bibinfo{person}{Pierre Weill-Tessier}, {and} \bibinfo{person}{Michael K\"{o}lling}.} \bibinfo{year}{2023}\natexlab{}.
\newblock \showarticletitle{Quick Fixes for Novice Programmers: Effective but Under-Utilised}. In \bibinfo{booktitle}{\emph{Proceedings of the 2023 Conference on United Kingdom \& Ireland Computing Education Research}} \emph{(\bibinfo{series}{UKICER '23})}. \bibinfo{publisher}{Association for Computing Machinery}, \bibinfo{address}{New York, NY, USA}, Article \bibinfo{articleno}{3}, \bibinfo{numpages}{7}~pages.
\newblock
\showISBNx{9798400708763}
\urldef\tempurl%
\url{https://doi.org/10.1145/3610969.3611117}
\showDOI{\tempurl}


\bibitem[Brown et~al\mbox{.}(2014)]%
        {Brown2014}
\bibfield{author}{\bibinfo{person}{Neil C.~C. Brown}, \bibinfo{person}{Michael K\"{o}lling}, \bibinfo{person}{Davin McCall}, {and} \bibinfo{person}{Ian Utting}.} \bibinfo{year}{2014}\natexlab{}.
\newblock \showarticletitle{Blackbox: A Large Scale Repository of Novice Programmers' Activity}. In \bibinfo{booktitle}{\emph{Proceedings of the 45th ACM Technical Symposium on Computer Science Education}} (Atlanta, Georgia, USA) \emph{(\bibinfo{series}{SIGCSE '14})}. \bibinfo{publisher}{Association for Computing Machinery}, \bibinfo{address}{New York, NY, USA}, \bibinfo{pages}{223–228}.
\newblock
\showISBNx{9781450326056}
\urldef\tempurl%
\url{https://doi.org/10.1145/2538862.2538924}
\showDOI{\tempurl}


\bibitem[Brysbaert(2019)]%
        {Brysbaert2019}
\bibfield{author}{\bibinfo{person}{Marc Brysbaert}.} \bibinfo{year}{2019}\natexlab{}.
\newblock \showarticletitle{How many words do we read per minute? A review and meta-analysis of reading rate}.
\newblock \bibinfo{journal}{\emph{Journal of Memory and Language}}  \bibinfo{volume}{109} (\bibinfo{year}{2019}), \bibinfo{pages}{104047}.
\newblock
\showISSN{0749-596X}
\urldef\tempurl%
\url{https://doi.org/10.1016/j.jml.2019.104047}
\showDOI{\tempurl}


\bibitem[Castro and Fisler(2020)]%
        {Castro2020}
\bibfield{author}{\bibinfo{person}{Francisco Enrique~Vicente Castro} {and} \bibinfo{person}{Kathi Fisler}.} \bibinfo{year}{2020}\natexlab{}.
\newblock \showarticletitle{Qualitative Analyses of Movements Between Task-Level and Code-Level Thinking of Novice Programmers}. In \bibinfo{booktitle}{\emph{Proceedings of the 51st ACM Technical Symposium on Computer Science Education}} (Portland, OR, USA) \emph{(\bibinfo{series}{SIGCSE '20})}. \bibinfo{publisher}{Association for Computing Machinery}, \bibinfo{address}{New York, NY, USA}, \bibinfo{pages}{487–493}.
\newblock
\showISBNx{9781450367936}
\urldef\tempurl%
\url{https://doi.org/10.1145/3328778.3366847}
\showDOI{\tempurl}


\bibitem[Crow et~al\mbox{.}(2018)]%
        {Crow2018}
\bibfield{author}{\bibinfo{person}{Tyne Crow}, \bibinfo{person}{Andrew Luxton-Reilly}, {and} \bibinfo{person}{Burkhard Wuensche}.} \bibinfo{year}{2018}\natexlab{}.
\newblock \showarticletitle{Intelligent Tutoring Systems for Programming Education: A Systematic Review}. In \bibinfo{booktitle}{\emph{Proceedings of the 20th Australasian Computing Education Conference}} (Brisbane, Queensland, Australia) \emph{(\bibinfo{series}{ACE '18})}. \bibinfo{publisher}{Association for Computing Machinery}, \bibinfo{address}{New York, NY, USA}, \bibinfo{pages}{53–62}.
\newblock
\showISBNx{9781450363402}
\urldef\tempurl%
\url{https://doi.org/10.1145/3160489.3160492}
\showDOI{\tempurl}


\bibitem[Cucuiat and Waite(2024)]%
        {Cucuiat2024}
\bibfield{author}{\bibinfo{person}{Veronica Cucuiat} {and} \bibinfo{person}{Jane Waite}.} \bibinfo{year}{2024}\natexlab{}.
\newblock \showarticletitle{Feedback Literacy: Holistic Analysis of Secondary Educators' Views of LLM Explanations of Program Error Messages}. In \bibinfo{booktitle}{\emph{Proceedings of the 2024 on Innovation and Technology in Computer Science Education V. 1}} (Milan, Italy) \emph{(\bibinfo{series}{ITiCSE 2024})}. \bibinfo{publisher}{Association for Computing Machinery}, \bibinfo{address}{New York, NY, USA}, \bibinfo{pages}{192–198}.
\newblock
\showISBNx{9798400706004}
\urldef\tempurl%
\url{https://doi.org/10.1145/3649217.3653595}
\showDOI{\tempurl}


\bibitem[Denny et~al\mbox{.}(2024a)]%
        {Denny2024}
\bibfield{author}{\bibinfo{person}{Paul Denny}, \bibinfo{person}{Juho Leinonen}, \bibinfo{person}{James Prather}, \bibinfo{person}{Andrew Luxton-Reilly}, \bibinfo{person}{Thezyrie Amarouche}, \bibinfo{person}{Brett~A. Becker}, {and} \bibinfo{person}{Brent~N. Reeves}.} \bibinfo{year}{2024}\natexlab{a}.
\newblock \showarticletitle{Prompt Problems: A New Programming Exercise for the Generative AI Era}. In \bibinfo{booktitle}{\emph{Proceedings of the 55th ACM Technical Symposium on Computer Science Education V. 1}} (Portland, OR, USA) \emph{(\bibinfo{series}{SIGCSE 2024})}. \bibinfo{publisher}{Association for Computing Machinery}, \bibinfo{address}{New York, NY, USA}, \bibinfo{pages}{296–302}.
\newblock
\showISBNx{9798400704239}
\urldef\tempurl%
\url{https://doi.org/10.1145/3626252.3630909}
\showDOI{\tempurl}


\bibitem[Denny et~al\mbox{.}(2024b)]%
        {denny2024desirable}
\bibfield{author}{\bibinfo{person}{Paul Denny}, \bibinfo{person}{Stephen MacNeil}, \bibinfo{person}{Jaromir Savelka}, \bibinfo{person}{Leo Porter}, {and} \bibinfo{person}{Andrew Luxton-Reilly}.} \bibinfo{year}{2024}\natexlab{b}.
\newblock \showarticletitle{Desirable Characteristics for AI Teaching Assistants in Programming Education}. In \bibinfo{booktitle}{\emph{Proceedings of the 2024 on Innovation and Technology in Computer Science Education V. 1}} (Milan, Italy) \emph{(\bibinfo{series}{ITiCSE 2024})}. \bibinfo{publisher}{Association for Computing Machinery}, \bibinfo{address}{New York, NY, USA}, \bibinfo{pages}{408–414}.
\newblock
\showISBNx{9798400706004}
\urldef\tempurl%
\url{https://doi.org/10.1145/3649217.3653574}
\showDOI{\tempurl}


\bibitem[Denny et~al\mbox{.}(2024c)]%
        {denny2024computing}
\bibfield{author}{\bibinfo{person}{Paul Denny}, \bibinfo{person}{James Prather}, \bibinfo{person}{Brett~A. Becker}, \bibinfo{person}{James Finnie-Ansley}, \bibinfo{person}{Arto Hellas}, \bibinfo{person}{Juho Leinonen}, \bibinfo{person}{Andrew Luxton-Reilly}, \bibinfo{person}{Brent~N. Reeves}, \bibinfo{person}{Eddie~Antonio Santos}, {and} \bibinfo{person}{Sami Sarsa}.} \bibinfo{year}{2024}\natexlab{c}.
\newblock \showarticletitle{Computing Education in the Era of Generative AI}.
\newblock \bibinfo{journal}{\emph{Commun. ACM}} \bibinfo{volume}{67}, \bibinfo{number}{2} (\bibinfo{date}{Jan.} \bibinfo{year}{2024}), \bibinfo{pages}{56–67}.
\newblock
\showISSN{0001-0782}
\urldef\tempurl%
\url{https://doi.org/10.1145/3624720}
\showDOI{\tempurl}


\bibitem[Denny et~al\mbox{.}(2021)]%
        {Denny2021}
\bibfield{author}{\bibinfo{person}{Paul Denny}, \bibinfo{person}{James Prather}, \bibinfo{person}{Brett~A. Becker}, \bibinfo{person}{Catherine Mooney}, \bibinfo{person}{John Homer}, \bibinfo{person}{Zachary~C Albrecht}, {and} \bibinfo{person}{Garrett~B. Powell}.} \bibinfo{year}{2021}\natexlab{}.
\newblock \showarticletitle{On Designing Programming Error Messages for Novices: Readability and Its Constituent Factors}. In \bibinfo{booktitle}{\emph{Proceedings of the 2021 CHI Conference on Human Factors in Computing Systems}} (Yokohama, Japan) \emph{(\bibinfo{series}{CHI '21})}. \bibinfo{publisher}{Association for Computing Machinery}, \bibinfo{address}{New York, NY, USA}, Article \bibinfo{articleno}{55}, \bibinfo{numpages}{15}~pages.
\newblock
\showISBNx{9781450380966}
\urldef\tempurl%
\url{https://doi.org/10.1145/3411764.3445696}
\showDOI{\tempurl}


\bibitem[Falkner and Falkner(2012)]%
        {Falkner2012}
\bibfield{author}{\bibinfo{person}{Nickolas~J.G. Falkner} {and} \bibinfo{person}{Katrina~E. Falkner}.} \bibinfo{year}{2012}\natexlab{}.
\newblock \showarticletitle{A Fast Measure for Identifying At-Risk Students in Computer Science}. In \bibinfo{booktitle}{\emph{Proceedings of the Ninth Annual International Conference on International Computing Education Research}} (Auckland, New Zealand) \emph{(\bibinfo{series}{ICER '12})}. \bibinfo{publisher}{Association for Computing Machinery}, \bibinfo{address}{New York, NY, USA}, \bibinfo{pages}{55–62}.
\newblock
\showISBNx{9781450316040}
\urldef\tempurl%
\url{https://doi.org/10.1145/2361276.2361288}
\showDOI{\tempurl}


\bibitem[Fiannaca et~al\mbox{.}(2023)]%
        {Fiannaca2023}
\bibfield{author}{\bibinfo{person}{Alexander~J. Fiannaca}, \bibinfo{person}{Chinmay Kulkarni}, \bibinfo{person}{Carrie~J Cai}, {and} \bibinfo{person}{Michael Terry}.} \bibinfo{year}{2023}\natexlab{}.
\newblock \showarticletitle{Programming without a Programming Language: Challenges and Opportunities for Designing Developer Tools for Prompt Programming}. In \bibinfo{booktitle}{\emph{Extended Abstracts of the 2023 CHI Conference on Human Factors in Computing Systems}} (Hamburg, Germany) \emph{(\bibinfo{series}{CHI EA '23})}. \bibinfo{publisher}{Association for Computing Machinery}, \bibinfo{address}{New York, NY, USA}, Article \bibinfo{articleno}{235}, \bibinfo{numpages}{7}~pages.
\newblock
\showISBNx{9781450394222}
\urldef\tempurl%
\url{https://doi.org/10.1145/3544549.3585737}
\showDOI{\tempurl}


\bibitem[Garner et~al\mbox{.}(2005)]%
        {Garner2005}
\bibfield{author}{\bibinfo{person}{Sandy Garner}, \bibinfo{person}{Patricia Haden}, {and} \bibinfo{person}{Anthony Robins}.} \bibinfo{year}{2005}\natexlab{}.
\newblock \showarticletitle{My Program is Correct but It Doesn't Run: A Preliminary Investigation of Novice Programmers' Problems}. In \bibinfo{booktitle}{\emph{Proceedings of the 7th Australasian Conference on Computing Education - Volume 42}} (Newcastle, New South Wales, Australia) \emph{(\bibinfo{series}{ACE '05})}. \bibinfo{publisher}{Australian Computer Society, Inc.}, \bibinfo{address}{AUS}, \bibinfo{pages}{173–180}.
\newblock
\showISBNx{1920682244}


\bibitem[Ghimire and Edwards(2024)]%
        {Ghimire2024}
\bibfield{author}{\bibinfo{person}{Aashish Ghimire} {and} \bibinfo{person}{John Edwards}.} \bibinfo{year}{2024}\natexlab{}.
\newblock \showarticletitle{Coding with AI: How Are Tools Like ChatGPT Being Used by Students in Foundational Programming Courses}. In \bibinfo{booktitle}{\emph{Artificial Intelligence in Education}}, \bibfield{editor}{\bibinfo{person}{Andrew~M. Olney}, \bibinfo{person}{Irene-Angelica Chounta}, \bibinfo{person}{Zitao Liu}, \bibinfo{person}{Olga~C. Santos}, {and} \bibinfo{person}{Ig~Ibert Bittencourt}} (Eds.). \bibinfo{publisher}{Springer Nature Switzerland}, \bibinfo{address}{Cham}, \bibinfo{pages}{259--267}.
\newblock
\showISBNx{978-3-031-64299-9}


\bibitem[Glassman et~al\mbox{.}(2016)]%
        {Glassman2016}
\bibfield{author}{\bibinfo{person}{Elena~L. Glassman}, \bibinfo{person}{Aaron Lin}, \bibinfo{person}{Carrie~J. Cai}, {and} \bibinfo{person}{Robert~C. Miller}.} \bibinfo{year}{2016}\natexlab{}.
\newblock \showarticletitle{Learnersourcing Personalized Hints}. In \bibinfo{booktitle}{\emph{Proceedings of the 19th ACM Conference on Computer-Supported Cooperative Work \& Social Computing}} (San Francisco, California, USA) \emph{(\bibinfo{series}{CSCW '16})}. \bibinfo{publisher}{Association for Computing Machinery}, \bibinfo{address}{New York, NY, USA}, \bibinfo{pages}{1626–1636}.
\newblock
\showISBNx{9781450335928}
\urldef\tempurl%
\url{https://doi.org/10.1145/2818048.2820011}
\showDOI{\tempurl}


\bibitem[Guo et~al\mbox{.}(2020)]%
        {Guo2020}
\bibfield{author}{\bibinfo{person}{Philip~J. Guo}, \bibinfo{person}{Julia~M. Markel}, {and} \bibinfo{person}{Xiong Zhang}.} \bibinfo{year}{2020}\natexlab{}.
\newblock \showarticletitle{Learnersourcing at Scale to Overcome Expert Blind Spots for Introductory Programming: A Three-Year Deployment Study on the Python Tutor Website}. In \bibinfo{booktitle}{\emph{Proceedings of the Seventh ACM Conference on Learning @ Scale}} (Virtual Event, USA) \emph{(\bibinfo{series}{L@S '20})}. \bibinfo{publisher}{Association for Computing Machinery}, \bibinfo{address}{New York, NY, USA}, \bibinfo{pages}{301–304}.
\newblock
\showISBNx{9781450379519}
\urldef\tempurl%
\url{https://doi.org/10.1145/3386527.3406733}
\showDOI{\tempurl}


\bibitem[Gusukuma et~al\mbox{.}(2017)]%
        {Gusukuma2017}
\bibfield{author}{\bibinfo{person}{Luke Gusukuma}, \bibinfo{person}{Dennis Kafura}, {and} \bibinfo{person}{Austin~Cory Bart}.} \bibinfo{year}{2017}\natexlab{}.
\newblock \showarticletitle{Authoring feedback for novice programmers in a block-based language}. In \bibinfo{booktitle}{\emph{2017 IEEE Blocks and Beyond Workshop (B\&B)}}. \bibinfo{pages}{37--40}.
\newblock
\urldef\tempurl%
\url{https://doi.org/10.1109/BLOCKS.2017.8120407}
\showDOI{\tempurl}


\bibitem[Hellas et~al\mbox{.}(2018)]%
        {Hellas2018}
\bibfield{author}{\bibinfo{person}{Arto Hellas}, \bibinfo{person}{Petri Ihantola}, \bibinfo{person}{Andrew Petersen}, \bibinfo{person}{Vangel~V. Ajanovski}, \bibinfo{person}{Mirela Gutica}, \bibinfo{person}{Timo Hynninen}, \bibinfo{person}{Antti Knutas}, \bibinfo{person}{Juho Leinonen}, \bibinfo{person}{Chris Messom}, {and} \bibinfo{person}{Soohyun~Nam Liao}.} \bibinfo{year}{2018}\natexlab{}.
\newblock \showarticletitle{Taxonomizing Features and Methods for Identifying At-Risk Students in Computing Courses}. In \bibinfo{booktitle}{\emph{Proceedings of the 23rd Annual ACM Conference on Innovation and Technology in Computer Science Education}} (Larnaca, Cyprus) \emph{(\bibinfo{series}{ITiCSE 2018})}. \bibinfo{publisher}{Association for Computing Machinery}, \bibinfo{address}{New York, NY, USA}, \bibinfo{pages}{364–365}.
\newblock
\showISBNx{9781450357074}
\urldef\tempurl%
\url{https://doi.org/10.1145/3197091.3205845}
\showDOI{\tempurl}


\bibitem[Hellas et~al\mbox{.}(2023)]%
        {Hellas2023}
\bibfield{author}{\bibinfo{person}{Arto Hellas}, \bibinfo{person}{Juho Leinonen}, \bibinfo{person}{Sami Sarsa}, \bibinfo{person}{Charles Koutcheme}, \bibinfo{person}{Lilja Kujanp\"{a}\"{a}}, {and} \bibinfo{person}{Juha Sorva}.} \bibinfo{year}{2023}\natexlab{}.
\newblock \showarticletitle{Exploring the Responses of Large Language Models to Beginner Programmers’ Help Requests}. In \bibinfo{booktitle}{\emph{Proceedings of the 2023 ACM Conference on International Computing Education Research - Volume 1}} (Chicago, IL, USA) \emph{(\bibinfo{series}{ICER '23})}. \bibinfo{publisher}{Association for Computing Machinery}, \bibinfo{address}{New York, NY, USA}, \bibinfo{pages}{93–105}.
\newblock
\showISBNx{9781450399760}
\urldef\tempurl%
\url{https://doi.org/10.1145/3568813.3600139}
\showDOI{\tempurl}


\bibitem[Hewitt et~al\mbox{.}(2024)]%
        {Ashokkumar2024}
\bibfield{author}{\bibinfo{person}{Luke Hewitt}, \bibinfo{person}{Ashwini Ashokkumar}, \bibinfo{person}{Isaias Ghezae}, {and} \bibinfo{person}{Robb Willer}.} \bibinfo{year}{2024}\natexlab{}.
\newblock \bibinfo{booktitle}{\emph{Predicting Results of Social Science Experiments Using Large Language Models}}.
\newblock \bibinfo{type}{{T}echnical {R}eport}. \bibinfo{institution}{Working Paper}.
\newblock
\urldef\tempurl%
\url{https://samim.io/dl/Predicting%20results%20of%20social%20science%20experiments%20using%20large%20language%20models.pdf}
\showURL{%
\tempurl}


\bibitem[Ho(1995)]%
        {Ho1995}
\bibfield{author}{\bibinfo{person}{Tin~Kam Ho}.} \bibinfo{year}{1995}\natexlab{}.
\newblock \showarticletitle{Random decision forests}. In \bibinfo{booktitle}{\emph{Proceedings of the Third International Conference on Document Analysis and Recognition (Volume 1) - Volume 1}} \emph{(\bibinfo{series}{ICDAR '95})}. \bibinfo{publisher}{IEEE Computer Society}, \bibinfo{address}{USA}, \bibinfo{pages}{278}.
\newblock
\showISBNx{0818671289}


\bibitem[Hutto and Gilbert(2014)]%
        {Hutto2014}
\bibfield{author}{\bibinfo{person}{C. Hutto} {and} \bibinfo{person}{Eric Gilbert}.} \bibinfo{year}{2014}\natexlab{}.
\newblock \showarticletitle{VADER: A Parsimonious Rule-Based Model for Sentiment Analysis of Social Media Text}.
\newblock \bibinfo{journal}{\emph{Proceedings of the International AAAI Conference on Web and Social Media}} \bibinfo{volume}{8}, \bibinfo{number}{1} (\bibinfo{date}{May} \bibinfo{year}{2014}), \bibinfo{pages}{216--225}.
\newblock
\urldef\tempurl%
\url{https://doi.org/10.1609/icwsm.v8i1.14550}
\showDOI{\tempurl}


\bibitem[Ichinco and Kelleher(2018)]%
        {Ichinco2018}
\bibfield{author}{\bibinfo{person}{Michelle Ichinco} {and} \bibinfo{person}{Caitlin Kelleher}.} \bibinfo{year}{2018}\natexlab{}.
\newblock \showarticletitle{Semi-Automatic Suggestion Generation for Young Novice Programmers in an Open-Ended Context}. In \bibinfo{booktitle}{\emph{Proceedings of the 17th ACM Conference on Interaction Design and Children}} (Trondheim, Norway) \emph{(\bibinfo{series}{IDC '18})}. \bibinfo{publisher}{Association for Computing Machinery}, \bibinfo{address}{New York, NY, USA}, \bibinfo{pages}{405–412}.
\newblock
\showISBNx{9781450351522}
\urldef\tempurl%
\url{https://doi.org/10.1145/3202185.3202762}
\showDOI{\tempurl}


\bibitem[Jeuring et~al\mbox{.}(2022)]%
        {Jeuring2022}
\bibfield{author}{\bibinfo{person}{Johan Jeuring}, \bibinfo{person}{Hieke Keuning}, \bibinfo{person}{Samiha Marwan}, \bibinfo{person}{Dennis Bouvier}, \bibinfo{person}{Cruz Izu}, \bibinfo{person}{Natalie Kiesler}, \bibinfo{person}{Teemu Lehtinen}, \bibinfo{person}{Dominic Lohr}, \bibinfo{person}{Andrew Peterson}, {and} \bibinfo{person}{Sami Sarsa}.} \bibinfo{year}{2022}\natexlab{}.
\newblock \showarticletitle{Towards Giving Timely Formative Feedback and Hints to Novice Programmers}. In \bibinfo{booktitle}{\emph{Proceedings of the 2022 Working Group Reports on Innovation and Technology in Computer Science Education}} (Dublin, Ireland) \emph{(\bibinfo{series}{ITiCSE-WGR '22})}. \bibinfo{publisher}{Association for Computing Machinery}, \bibinfo{address}{New York, NY, USA}, \bibinfo{pages}{95–115}.
\newblock
\showISBNx{9798400700101}
\urldef\tempurl%
\url{https://doi.org/10.1145/3571785.3574124}
\showDOI{\tempurl}


\bibitem[Jones and Davies(2024)]%
        {Jones2023}
\bibfield{author}{\bibinfo{person}{Ian Jones} {and} \bibinfo{person}{Ben Davies}.} \bibinfo{year}{2024}\natexlab{}.
\newblock \showarticletitle{Comparative judgement in education research}.
\newblock \bibinfo{journal}{\emph{International Journal of Research \& Method in Education}} \bibinfo{volume}{47}, \bibinfo{number}{2} (\bibinfo{year}{2024}), \bibinfo{pages}{170--181}.
\newblock
\urldef\tempurl%
\url{https://doi.org/10.1080/1743727X.2023.2242273}
\showDOI{\tempurl}
\showeprint{https://doi.org/10.1080/1743727X.2023.2242273}


\bibitem[Joshi et~al\mbox{.}(2023)]%
        {Joshi2023}
\bibfield{author}{\bibinfo{person}{Ishika Joshi}, \bibinfo{person}{Ritvik Budhiraja}, \bibinfo{person}{Pranav~Deepak Tanna}, \bibinfo{person}{Lovenya Jain}, \bibinfo{person}{Mihika Deshpande}, \bibinfo{person}{Arjun Srivastava}, \bibinfo{person}{Srinivas Rallapalli}, \bibinfo{person}{Harshal~D Akolekar}, \bibinfo{person}{Jagat~Sesh Challa}, {and} \bibinfo{person}{Dhruv Kumar}.} \bibinfo{year}{2023}\natexlab{}.
\newblock \bibinfo{title}{"With Great Power Comes Great Responsibility!": Student and Instructor Perspectives on the influence of LLMs on Undergraduate Engineering Education}.
\newblock
\newblock
\showeprint[arxiv]{2309.10694}~[cs.HC]


\bibitem[Kazemitabaar et~al\mbox{.}(2023)]%
        {Kazemitabaar2023}
\bibfield{author}{\bibinfo{person}{Majeed Kazemitabaar}, \bibinfo{person}{Justin Chow}, \bibinfo{person}{Carl Ka~To Ma}, \bibinfo{person}{Barbara~J. Ericson}, \bibinfo{person}{David Weintrop}, {and} \bibinfo{person}{Tovi Grossman}.} \bibinfo{year}{2023}\natexlab{}.
\newblock \showarticletitle{Studying the Effect of AI Code Generators on Supporting Novice Learners in Introductory Programming}. In \bibinfo{booktitle}{\emph{Proceedings of the 2023 CHI Conference on Human Factors in Computing Systems}} (Hamburg, Germany) \emph{(\bibinfo{series}{CHI '23})}. \bibinfo{publisher}{Association for Computing Machinery}, \bibinfo{address}{New York, NY, USA}, Article \bibinfo{articleno}{455}, \bibinfo{numpages}{23}~pages.
\newblock
\showISBNx{9781450394215}
\urldef\tempurl%
\url{https://doi.org/10.1145/3544548.3580919}
\showDOI{\tempurl}


\bibitem[Kendon et~al\mbox{.}(2023)]%
        {Tyson2023}
\bibfield{author}{\bibinfo{person}{Tyson Kendon}, \bibinfo{person}{Leanne Wu}, {and} \bibinfo{person}{John Aycock}.} \bibinfo{year}{2023}\natexlab{}.
\newblock \showarticletitle{AI-Generated Code Not Considered Harmful}. In \bibinfo{booktitle}{\emph{Proceedings of the 25th Western Canadian Conference on Computing Education}} (Vancouver, BC, Canada) \emph{(\bibinfo{series}{WCCCE '23})}. \bibinfo{publisher}{Association for Computing Machinery}, \bibinfo{address}{New York, NY, USA}, Article \bibinfo{articleno}{3}, \bibinfo{numpages}{7}~pages.
\newblock
\showISBNx{9798400707896}
\urldef\tempurl%
\url{https://doi.org/10.1145/3593342.3593349}
\showDOI{\tempurl}


\bibitem[Keuning et~al\mbox{.}(2018)]%
        {Keuning2018}
\bibfield{author}{\bibinfo{person}{Hieke Keuning}, \bibinfo{person}{Johan Jeuring}, {and} \bibinfo{person}{Bastiaan Heeren}.} \bibinfo{year}{2018}\natexlab{}.
\newblock \showarticletitle{A Systematic Literature Review of Automated Feedback Generation for Programming Exercises}.
\newblock \bibinfo{journal}{\emph{ACM Trans. Comput. Educ.}} \bibinfo{volume}{19}, \bibinfo{number}{1}, Article \bibinfo{articleno}{3} (\bibinfo{date}{Sept.} \bibinfo{year}{2018}), \bibinfo{numpages}{43}~pages.
\newblock
\urldef\tempurl%
\url{https://doi.org/10.1145/3231711}
\showDOI{\tempurl}


\bibitem[Kiesler et~al\mbox{.}(2023)]%
        {Kiesler2023b}
\bibfield{author}{\bibinfo{person}{Natalie Kiesler}, \bibinfo{person}{Dominic Lohr}, {and} \bibinfo{person}{Hieke Keuning}.} \bibinfo{year}{2023}\natexlab{}.
\newblock \showarticletitle{Exploring the Potential of Large Language Models to Generate Formative Programming Feedback}. In \bibinfo{booktitle}{\emph{2023 IEEE Frontiers in Education Conference (FIE)}}. \bibinfo{pages}{1--5}.
\newblock
\urldef\tempurl%
\url{https://doi.org/10.1109/FIE58773.2023.10343457}
\showDOI{\tempurl}


\bibitem[Kiesler and Schiffner(2023)]%
        {Kiesler2023}
\bibfield{author}{\bibinfo{person}{Natalie Kiesler} {and} \bibinfo{person}{Daniel Schiffner}.} \bibinfo{year}{2023}\natexlab{}.
\newblock \bibinfo{title}{Large Language Models in Introductory Programming Education: ChatGPT's Performance and Implications for Assessments}.
\newblock
\newblock
\showeprint[arxiv]{2308.08572}~[cs.SE]
\urldef\tempurl%
\url{https://arxiv.org/abs/2308.08572}
\showURL{%
\tempurl}


\bibitem[Kincaid(1975)]%
        {Kincaid1975}
\bibfield{author}{\bibinfo{person}{JP Kincaid}.} \bibinfo{year}{1975}\natexlab{}.
\newblock \showarticletitle{Derivation of new readability formulas (automated readability index, fog count and flesch reading ease formula) for navy enlisted personnel}.
\newblock \bibinfo{journal}{\emph{Chief of Naval Technical Training}} (\bibinfo{year}{1975}).
\newblock


\bibitem[Kolen and Brennan(2016)]%
        {Kolen2016}
\bibfield{author}{\bibinfo{person}{MJ Kolen} {and} \bibinfo{person}{RL Brennan}.} \bibinfo{year}{2016}\natexlab{}.
\newblock \showarticletitle{‘No More Marking’: An online tool for comparative judgement}.
\newblock \bibinfo{journal}{\emph{ISSN 1756-509X}} (\bibinfo{year}{2016}), \bibinfo{pages}{12}.
\newblock


\bibitem[Koutcheme et~al\mbox{.}(2024a)]%
        {Koutcheme2024b}
\bibfield{author}{\bibinfo{person}{Charles Koutcheme}, \bibinfo{person}{Nicola Dainese}, \bibinfo{person}{Arto Hellas}, \bibinfo{person}{Sami Sarsa}, \bibinfo{person}{Juho Leinonen}, \bibinfo{person}{Syed Ashraf}, {and} \bibinfo{person}{Paul Denny}.} \bibinfo{year}{2024}\natexlab{a}.
\newblock \bibinfo{title}{Evaluating Language Models for Generating and Judging Programming Feedback}.
\newblock
\newblock
\showeprint[arxiv]{2407.04873}~[cs.AI]
\urldef\tempurl%
\url{https://arxiv.org/abs/2407.04873}
\showURL{%
\tempurl}


\bibitem[Koutcheme et~al\mbox{.}(2024b)]%
        {Koutcheme2024}
\bibfield{author}{\bibinfo{person}{Charles Koutcheme}, \bibinfo{person}{Nicola Dainese}, \bibinfo{person}{Sami Sarsa}, \bibinfo{person}{Arto Hellas}, \bibinfo{person}{Juho Leinonen}, {and} \bibinfo{person}{Paul Denny}.} \bibinfo{year}{2024}\natexlab{b}.
\newblock \showarticletitle{Open Source Language Models Can Provide Feedback: Evaluating LLMs' Ability to Help Students Using GPT-4-As-A-Judge}. In \bibinfo{booktitle}{\emph{Proceedings of the 2024 on Innovation and Technology in Computer Science Education V. 1}} (Milan, Italy) \emph{(\bibinfo{series}{ITiCSE 2024})}. \bibinfo{publisher}{Association for Computing Machinery}, \bibinfo{address}{New York, NY, USA}, \bibinfo{pages}{52–58}.
\newblock
\showISBNx{9798400706004}
\urldef\tempurl%
\url{https://doi.org/10.1145/3649217.3653612}
\showDOI{\tempurl}


\bibitem[Lau and Guo(2023)]%
        {Lau2023}
\bibfield{author}{\bibinfo{person}{Sam Lau} {and} \bibinfo{person}{Philip Guo}.} \bibinfo{year}{2023}\natexlab{}.
\newblock \showarticletitle{From "Ban It Till We Understand It" to "Resistance is Futile": How University Programming Instructors Plan to Adapt as More Students Use AI Code Generation and Explanation Tools Such as ChatGPT and GitHub Copilot}. In \bibinfo{booktitle}{\emph{Proceedings of the 2023 ACM Conference on International Computing Education Research - Volume 1}} (Chicago, IL, USA) \emph{(\bibinfo{series}{ICER '23})}. \bibinfo{publisher}{Association for Computing Machinery}, \bibinfo{address}{New York, NY, USA}, \bibinfo{pages}{106–121}.
\newblock
\showISBNx{9781450399760}
\urldef\tempurl%
\url{https://doi.org/10.1145/3568813.3600138}
\showDOI{\tempurl}


\bibitem[Leinonen et~al\mbox{.}(2023a)]%
        {Leinonen2023}
\bibfield{author}{\bibinfo{person}{Juho Leinonen}, \bibinfo{person}{Paul Denny}, \bibinfo{person}{Stephen MacNeil}, \bibinfo{person}{Sami Sarsa}, \bibinfo{person}{Seth Bernstein}, \bibinfo{person}{Joanne Kim}, \bibinfo{person}{Andrew Tran}, {and} \bibinfo{person}{Arto Hellas}.} \bibinfo{year}{2023}\natexlab{a}.
\newblock \showarticletitle{Comparing Code Explanations Created by Students and Large Language Models}. In \bibinfo{booktitle}{\emph{Proceedings of the 2023 Conference on Innovation and Technology in Computer Science Education V. 1}} (Turku, Finland) \emph{(\bibinfo{series}{ITiCSE 2023})}. \bibinfo{publisher}{Association for Computing Machinery}, \bibinfo{address}{New York, NY, USA}, \bibinfo{pages}{124–130}.
\newblock
\showISBNx{9798400701382}
\urldef\tempurl%
\url{https://doi.org/10.1145/3587102.3588785}
\showDOI{\tempurl}


\bibitem[Leinonen et~al\mbox{.}(2023b)]%
        {Leinonen2023b}
\bibfield{author}{\bibinfo{person}{Juho Leinonen}, \bibinfo{person}{Arto Hellas}, \bibinfo{person}{Sami Sarsa}, \bibinfo{person}{Brent Reeves}, \bibinfo{person}{Paul Denny}, \bibinfo{person}{James Prather}, {and} \bibinfo{person}{Brett~A. Becker}.} \bibinfo{year}{2023}\natexlab{b}.
\newblock \showarticletitle{Using Large Language Models to Enhance Programming Error Messages}. In \bibinfo{booktitle}{\emph{Proceedings of the 54th ACM Technical Symposium on Computer Science Education V. 1}} (Toronto ON, Canada) \emph{(\bibinfo{series}{SIGCSE 2023})}. \bibinfo{publisher}{Association for Computing Machinery}, \bibinfo{address}{New York, NY, USA}, \bibinfo{pages}{563–569}.
\newblock
\showISBNx{9781450394314}
\urldef\tempurl%
\url{https://doi.org/10.1145/3545945.3569770}
\showDOI{\tempurl}


\bibitem[Liffiton et~al\mbox{.}(2024)]%
        {Liffiton2023}
\bibfield{author}{\bibinfo{person}{Mark Liffiton}, \bibinfo{person}{Brad~E Sheese}, \bibinfo{person}{Jaromir Savelka}, {and} \bibinfo{person}{Paul Denny}.} \bibinfo{year}{2024}\natexlab{}.
\newblock \bibinfo{title}{CodeHelp: Using Large Language Models with Guardrails for Scalable Support in Programming Classes}.
\newblock , \bibinfo{numpages}{11}~pages.
\newblock
\showISBNx{9798400716539}
\urldef\tempurl%
\url{https://doi.org/10.1145/3631802.3631830}
\showDOI{\tempurl}


\bibitem[Lippert et~al\mbox{.}(2024)]%
        {Lippert2024}
\bibfield{author}{\bibinfo{person}{Steffen Lippert}, \bibinfo{person}{Anna Dreber}, \bibinfo{person}{Magnus Johannesson}, \bibinfo{person}{Warren Tierney}, \bibinfo{person}{Wilson Cyrus-Lai}, \bibinfo{person}{Eric~Luis Uhlmann}, \bibinfo{person}{Emotion~Expression Collaboration}, {and} \bibinfo{person}{Thomas Pfeiffer}.} \bibinfo{year}{2024}\natexlab{}.
\newblock \showarticletitle{Can large language models help predict results from a complex behavioural science study?}
\newblock \bibinfo{journal}{\emph{Royal Society Open Science}} \bibinfo{volume}{11}, \bibinfo{number}{9} (\bibinfo{year}{2024}), \bibinfo{pages}{240682}.
\newblock


\bibitem[Lohr et~al\mbox{.}(2024)]%
        {Lohr2024}
\bibfield{author}{\bibinfo{person}{Dominic Lohr}, \bibinfo{person}{Natalie Kiesler}, \bibinfo{person}{Hieke Keuning}, {and} \bibinfo{person}{Johan Jeuring}.} \bibinfo{year}{2024}\natexlab{}.
\newblock \showarticletitle{"Let Them Try to Figure It Out First" - Reasons Why Experts (Do Not) Provide Feedback to Novice Programmers}. In \bibinfo{booktitle}{\emph{Proceedings of the 2024 on Innovation and Technology in Computer Science Education V. 1}} (Milan, Italy) \emph{(\bibinfo{series}{ITiCSE 2024})}. \bibinfo{publisher}{Association for Computing Machinery}, \bibinfo{address}{New York, NY, USA}, \bibinfo{pages}{38–44}.
\newblock
\showISBNx{9798400706004}
\urldef\tempurl%
\url{https://doi.org/10.1145/3649217.3653530}
\showDOI{\tempurl}


\bibitem[MacNeil et~al\mbox{.}(2023)]%
        {macneil2023experiences}
\bibfield{author}{\bibinfo{person}{Stephen MacNeil}, \bibinfo{person}{Andrew Tran}, \bibinfo{person}{Arto Hellas}, \bibinfo{person}{Joanne Kim}, \bibinfo{person}{Sami Sarsa}, \bibinfo{person}{Paul Denny}, \bibinfo{person}{Seth Bernstein}, {and} \bibinfo{person}{Juho Leinonen}.} \bibinfo{year}{2023}\natexlab{}.
\newblock \showarticletitle{Experiences from Using Code Explanations Generated by Large Language Models in a Web Software Development E-Book}. In \bibinfo{booktitle}{\emph{Proceedings of the 54th ACM Technical Symposium on Computer Science Education V. 1}} (Toronto ON, Canada) \emph{(\bibinfo{series}{SIGCSE 2023})}. \bibinfo{publisher}{Association for Computing Machinery}, \bibinfo{address}{New York, NY, USA}, \bibinfo{pages}{931–937}.
\newblock
\showISBNx{9781450394314}
\urldef\tempurl%
\url{https://doi.org/10.1145/3545945.3569785}
\showDOI{\tempurl}


\bibitem[Mahon et~al\mbox{.}(2023)]%
        {Mahon2023}
\bibfield{author}{\bibinfo{person}{Joyce Mahon}, \bibinfo{person}{Brian Mac~Namee}, {and} \bibinfo{person}{Brett~A. Becker}.} \bibinfo{year}{2023}\natexlab{}.
\newblock \showarticletitle{No More Pencils No More Books: Capabilities of Generative AI on Irish and UK Computer Science School Leaving Examinations}. In \bibinfo{booktitle}{\emph{Proceedings of the 2023 Conference on United Kingdom \& Ireland Computing Education Research}} (Swansea, Wales Uk) \emph{(\bibinfo{series}{UKICER '23})}. \bibinfo{publisher}{Association for Computing Machinery}, \bibinfo{address}{New York, NY, USA}, Article \bibinfo{articleno}{2}, \bibinfo{numpages}{7}~pages.
\newblock
\showISBNx{9798400708763}
\urldef\tempurl%
\url{https://doi.org/10.1145/3610969.3610982}
\showDOI{\tempurl}


\bibitem[Mailach et~al\mbox{.}(2024)]%
        {Mailach2024}
\bibfield{author}{\bibinfo{person}{Alina Mailach}, \bibinfo{person}{Dominik Gorgosch}, \bibinfo{person}{Norbert Siegmund}, {and} \bibinfo{person}{Janet Siegmund}.} \bibinfo{year}{2024}\natexlab{}.
\newblock \showarticletitle{“Ok Pal, We Have to Code That Now”: Interaction Patterns of Programming Beginners with a Conversational Chatbot}.
\newblock \bibinfo{journal}{\emph{Empirical Software Engineering (EMSE)}} (\bibinfo{year}{2024}), \bibinfo{pages}{to appear}.
\newblock


\bibitem[Marwan et~al\mbox{.}(2019a)]%
        {Marwan2019b}
\bibfield{author}{\bibinfo{person}{Samiha Marwan}, \bibinfo{person}{Joseph Jay~Williams}, {and} \bibinfo{person}{Thomas Price}.} \bibinfo{year}{2019}\natexlab{a}.
\newblock \showarticletitle{An Evaluation of the Impact of Automated Programming Hints on Performance and Learning}. In \bibinfo{booktitle}{\emph{Proceedings of the 2019 ACM Conference on International Computing Education Research}} (Toronto ON, Canada) \emph{(\bibinfo{series}{ICER '19})}. \bibinfo{publisher}{Association for Computing Machinery}, \bibinfo{address}{New York, NY, USA}, \bibinfo{pages}{61–70}.
\newblock
\showISBNx{9781450361859}
\urldef\tempurl%
\url{https://doi.org/10.1145/3291279.3339420}
\showDOI{\tempurl}


\bibitem[Marwan et~al\mbox{.}(2019b)]%
        {Marwan2019}
\bibfield{author}{\bibinfo{person}{Samiha Marwan}, \bibinfo{person}{Nicholas Lytle}, \bibinfo{person}{Joseph~Jay Williams}, {and} \bibinfo{person}{Thomas Price}.} \bibinfo{year}{2019}\natexlab{b}.
\newblock \showarticletitle{The Impact of Adding Textual Explanations to Next-Step Hints in a Novice Programming Environment}. In \bibinfo{booktitle}{\emph{Proceedings of the 2019 ACM Conference on Innovation and Technology in Computer Science Education}} (Aberdeen, Scotland Uk) \emph{(\bibinfo{series}{ITiCSE '19})}. \bibinfo{publisher}{Association for Computing Machinery}, \bibinfo{address}{New York, NY, USA}, \bibinfo{pages}{520–526}.
\newblock
\showISBNx{9781450368957}
\urldef\tempurl%
\url{https://doi.org/10.1145/3304221.3319759}
\showDOI{\tempurl}


\bibitem[Marwan and Price(2023)]%
        {Marwan2023}
\bibfield{author}{\bibinfo{person}{Samiha Marwan} {and} \bibinfo{person}{Thomas~W. Price}.} \bibinfo{year}{2023}\natexlab{}.
\newblock \showarticletitle{ISnap: Evolution and Evaluation of a Data-Driven Hint System for Block-Based Programming}.
\newblock \bibinfo{journal}{\emph{IEEE Trans. Learn. Technol.}} \bibinfo{volume}{16}, \bibinfo{number}{3.2} (\bibinfo{date}{June} \bibinfo{year}{2023}), \bibinfo{pages}{399–413}.
\newblock
\showISSN{1939-1382}
\urldef\tempurl%
\url{https://doi.org/10.1109/TLT.2022.3223577}
\showDOI{\tempurl}


\bibitem[McBroom et~al\mbox{.}(2021)]%
        {McBroom2021}
\bibfield{author}{\bibinfo{person}{Jessica McBroom}, \bibinfo{person}{Irena Koprinska}, {and} \bibinfo{person}{Kalina Yacef}.} \bibinfo{year}{2021}\natexlab{}.
\newblock \showarticletitle{A Survey of Automated Programming Hint Generation: The HINTS Framework}.
\newblock \bibinfo{journal}{\emph{ACM Comput. Surv.}} \bibinfo{volume}{54}, \bibinfo{number}{8}, Article \bibinfo{articleno}{172} (\bibinfo{date}{oct} \bibinfo{year}{2021}), \bibinfo{numpages}{27}~pages.
\newblock
\showISSN{0360-0300}
\urldef\tempurl%
\url{https://doi.org/10.1145/3469885}
\showDOI{\tempurl}


\bibitem[McLean et~al\mbox{.}(2015)]%
        {Mclean2015}
\bibfield{author}{\bibinfo{person}{Angela~J McLean}, \bibinfo{person}{Carol~H Bond}, {and} \bibinfo{person}{Helen~D Nicholson}.} \bibinfo{year}{2015}\natexlab{}.
\newblock \showarticletitle{An anatomy of feedback: a phenomenographic investigation of undergraduate students' conceptions of feedback}.
\newblock \bibinfo{journal}{\emph{Studies in Higher Education}} \bibinfo{volume}{40}, \bibinfo{number}{5} (\bibinfo{year}{2015}), \bibinfo{pages}{921--932}.
\newblock


\bibitem[Nam et~al\mbox{.}(2024)]%
        {Nam2024}
\bibfield{author}{\bibinfo{person}{Daye Nam}, \bibinfo{person}{Andrew Macvean}, \bibinfo{person}{Vincent Hellendoorn}, \bibinfo{person}{Bogdan Vasilescu}, {and} \bibinfo{person}{Brad Myers}.} \bibinfo{year}{2024}\natexlab{}.
\newblock \showarticletitle{Using an LLM to Help With Code Understanding}. In \bibinfo{booktitle}{\emph{Proceedings of the IEEE/ACM 46th International Conference on Software Engineering}} (Lisbon, Portugal) \emph{(\bibinfo{series}{ICSE '24})}. \bibinfo{publisher}{Association for Computing Machinery}, \bibinfo{address}{New York, NY, USA}, Article \bibinfo{articleno}{97}, \bibinfo{numpages}{13}~pages.
\newblock
\showISBNx{9798400702174}
\urldef\tempurl%
\url{https://doi.org/10.1145/3597503.3639187}
\showDOI{\tempurl}


\bibitem[Nguyen and Allan(2024)]%
        {Nguyen2024b}
\bibfield{author}{\bibinfo{person}{Ha Nguyen} {and} \bibinfo{person}{Vicki Allan}.} \bibinfo{year}{2024}\natexlab{}.
\newblock \showarticletitle{Using GPT-4 to Provide Tiered, Formative Code Feedback}. In \bibinfo{booktitle}{\emph{Proceedings of the 55th ACM Technical Symposium on Computer Science Education V. 1}} (Portland, OR, USA) \emph{(\bibinfo{series}{SIGCSE 2024})}. \bibinfo{publisher}{Association for Computing Machinery}, \bibinfo{address}{New York, NY, USA}, \bibinfo{pages}{958–964}.
\newblock
\showISBNx{9798400704239}
\urldef\tempurl%
\url{https://doi.org/10.1145/3626252.3630960}
\showDOI{\tempurl}


\bibitem[Nguyen et~al\mbox{.}(2024)]%
        {Nguyen2024}
\bibfield{author}{\bibinfo{person}{Sydney Nguyen}, \bibinfo{person}{Hannah~McLean Babe}, \bibinfo{person}{Yangtian Zi}, \bibinfo{person}{Arjun Guha}, \bibinfo{person}{Carolyn~Jane Anderson}, {and} \bibinfo{person}{Molly~Q Feldman}.} \bibinfo{year}{2024}\natexlab{}.
\newblock \showarticletitle{How Beginning Programmers and Code LLMs (Mis)read Each Other}. In \bibinfo{booktitle}{\emph{Proceedings of the 2024 CHI Conference on Human Factors in Computing Systems}} (Honolulu, HI, USA) \emph{(\bibinfo{series}{CHI '24})}. \bibinfo{publisher}{Association for Computing Machinery}, \bibinfo{address}{New York, NY, USA}, Article \bibinfo{articleno}{651}, \bibinfo{numpages}{26}~pages.
\newblock
\showISBNx{9798400703300}
\urldef\tempurl%
\url{https://doi.org/10.1145/3613904.3642706}
\showDOI{\tempurl}


\bibitem[Oberm\"{u}ller et~al\mbox{.}(2021)]%
        {Obermueller2021}
\bibfield{author}{\bibinfo{person}{Florian Oberm\"{u}ller}, \bibinfo{person}{Ute Heuer}, {and} \bibinfo{person}{Gordon Fraser}.} \bibinfo{year}{2021}\natexlab{}.
\newblock \showarticletitle{Guiding Next-Step Hint Generation Using Automated Tests}. In \bibinfo{booktitle}{\emph{Proceedings of the 26th ACM Conference on Innovation and Technology in Computer Science Education V. 1}} (Virtual Event, Germany) \emph{(\bibinfo{series}{ITiCSE '21})}. \bibinfo{publisher}{Association for Computing Machinery}, \bibinfo{address}{New York, NY, USA}, \bibinfo{pages}{220–226}.
\newblock
\showISBNx{9781450382144}
\urldef\tempurl%
\url{https://doi.org/10.1145/3430665.3456344}
\showDOI{\tempurl}


\bibitem[Pankiewicz and Baker(2024)]%
        {Pankiewicz2024}
\bibfield{author}{\bibinfo{person}{Maciej Pankiewicz} {and} \bibinfo{person}{Ryan~S. Baker}.} \bibinfo{year}{2024}\natexlab{}.
\newblock \showarticletitle{Navigating Compiler Errors with AI Assistance - A Study of GPT Hints in an Introductory Programming Course}. In \bibinfo{booktitle}{\emph{Proceedings of the 2024 on Innovation and Technology in Computer Science Education V. 1}} (Milan, Italy) \emph{(\bibinfo{series}{ITiCSE 2024})}. \bibinfo{publisher}{Association for Computing Machinery}, \bibinfo{address}{New York, NY, USA}, \bibinfo{pages}{94–100}.
\newblock
\showISBNx{9798400706004}
\urldef\tempurl%
\url{https://doi.org/10.1145/3649217.3653608}
\showDOI{\tempurl}


\bibitem[Phothilimthana and Sridhara(2017)]%
        {Phothilimthana2017}
\bibfield{author}{\bibinfo{person}{Phitchaya~Mangpo Phothilimthana} {and} \bibinfo{person}{Sumukh Sridhara}.} \bibinfo{year}{2017}\natexlab{}.
\newblock \showarticletitle{High-Coverage Hint Generation for Massive Courses: Do Automated Hints Help CS1 Students?}. In \bibinfo{booktitle}{\emph{Proceedings of the 2017 ACM Conference on Innovation and Technology in Computer Science Education}} (Bologna, Italy) \emph{(\bibinfo{series}{ITiCSE '17})}. \bibinfo{publisher}{Association for Computing Machinery}, \bibinfo{address}{New York, NY, USA}, \bibinfo{pages}{182–187}.
\newblock
\showISBNx{9781450347044}
\urldef\tempurl%
\url{https://doi.org/10.1145/3059009.3059058}
\showDOI{\tempurl}


\bibitem[Pollitt(2012)]%
        {Pollitt2012}
\bibfield{author}{\bibinfo{person}{Alastair Pollitt}.} \bibinfo{year}{2012}\natexlab{}.
\newblock \showarticletitle{The method of Adaptive Comparative Judgement}.
\newblock \bibinfo{journal}{\emph{Assessment in Education: Principles, Policy \& Practice}} \bibinfo{volume}{19}, \bibinfo{number}{3} (\bibinfo{year}{2012}), \bibinfo{pages}{281--300}.
\newblock
\urldef\tempurl%
\url{https://doi.org/10.1080/0969594X.2012.665354}
\showDOI{\tempurl}
\showeprint{https://doi.org/10.1080/0969594X.2012.665354}


\bibitem[Pozdniakov et~al\mbox{.}(2024)]%
        {pozdniakov2024large}
\bibfield{author}{\bibinfo{person}{Stanislav Pozdniakov}, \bibinfo{person}{Jonathan Brazil}, \bibinfo{person}{Solmaz Abdi}, \bibinfo{person}{Aneesha Bakharia}, \bibinfo{person}{Shazia Sadiq}, \bibinfo{person}{Dragan Gašević}, \bibinfo{person}{Paul Denny}, {and} \bibinfo{person}{Hassan Khosravi}.} \bibinfo{year}{2024}\natexlab{}.
\newblock \showarticletitle{Large language models meet user interfaces: The case of provisioning feedback}.
\newblock \bibinfo{journal}{\emph{Computers and Education: Artificial Intelligence}}  \bibinfo{volume}{7} (\bibinfo{year}{2024}), \bibinfo{pages}{100289}.
\newblock
\showISSN{2666-920X}
\urldef\tempurl%
\url{https://doi.org/10.1016/j.caeai.2024.100289}
\showDOI{\tempurl}


\bibitem[Prather et~al\mbox{.}(2023a)]%
        {Prather2023b}
\bibfield{author}{\bibinfo{person}{James Prather}, \bibinfo{person}{Paul Denny}, \bibinfo{person}{Brett~A. Becker}, \bibinfo{person}{Robert Nix}, \bibinfo{person}{Brent~N. Reeves}, \bibinfo{person}{Arisoa~S. Randrianasolo}, {and} \bibinfo{person}{Garrett Powell}.} \bibinfo{year}{2023}\natexlab{a}.
\newblock \showarticletitle{First Steps Towards Predicting the Readability of Programming Error Messages}. In \bibinfo{booktitle}{\emph{Proceedings of the 54th ACM Technical Symposium on Computer Science Education V. 1}} (Toronto ON, Canada) \emph{(\bibinfo{series}{SIGCSE 2023})}. \bibinfo{publisher}{Association for Computing Machinery}, \bibinfo{address}{New York, NY, USA}, \bibinfo{pages}{549–555}.
\newblock
\showISBNx{9781450394314}
\urldef\tempurl%
\url{https://doi.org/10.1145/3545945.3569791}
\showDOI{\tempurl}


\bibitem[Prather et~al\mbox{.}(2018)]%
        {Prather2018}
\bibfield{author}{\bibinfo{person}{James Prather}, \bibinfo{person}{Raymond Pettit}, \bibinfo{person}{Kayla McMurry}, \bibinfo{person}{Alani Peters}, \bibinfo{person}{John Homer}, {and} \bibinfo{person}{Maxine Cohen}.} \bibinfo{year}{2018}\natexlab{}.
\newblock \showarticletitle{Metacognitive Difficulties Faced by Novice Programmers in Automated Assessment Tools}. In \bibinfo{booktitle}{\emph{Proceedings of the 2018 ACM Conference on International Computing Education Research}} (Espoo, Finland) \emph{(\bibinfo{series}{ICER '18})}. \bibinfo{publisher}{Association for Computing Machinery}, \bibinfo{address}{New York, NY, USA}, \bibinfo{pages}{41–50}.
\newblock
\showISBNx{9781450356282}
\urldef\tempurl%
\url{https://doi.org/10.1145/3230977.3230981}
\showDOI{\tempurl}


\bibitem[Prather et~al\mbox{.}(2017)]%
        {Prather2017}
\bibfield{author}{\bibinfo{person}{James Prather}, \bibinfo{person}{Raymond Pettit}, \bibinfo{person}{Kayla~Holcomb McMurry}, \bibinfo{person}{Alani Peters}, \bibinfo{person}{John Homer}, \bibinfo{person}{Nevan Simone}, {and} \bibinfo{person}{Maxine Cohen}.} \bibinfo{year}{2017}\natexlab{}.
\newblock \showarticletitle{On Novices' Interaction with Compiler Error Messages: A Human Factors Approach}. In \bibinfo{booktitle}{\emph{Proceedings of the 2017 ACM Conference on International Computing Education Research}} (Tacoma, Washington, USA) \emph{(\bibinfo{series}{ICER '17})}. \bibinfo{publisher}{Association for Computing Machinery}, \bibinfo{address}{New York, NY, USA}, \bibinfo{pages}{74–82}.
\newblock
\showISBNx{9781450349680}
\urldef\tempurl%
\url{https://doi.org/10.1145/3105726.3106169}
\showDOI{\tempurl}


\bibitem[Prather et~al\mbox{.}(2023b)]%
        {Prather2023}
\bibfield{author}{\bibinfo{person}{James Prather}, \bibinfo{person}{Brent~N. Reeves}, \bibinfo{person}{Paul Denny}, \bibinfo{person}{Brett~A. Becker}, \bibinfo{person}{Juho Leinonen}, \bibinfo{person}{Andrew Luxton-Reilly}, \bibinfo{person}{Garrett Powell}, \bibinfo{person}{James Finnie-Ansley}, {and} \bibinfo{person}{Eddie~Antonio Santos}.} \bibinfo{year}{2023}\natexlab{b}.
\newblock \showarticletitle{“It’s Weird That it Knows What I Want”: Usability and Interactions with Copilot for Novice Programmers}.
\newblock \bibinfo{journal}{\emph{ACM Trans. Comput.-Hum. Interact.}} \bibinfo{volume}{31}, \bibinfo{number}{1}, Article \bibinfo{articleno}{4} (\bibinfo{date}{Nov.} \bibinfo{year}{2023}), \bibinfo{numpages}{31}~pages.
\newblock
\showISSN{1073-0516}
\urldef\tempurl%
\url{https://doi.org/10.1145/3617367}
\showDOI{\tempurl}


\bibitem[Prather et~al\mbox{.}(2024)]%
        {Prather2024}
\bibfield{author}{\bibinfo{person}{James Prather}, \bibinfo{person}{Brent~N Reeves}, \bibinfo{person}{Juho Leinonen}, \bibinfo{person}{Stephen MacNeil}, \bibinfo{person}{Arisoa~S Randrianasolo}, \bibinfo{person}{Brett~A. Becker}, \bibinfo{person}{Bailey Kimmel}, \bibinfo{person}{Jared Wright}, {and} \bibinfo{person}{Ben Briggs}.} \bibinfo{year}{2024}\natexlab{}.
\newblock \showarticletitle{The Widening Gap: The Benefits and Harms of Generative AI for Novice Programmers}. In \bibinfo{booktitle}{\emph{Proceedings of the 2024 ACM Conference on International Computing Education Research - Volume 1}} (Melbourne, VIC, Australia) \emph{(\bibinfo{series}{ICER '24})}. \bibinfo{publisher}{Association for Computing Machinery}, \bibinfo{address}{New York, NY, USA}, \bibinfo{pages}{469–486}.
\newblock
\showISBNx{9798400704758}
\urldef\tempurl%
\url{https://doi.org/10.1145/3632620.3671116}
\showDOI{\tempurl}


\bibitem[Price et~al\mbox{.}(2021)]%
        {Price2021}
\bibfield{author}{\bibinfo{person}{Thomas~W. Price}, \bibinfo{person}{Samiha Marwan}, {and} \bibinfo{person}{Joseph~Jay Williams}.} \bibinfo{year}{2021}\natexlab{}.
\newblock \showarticletitle{Exploring Design Choices in Data-Driven Hints for Python Programming Homework}. In \bibinfo{booktitle}{\emph{Proceedings of the Eighth ACM Conference on Learning @ Scale}} (Virtual Event, Germany) \emph{(\bibinfo{series}{L@S '21})}. \bibinfo{publisher}{Association for Computing Machinery}, \bibinfo{address}{New York, NY, USA}, \bibinfo{pages}{283–286}.
\newblock
\showISBNx{9781450382151}
\urldef\tempurl%
\url{https://doi.org/10.1145/3430895.3460159}
\showDOI{\tempurl}


\bibitem[Price et~al\mbox{.}(2020)]%
        {Price2020}
\bibfield{author}{\bibinfo{person}{Thomas~W. Price}, \bibinfo{person}{Samiha Marwan}, \bibinfo{person}{Michael Winters}, {and} \bibinfo{person}{Joseph~Jay Williams}.} \bibinfo{year}{2020}\natexlab{}.
\newblock \showarticletitle{An Evaluation of Data-Driven Programming Hints in a Classroom Setting}. In \bibinfo{booktitle}{\emph{Artificial Intelligence in Education}}, \bibfield{editor}{\bibinfo{person}{Ig~Ibert Bittencourt}, \bibinfo{person}{Mutlu Cukurova}, \bibinfo{person}{Kasia Muldner}, \bibinfo{person}{Rose Luckin}, {and} \bibinfo{person}{Eva Mill{\'a}n}} (Eds.). \bibinfo{publisher}{Springer International Publishing}, \bibinfo{address}{Cham}, \bibinfo{pages}{246--251}.
\newblock
\showISBNx{978-3-030-52240-7}


\bibitem[Raman and Kumar(2022)]%
        {Raman2022}
\bibfield{author}{\bibinfo{person}{Arun Raman} {and} \bibinfo{person}{Viraj Kumar}.} \bibinfo{year}{2022}\natexlab{}.
\newblock \showarticletitle{Programming Pedagogy and Assessment in the Era of AI/ML: A Position Paper}. In \bibinfo{booktitle}{\emph{Proceedings of the 15th Annual ACM India Compute Conference}} (Jaipur, India) \emph{(\bibinfo{series}{COMPUTE '22})}. \bibinfo{publisher}{Association for Computing Machinery}, \bibinfo{address}{New York, NY, USA}, \bibinfo{pages}{29–34}.
\newblock
\showISBNx{9781450397759}
\urldef\tempurl%
\url{https://doi.org/10.1145/3561833.3561843}
\showDOI{\tempurl}


\bibitem[Rietzschel et~al\mbox{.}(2006)]%
        {Rietzschel2006}
\bibfield{author}{\bibinfo{person}{Eric~F Rietzschel}, \bibinfo{person}{Bernard~A Nijstad}, {and} \bibinfo{person}{Wolfgang Stroebe}.} \bibinfo{year}{2006}\natexlab{}.
\newblock \showarticletitle{Productivity is not enough: A comparison of interactive and nominal brainstorming groups on idea generation and selection}.
\newblock \bibinfo{journal}{\emph{Journal of Experimental Social Psychology}} \bibinfo{volume}{42}, \bibinfo{number}{2} (\bibinfo{year}{2006}), \bibinfo{pages}{244--251}.
\newblock


\bibitem[Roest et~al\mbox{.}(2024)]%
        {Roest2024}
\bibfield{author}{\bibinfo{person}{Lianne Roest}, \bibinfo{person}{Hieke Keuning}, {and} \bibinfo{person}{Johan Jeuring}.} \bibinfo{year}{2024}\natexlab{}.
\newblock \showarticletitle{Next-Step Hint Generation for Introductory Programming Using Large Language Models}. In \bibinfo{booktitle}{\emph{Proceedings of the 26th Australasian Computing Education Conference}} (Sydney, NSW, Australia) \emph{(\bibinfo{series}{ACE '24})}. \bibinfo{publisher}{Association for Computing Machinery}, \bibinfo{address}{New York, NY, USA}, \bibinfo{pages}{144–153}.
\newblock
\showISBNx{9798400716195}
\urldef\tempurl%
\url{https://doi.org/10.1145/3636243.3636259}
\showDOI{\tempurl}


\bibitem[San~Verhavert and Maeyer(2019)]%
        {Verhavert2019}
\bibfield{author}{\bibinfo{person}{Vincent~Donche San~Verhavert, Renske~Bouwer} {and} \bibinfo{person}{Sven~De Maeyer}.} \bibinfo{year}{2019}\natexlab{}.
\newblock \showarticletitle{A meta-analysis on the reliability of comparative judgement}.
\newblock \bibinfo{journal}{\emph{Assessment in Education: Principles, Policy \& Practice}} \bibinfo{volume}{26}, \bibinfo{number}{5} (\bibinfo{year}{2019}), \bibinfo{pages}{541--562}.
\newblock
\urldef\tempurl%
\url{https://doi.org/10.1080/0969594X.2019.1602027}
\showDOI{\tempurl}
\showeprint{https://doi.org/10.1080/0969594X.2019.1602027}


\bibitem[Scholl and Kiesler(2024)]%
        {Scholl2024}
\bibfield{author}{\bibinfo{person}{Andreas Scholl} {and} \bibinfo{person}{Natalie Kiesler}.} \bibinfo{year}{2024}\natexlab{}.
\newblock \bibinfo{title}{How Novice Programmers Use and Experience ChatGPT when Solving Programming Exercises in an Introductory Course}.
\newblock
\newblock
\showeprint[arxiv]{2407.20792}~[cs.AI]
\urldef\tempurl%
\url{https://arxiv.org/abs/2407.20792}
\showURL{%
\tempurl}


\bibitem[Sheard et~al\mbox{.}(2024)]%
        {Sheard2024}
\bibfield{author}{\bibinfo{person}{Judy Sheard}, \bibinfo{person}{Paul Denny}, \bibinfo{person}{Arto Hellas}, \bibinfo{person}{Juho Leinonen}, \bibinfo{person}{Lauri Malmi}, {and} \bibinfo{person}{Simon}.} \bibinfo{year}{2024}\natexlab{}.
\newblock \showarticletitle{Instructor Perceptions of AI Code Generation Tools - A Multi-Institutional Interview Study}. In \bibinfo{booktitle}{\emph{Proceedings of the 55th ACM Technical Symposium on Computer Science Education V. 1}} (Portland, OR, USA) \emph{(\bibinfo{series}{SIGCSE 2024})}. \bibinfo{publisher}{Association for Computing Machinery}, \bibinfo{address}{New York, NY, USA}, \bibinfo{pages}{1223–1229}.
\newblock
\showISBNx{9798400704239}
\urldef\tempurl%
\url{https://doi.org/10.1145/3626252.3630880}
\showDOI{\tempurl}


\bibitem[Sheese et~al\mbox{.}(2024)]%
        {Sheese2024}
\bibfield{author}{\bibinfo{person}{Brad Sheese}, \bibinfo{person}{Mark Liffiton}, \bibinfo{person}{Jaromir Savelka}, {and} \bibinfo{person}{Paul Denny}.} \bibinfo{year}{2024}\natexlab{}.
\newblock \showarticletitle{Patterns of Student Help-Seeking When Using a Large Language Model-Powered Programming Assistant}. In \bibinfo{booktitle}{\emph{Proceedings of the 26th Australasian Computing Education Conference}} (Sydney, NSW, Australia) \emph{(\bibinfo{series}{ACE '24})}. \bibinfo{publisher}{Association for Computing Machinery}, \bibinfo{address}{New York, NY, USA}, \bibinfo{pages}{49–57}.
\newblock
\showISBNx{9798400716195}
\urldef\tempurl%
\url{https://doi.org/10.1145/3636243.3636249}
\showDOI{\tempurl}


\bibitem[Smith and Rixner(2019)]%
        {Smith2019}
\bibfield{author}{\bibinfo{person}{Rebecca Smith} {and} \bibinfo{person}{Scott Rixner}.} \bibinfo{year}{2019}\natexlab{}.
\newblock \showarticletitle{The Error Landscape: Characterizing the Mistakes of Novice Programmers}. In \bibinfo{booktitle}{\emph{Proceedings of the 50th ACM Technical Symposium on Computer Science Education}} (Minneapolis, MN, USA) \emph{(\bibinfo{series}{SIGCSE '19})}. \bibinfo{publisher}{Association for Computing Machinery}, \bibinfo{address}{New York, NY, USA}, \bibinfo{pages}{538–544}.
\newblock
\showISBNx{9781450358903}
\urldef\tempurl%
\url{https://doi.org/10.1145/3287324.3287394}
\showDOI{\tempurl}


\bibitem[Suzuki et~al\mbox{.}(2017)]%
        {Suzuki2017}
\bibfield{author}{\bibinfo{person}{Ryo Suzuki}, \bibinfo{person}{Gustavo Soares}, \bibinfo{person}{Elena Glassman}, \bibinfo{person}{Andrew Head}, \bibinfo{person}{Loris D'Antoni}, {and} \bibinfo{person}{Bj\"{o}rn Hartmann}.} \bibinfo{year}{2017}\natexlab{}.
\newblock \showarticletitle{Exploring the Design Space of Automatically Synthesized Hints for Introductory Programming Assignments}. In \bibinfo{booktitle}{\emph{Proceedings of the 2017 CHI Conference Extended Abstracts on Human Factors in Computing Systems}} (Denver, Colorado, USA) \emph{(\bibinfo{series}{CHI EA '17})}. \bibinfo{publisher}{Association for Computing Machinery}, \bibinfo{address}{New York, NY, USA}, \bibinfo{pages}{2951–2958}.
\newblock
\showISBNx{9781450346566}
\urldef\tempurl%
\url{https://doi.org/10.1145/3027063.3053187}
\showDOI{\tempurl}


\bibitem[Whalley et~al\mbox{.}(2023)]%
        {Whalley2023}
\bibfield{author}{\bibinfo{person}{Jacqueline Whalley}, \bibinfo{person}{Amber Settle}, {and} \bibinfo{person}{Andrew Luxton-Reilly}.} \bibinfo{year}{2023}\natexlab{}.
\newblock \showarticletitle{A Think-Aloud Study of Novice Debugging}.
\newblock \bibinfo{journal}{\emph{ACM Trans. Comput. Educ.}} \bibinfo{volume}{23}, \bibinfo{number}{2}, Article \bibinfo{articleno}{28} (\bibinfo{date}{jun} \bibinfo{year}{2023}), \bibinfo{numpages}{38}~pages.
\newblock
\urldef\tempurl%
\url{https://doi.org/10.1145/3589004}
\showDOI{\tempurl}


\bibitem[Wiggins et~al\mbox{.}(2021)]%
        {Wiggins2021}
\bibfield{author}{\bibinfo{person}{Joseph~B. Wiggins}, \bibinfo{person}{Fahmid~M. Fahid}, \bibinfo{person}{Andrew Emerson}, \bibinfo{person}{Madeline Hinckle}, \bibinfo{person}{Andy Smith}, \bibinfo{person}{Kristy~Elizabeth Boyer}, \bibinfo{person}{Bradford Mott}, \bibinfo{person}{Eric Wiebe}, {and} \bibinfo{person}{James Lester}.} \bibinfo{year}{2021}\natexlab{}.
\newblock \showarticletitle{Exploring Novice Programmers' Hint Requests in an Intelligent Block-Based Coding Environment}. In \bibinfo{booktitle}{\emph{Proceedings of the 52nd ACM Technical Symposium on Computer Science Education}} (Virtual Event, USA) \emph{(\bibinfo{series}{SIGCSE '21})}. \bibinfo{publisher}{Association for Computing Machinery}, \bibinfo{address}{New York, NY, USA}, \bibinfo{pages}{52–58}.
\newblock
\showISBNx{9781450380621}
\urldef\tempurl%
\url{https://doi.org/10.1145/3408877.3432538}
\showDOI{\tempurl}


\bibitem[Xiao et~al\mbox{.}(2024)]%
        {Xiao2024}
\bibfield{author}{\bibinfo{person}{Ruiwei Xiao}, \bibinfo{person}{Xinying Hou}, {and} \bibinfo{person}{John Stamper}.} \bibinfo{year}{2024}\natexlab{}.
\newblock \showarticletitle{Exploring How Multiple Levels of GPT-Generated Programming Hints Support or Disappoint Novices}. In \bibinfo{booktitle}{\emph{Extended Abstracts of the CHI Conference on Human Factors in Computing Systems}} (Honolulu, HI, USA) \emph{(\bibinfo{series}{CHI EA '24})}. \bibinfo{publisher}{Association for Computing Machinery}, \bibinfo{address}{New York, NY, USA}, Article \bibinfo{articleno}{142}, \bibinfo{numpages}{10}~pages.
\newblock
\showISBNx{9798400703317}
\urldef\tempurl%
\url{https://doi.org/10.1145/3613905.3650937}
\showDOI{\tempurl}


\bibitem[Xue et~al\mbox{.}(2024)]%
        {Xue2024}
\bibfield{author}{\bibinfo{person}{Yuankai Xue}, \bibinfo{person}{Hanlin Chen}, \bibinfo{person}{Gina~R. Bai}, \bibinfo{person}{Robert Tairas}, {and} \bibinfo{person}{Yu Huang}.} \bibinfo{year}{2024}\natexlab{}.
\newblock \showarticletitle{Does ChatGPT Help With Introductory Programming?An Experiment of Students Using ChatGPT in CS1}. In \bibinfo{booktitle}{\emph{Proceedings of the 46th International Conference on Software Engineering: Software Engineering Education and Training}} (Lisbon, Portugal) \emph{(\bibinfo{series}{ICSE-SEET '24})}. \bibinfo{publisher}{Association for Computing Machinery}, \bibinfo{address}{New York, NY, USA}, \bibinfo{pages}{331–341}.
\newblock
\showISBNx{9798400704987}
\urldef\tempurl%
\url{https://doi.org/10.1145/3639474.3640076}
\showDOI{\tempurl}


\bibitem[Zamfirescu-Pereira et~al\mbox{.}(2023)]%
        {Zamfirescu2023}
\bibfield{author}{\bibinfo{person}{J.D. Zamfirescu-Pereira}, \bibinfo{person}{Richmond~Y. Wong}, \bibinfo{person}{Bjoern Hartmann}, {and} \bibinfo{person}{Qian Yang}.} \bibinfo{year}{2023}\natexlab{}.
\newblock \showarticletitle{Why Johnny Can’t Prompt: How Non-AI Experts Try (and Fail) to Design LLM Prompts}. In \bibinfo{booktitle}{\emph{Proceedings of the 2023 CHI Conference on Human Factors in Computing Systems}} (Hamburg, Germany) \emph{(\bibinfo{series}{CHI '23})}. \bibinfo{publisher}{Association for Computing Machinery}, \bibinfo{address}{New York, NY, USA}, Article \bibinfo{articleno}{437}, \bibinfo{numpages}{21}~pages.
\newblock
\showISBNx{9781450394215}
\urldef\tempurl%
\url{https://doi.org/10.1145/3544548.3581388}
\showDOI{\tempurl}


\end{thebibliography}

\end{document}